\documentclass[final,leqno]{siamltex}
\usepackage{times}
\usepackage[dvips,letterpaper=true,colorlinks=true,linkcolor=red,filecolor=green,citecolor=red,pdfpagemode=None]{hyperref}

\usepackage[]{amsmath,amssymb,epsfig}
\usepackage{subfigure}
\usepackage{graphicx}

\title{Extracting spatial information from networks with low-order eigenvectors}
\author{
Mihai Cucuringu%
        \footnotemark[2]
\and Vincent D. Blondel
       \footnotemark[3]
\and Paul Van Dooren%
        \footnotemark[3]
}

\begin{document}
\maketitle
\renewcommand{\thefootnote}{\fnsymbol{footnote}}
\footnotetext[2]{Program in Applied and Computational Mathematics (PACM), Princeton University, Fine Hall, Washington Road, Princeton NJ 08544-1000 USA, email: mcucurin@math.princeton.edu}
\footnotetext[3]{Department of Applied Mathematics, Universit\'e catholique de Louvain, B-1348 Louvain-la-Neuve, Belgium, emails: \{vincent.blondel, paul.vandooren\}@uclouvain.be }

\begin{abstract}
We consider the problem of inferring meaningful spatial information in networks from incomplete information on the connection intensity between the nodes of the network. We consider two spatially distributed networks: a population migration flow network within the US, and a network of mobile phone calls between cities in Belgium. For both networks we use the eigenvectors of the Laplacian matrix constructed from the link intensities to obtain informative visualizations and capture natural geographical subdivisions.
We observe that some low order eigenvectors localize very well and seem to reveal small geographically cohesive regions that match remarkably well with  political and administrative boundaries. We discuss possible explanations for this observation by describing diffusion maps and localized eigenfunctions. In addition, we discuss a possible connection with the weighted graph cut problem, and provide numerical evidence supporting the idea that lower order eigenvectors point out local cuts in the network. However, we do not provide a formal and rigorous justification for our observations.
\end{abstract}
\pagestyle{myheadings}\thispagestyle{plain}
\markboth{Mihai Cucuringu, Paul Van Dooren, Vincent D. Blondel}{Extracting spatial information  from networks with low-order eigenvectors}

\section{Introduction}
\label{intro}

Extensive  research over the last decades has greatly increased our understanding of the topology and the spatial distribution of many social, biological and technological networks. This paper considers the problem of inferring meaningful spatial and structural information from incomplete data sets  of pairwise interactions between nodes in a network.

The way people interact in many aspects of everyday life often reflect surprisingly well geopolitical boundaries. This inhomogeneity of connections in networks leads to  natural divisions, and identifying such divisions can provide valuable insight into how interactions in a network are influenced by its topology. The problem of finding the so-called network communities, i.e., groups of tightly connected nodes, has been extensively studied in recent years and many community detection algorithms exist with different levels of success \cite{fortunato}. In this paper, we consider two particular networks:  a county-to-county migration network constructed from 1995-2000 US Census data, and a city-to-city communication network built from mobile phone data over a six months period in Belgium. Communities in these networks emerge naturally and are revealed, often at different scales, by the eigenvectors of a normalized matrix constructed from the weighted adjacency matrix of the network. We discuss possible explanations for this observation by describing diffusion maps and localized eigenfunctions.

In the remaining part of this introduction we report on some related contributions that deal with communities in networks and spectrum of matrices. However, in none of these contributions we were able to find an explanation of why low order eigenvectors localize so well and seem to identify meaningful geographical boundaries.

One example of a recent study that is related to our work both in terms of the technique and  end goal, is a paper by Ratti et al. \cite{britain}. Starting from measures of the communication intensities between counties in the UK, the authors propose a spectral modularity\footnote{Many popular methods for community detection in networks are based on the optimization of the modularity function, a measure of the quality of a network partition into communities.} optimization algorithm that partitions the country into small non-overlapping geographically cohesive regions that correspond remarkably well with administrative regions.

In  \cite{shimalik}, the authors Shi and Malik develop a spectral-based algorithm that solves the perceptual grouping problem in computer vision by treating the task of image segmentation as a graph partitioning problem.  Their approach is to segment the graph by introducing a new global criterion called normalized cut, that measures not just the dissimilarity between different groups but also the total similarity within the groups themselves. They successfully extract global impressions of a scene and provide a hierarchical description of it.

In another recent paper \cite{eigenplaces}, the authors connect mobile data from Telecom Italia Mobile to a series of human activities derived from data on commercial premises advertised through the Italian version of ``Yellow Pages''. The eigendecomposition of a specific correlation matrix provides a top eigenvector which clearly indicates a common underlying pattern to mobile phone usage in Rome, while the second and third eigenvectors indicate spatial variation that is very suggestive of temporally-related and activity-related patterns.

Another line of work where lower order eigenvectors provide useful information
comes from the community detection literature. Newman \cite{Newman06} shows that the modularity of a network can be expressed in terms of the top eigenvalues and eigenvectors of  a matrix called the modularity matrix, which plays a role in the maximization of the modularity equivalent to that played by the Laplacian in standard spectral partitioning. In related work, Richardson et al. \cite{tripart} extend previously available methods for spectral optimization of modularity by introducing a computationally efficient algorithm for spectral tripartitioning of a network using the top two eigenvectors of the modularity matrix.

Recent work \cite{expert}, co-authored by one of the authors of this paper, investigates the constraints imposed by space on the network topology, and focuses on community detection by proposing a modularity function adapted to spatial networks. The proposed methods were tested on a large mobile phone network and computer-generated benchmarks, and showed that it is possible to factor out the effect of space in order to reveal more clearly any hidden structural similarities between the nodes.

Finally, we point out a recent paper of Onnela et al. \cite{onnela} who investigate social networks of individuals whose most frequent geographic locations are known. The authors classify the members into groups using community detection algorithms, and explore the relationship between their topological and geographic positions.

This paper is organized as follows: Section \ref{diffmaps} is an introduction to the diffusion map technique and some of its underlying theory.  Section \ref{USSection} contains the results of numerical simulations in which we applied diffusion maps and eigenvector colorings to the US migration data set. In Section \ref{BelgiumSection} we present the outcome of similar experiments on the Belgium mobile phone data set. In Section \ref{localization}, we explore the connection with localized eigenfunctions, a phenomenon observed before in the mathematics and physics community. Finally, the last section is a summary and a discussion of possible extensions of our approach and its usefulness in other applications.

\section{Diffusion Maps and Eigenvector Colorings}
\label{diffmaps}

This section is a brief introduction to the diffusion maps literature and references therein.
We also clarify the notion of eigenvector localizations and eigenvector coloring, that we use in subsequent sections. Diffusion maps were introduced in S. Lafon's  Ph.D. Thesis \cite{dif2} in 2004 as a dimensionality reduction tool, and connected data analysis and clustering techniques based on eigenvectors of similarity matrices with the geometric structure of non-linear manifolds.
In recent years, diffusion maps have gained a lot of popularity.
A nonexhaustive list of references to its underlying theory and applications  includes \cite{dif5,dif4,dif1,dif3,dif2}. Often called Laplacian eigenmaps, these manifold learning techniques identify significant variables that live in a lower dimensional space, while preserving the local proximity between data points. Consider a set of $N$ points $ V = \{ x_1, x_2, \ldots, x_N \} $ in an $n$-dimensional space $\mathbb{R}^n$, where each point (typically) characterizes an image (or an audio stream, text string, etc.). If two images $x_i$ and $x_j$ are similar, then $||x_i - x_j||$ is small.
A popular measure of similarity between points in $\mathbb{R}^n$ is defined using the Gaussian kernel
$   w_{ij} = e^{ - ||x_i - x_j||^2/\epsilon} $,
for some constant $\epsilon$, so that the closer $x_i$ is from $x_j$, the larger $w_{ij}$. The matrix $W=(w_{ij})_{1 \leq i,j \leq N}$ is symmetric and has positive coefficients. To normalize $W$, we define the diagonal matrix $D$, with $D_{ii} = \sum_{j=1}^{N} w_{ij}$ and define $A$ by
\begin{equation}
A = D^{-1}W
\label{ALap}
\nonumber
\end{equation} 
such that every row of $A$ sums to 1.

Next, one may also define the symmetric matrix $ S =  D^{-1/2} W D^{-1/2}$, which can also be written as $S=D^{1/2}AD^{-1/2}$ and hence is similar to $A$. As a symmetric matrix, $S$ has an orthogonal basis of eigenvectors $v_0, v_1,\ldots, v_{N-1}$ associated to the $N$ real ordered eigenvalues $1= \lambda_0 \geq \lambda_1 \geq \ldots  \geq \lambda_{N-1}$. If we decompose $S$ as $S=V \Lambda V^{T}$ with $V V^{T} = V^{T} V = I$ and $\Lambda = Diag(\lambda_0, \lambda_1, \ldots ,\lambda_{N-1})$, then $A$ becomes
$ A = \Psi \Lambda \Phi^{T} $
where $\Psi = D^{-1/2} V$ and $\Phi =  D^{1/2} V$. Therefore $ A\Psi=\Psi\Lambda$ and the columns of $\Psi$ form  a $D$-orthogonal basis of eigenvectors  with columns $\psi_0, \psi_1, \ldots ,\psi_{N-1}$ (i.e.  $ \left\langle { \psi_i, D \psi_j} \right\rangle  = 0,  \forall i \neq j$) associated to the $N$ real eigenvalues $\lambda_0, \lambda_1, \ldots ,\lambda_{N-1}$ such that $A \psi_i = \lambda_i \psi_i$, for $i=0,1, \ldots N-1$. Also, $\Phi ^{T} A  = \Lambda \Phi^T$ implies that the columns of $\Phi$ are left eigenvectors of $A$, which we denote by $\phi_0, \phi_1, \ldots, \phi_{N-1}$. Since
$\Phi^T\Psi=I$, it follows that the vectors $\phi_i$ and $\psi_j$ are bi-orthonormal
	$ \left\langle {\phi_i, \psi_j} \right\rangle  = \delta_{i,j}$.

Note that since $A$ is a row-stochastic matrix, $\lambda_0=1$ and $\psi_0 = (1, 1, \ldots, 1)^{T}$, and we disregard this trivial eigenvalue/eigenvector pair as irrelevant. Using the  stochasticity  of $A$, we can interpret it as a random walk matrix on a weighted graph $ G = (V,E,W) $, where the set of nodes consists of the points $x_i$, and there is an edge between nodes $i$ and $j$ if and ony if $w_{ij}>0$. Taking this perspective, $A_{ij}$ denotes the transition probability from point $x_i$ to $x_j$ in one step time $\Delta t = \epsilon $
\begin{equation}
Pr\{ x(t+\epsilon) = x_j  | x(t)=x_i \} = A_{ij}.
\nonumber
\end{equation}
The parameter $\epsilon$ can now be interpreted in two ways. On the one hand, it is the squared radius of the neighborhood used to infer local geometric and density information, in particular $w_{ij}$ is $O(1)$ when $x_i$ and $x_j$ are in a ball of radius $\sqrt{\epsilon}$, but it is exponentially small for points that are more than $\sqrt{\epsilon}$ apart. On the other hand, $\epsilon$ represents the the discrete time step at which the random walk jumps from one point to another.

Interpreting the eigenvectors as functions over our data set, the \emph{diffusion map} (also called \emph{Laplacian eigenmap)} maps points from the original space to the first $k$ eigenvectors, $\mathcal{L} : V  \mapsto \mathbb{R}^k $, is defined as
\begin{equation}
\begin{aligned}
  \mathcal{L}_t(x_j) = (\lambda_1^t \psi_1(j), \lambda_2^t \psi_2(j), \ldots ,\lambda_k^t \psi_k(j))
\end{aligned}
\label{difmap}
\end{equation}
where the meaning of the exponent $t$ will be made clear in what follows.


Using the left and right eigenvectors denoted earlier, we now write the entries of $A$ as $A_{ij} = \sum_{r=0}^{N-1} \lambda_r \phi_r(i) \psi_r(j)$, and note that $A_{ij}^t = \sum_{r=0}^{N-1} \lambda_r^t \phi_r(i) \psi_r(j)$. However, recall that the probability distribution of a random walk landing at location $x_j$ after exactly $t$ steps, given that is starts at point $x_i$ is precisely given by the expression $A_{ij}^t = Pr\{ x(t) = x_j  | x(0)=x_i \}$.  Given the random walk interpretation, it is natural to quantify the similarity between two points according to the evolution of their probability distributions
$$ D_t^2(i,j) = \sum_{k=1}^{N} (A_{ik}^{t} - A_{jk}^{t})^2 \frac{1}{d_k}, $$
where the weight $\frac{1}{d_k}$ takes into account the empirical local density of the points by giving larger weight to the vertices of lower degree. Since $D_t(i,j)$ naturally depends on the random walk on the graph, it is denoted as the \textit{diffusion distance} at time $t$. In the diffusion map introduced above, it is a matter of choice to tune the parameter $t$ corresponding to the number of time steps of the random walk. Note that we used $t=1$ in the diffusion maps embeddings throughout our simulations, and that using different values of $t$ corresponds to rescaling the axis. The Euclidean distance between two points in the diffusion map space introduced in (\ref{difmap}) is given by
\begin{equation}
 || \mathcal{L}(x_i)  - \mathcal{L}(x_j) || ^ 2 = \sum_{r=1}^{N-1} \left( \lambda_r^t \psi_r(i) - \lambda_r^t \psi_r(j) \right) ^2.	
\label{eucdif}
\end{equation}
Notice that the first eigenvalue $\lambda_0$ does not enter this expression, since it cancels out. Moreover, as shown in \cite{dif6}, the expression (\ref{eucdif}) equals the diffusion distance $D_t^2(i,j)$, when $k=N-1$, i.e., when all $N-1$ eigenvectors are considered. For ease of visualization, we used the top $k=2$ eigenvectors for the projections shown in Figures \ref{fig:census_2D_dif}, \ref{fig:census_2D_dif_NordSud} and \ref{fig:mob_2D_dif}.


Finally, we denote by $\mathcal{C}_i$ the coloring of the $N$ data points given by the eigenvector $\psi_{i}$, where the color of point $x_k \in V$ is given by the $j$-th entry in $\psi_{i}$, i.e.
\begin{equation}
\mathcal{C}_i(x_k) = \psi_{i}(k), \text{ for all } i=1, \ldots, N \text{ and } k=0, \ldots, N-1.
\label{coloreq}
\nonumber
\end{equation}
We refer to $\mathcal{C}_i$ as an \textit{eigenvector coloring}\footnote{Not to be confused with the ``coloring" terminology from graph theory, where the colors are integers.} of order $i$. The top left plot in Figure \ref{fig:US_K1_top18} shows the eigenvector coloring of order $k=1$, together with the associated colorbar. In practice,  only the first $k$ eigenvectors are used in the diffusion map introduced in (\ref{difmap}), with $ k << N-1$ chosen such that $\lambda_1^t \geq \lambda_2^t \ldots \geq \lambda_k^t > \delta$ but $\lambda_{k+1}^t < \delta$, where $\delta$ is a chosen tolerance. Typically, only the top few eigenvectors of $A$ are expected to contain meaningful information, but as illustrated by the eigenvector colorings shown in this paper, one can extract relevant information from eigenvectors of much lower order.  The phenomenon of \textit{eigenvector localization} occurs when most of the components of an eigenvector are zero or close to zero, and almost all the mass is localized on a relatively small subset of nodes. On the contrary, delocalized eigenvectors have most of their components small and of roughly the same magnitude. Furthermore, note there is no issue with the fact that the eigenvectors are defined up to a scalar. Since each of them is normalized and real, we can just consider eigenvectors of different sign, however this can only reverse the color map used, and does not change the localization phenomenon.

\section{US Census Migration Data}
\label{USSection}

\begin{figure}[h!t]
\centering
\subfigure[Map of USA, colored by longitude]{
\includegraphics[width=0.47\textwidth]{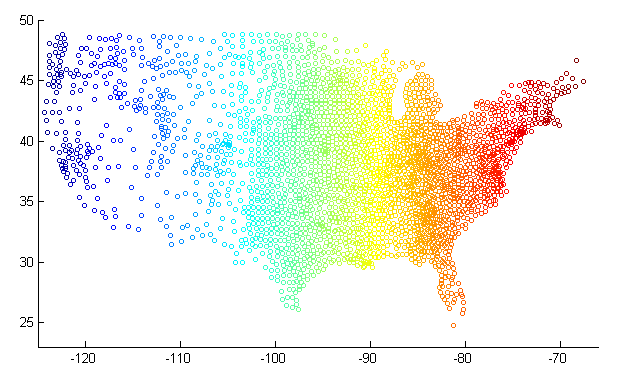}}
\subfigure[ Kernel $W^{(1)}_{ij} = \frac{ M_{ij}^2 }{P_i P_j}$]{
\includegraphics[width=0.47\textwidth]{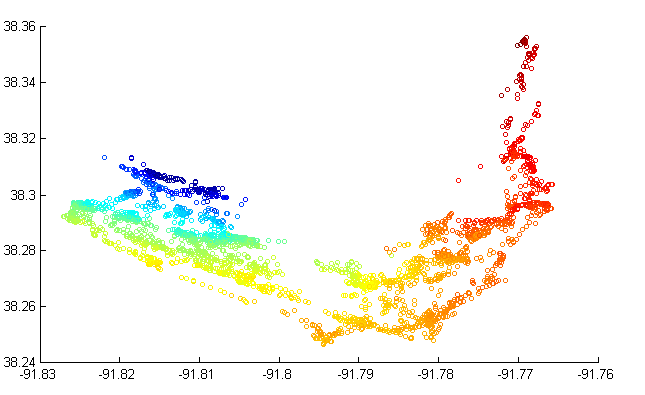}}
\subfigure[ Kernel $W^{(2)}_{ij} = \frac{ M_{ij} }{P_i + P_j}$]{
\includegraphics[width=0.47\textwidth]{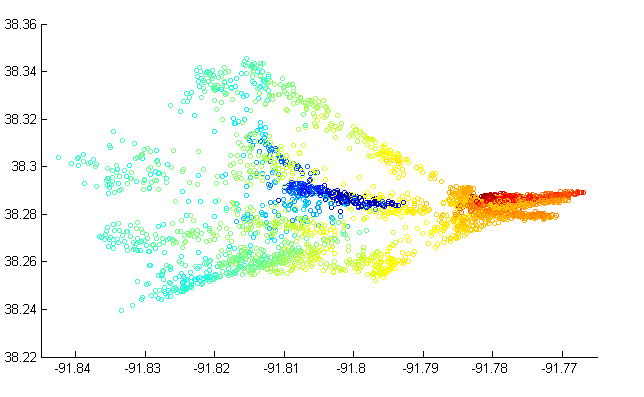}}
\subfigure[ Kernel $W^{(3)}_{ij} =  \frac{ M_{ij}}{P_i P_j}$]{
\includegraphics[width=0.47\textwidth]{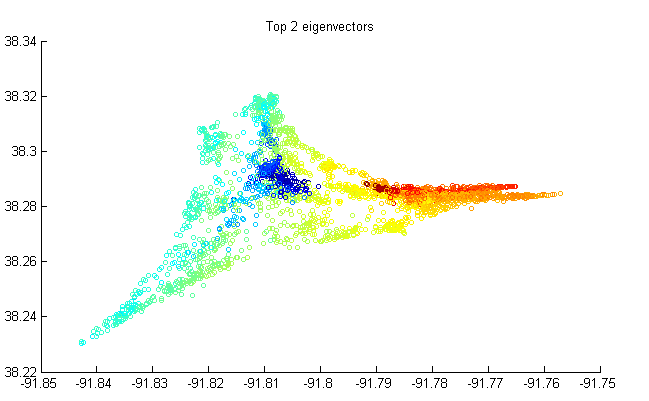}}
\caption[x]{Diffusion map reconstructions from the top two eigenvectors, for various kernels, with nodes colored by longitude.}
\label{fig:census_2D_dif}
\end{figure}

\begin{figure}[h!t]
\centering
\subfigure[Map of USA, colored by latitude]{
\includegraphics[width=0.48\textwidth]{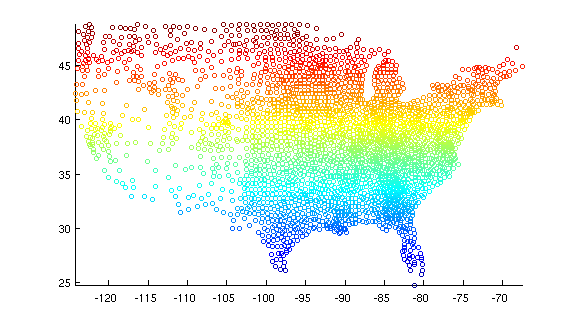}}
\subfigure[ Kernel $W^{(1)}_{ij} = \frac{ M_{ij}^2 }{P_i P_j}$]{
\includegraphics[width=0.48\textwidth]{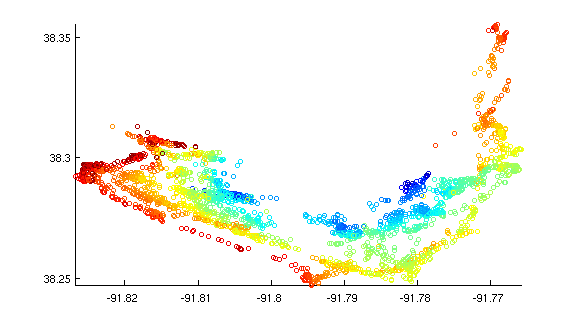}}
\subfigure[ Kernel $W^{(2)}_{ij} = \frac{ M_{ij} }{P_i + P_j}$]{
\includegraphics[width=0.48\textwidth]{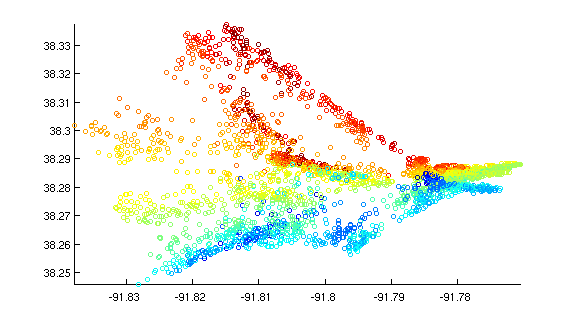}}
\subfigure[ Kernel $W^{(3)}_{ij} = 5500 \frac{ M_{ij}}{P_i P_j}$]{
\includegraphics[width=0.48\textwidth]{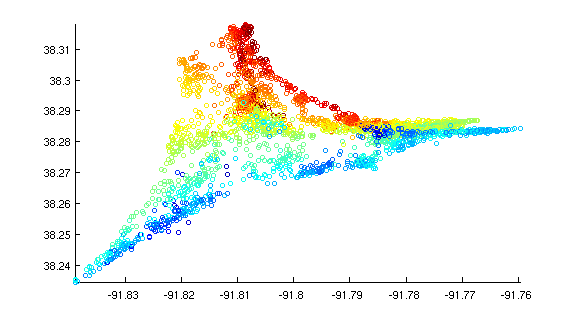}}
\caption[x]{Diffusion map reconstructions from the top two eigenvectors, for various kernels, with nodes colored by latitude.}
\label{fig:census_2D_dif_NordSud}
\end{figure}

\begin{figure}[h!t]
\centering
\subfigure[  $ A = D^{-1} W^{(1)}$]{
\includegraphics[width=0.31\textwidth]{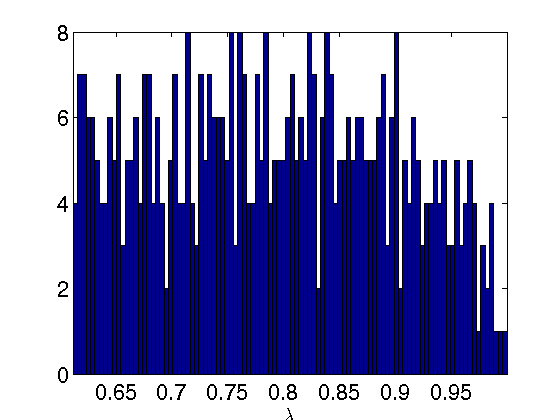}}
\subfigure[  $ A = D^{-1} W^{(2)}$]{
\includegraphics[width=0.31\textwidth]{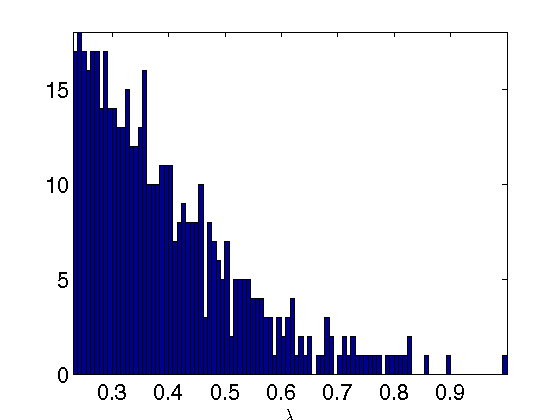}}
\subfigure[  $ A = D^{-1} W^{(3)}$]{
\includegraphics[width=0.31\textwidth]{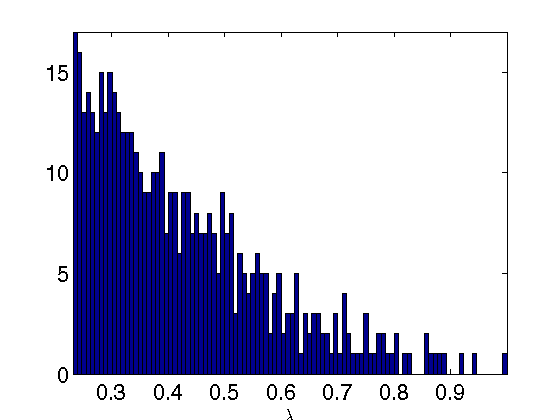}}
\caption[x]{Histogram of the top 500 eigenvalues of matrix $A$ for different kernels.}
\label{fig:USspectra}
\end{figure}

We apply the diffusion map technique to the 2000 US Census that reports the number of people that migrated from every county to every other county in the US during the 1995-2000 time frame
\cite{census,census_rep}. We denote by $M = (M_{ij})_{1 \leq i,j \leq N}$ the total number of people that migrated between county $i$ and county $j$ (so $M_{ij} = M_{ji}$), where $N=3107$  denotes the number of counties in mainland US. We let $P_i$ denote the population of county $i$. Figures \ref{fig:census_2D_dif} and \ref{fig:census_2D_dif_NordSud}  show the results of the diffusion map technique for longitude and  latitude colorings when the following kernels are used: $W^{(1)}_{ij} = \frac{ M_{ij}^2 }{P_i P_j}$, $W^{(2)}_{ij} = \frac{ M_{ij} }{P_i + P_j}$, and $W^{(3)}_{ij} = 5500 \frac{ M_{ij}}{P_i P_j}$.
The diffusion map resulting from these kernels place the Midwest closer to the west coast (Figure \ref{fig:census_2D_dif}), but further from the east coast. Similarly, the colorings based on latitude reveal the north-south separation.
Note that kernel $W^{(1)}$ does a better job at separating the east and west coasts, Figure \ref{fig:census_2D_dif} (b), while kernel  $W^{(2)}$ highlights best the separation between north and south as shown in Figure \ref{fig:census_2D_dif_NordSud} (c). Figure \ref{fig:USspectra} shows the histogram of the top $500$ eigenvalues of the diffusion matrix $A$, when different kernels are used.

\begin{figure}[h!t]
\begin{center}
\includegraphics[width=0.32 \textwidth]{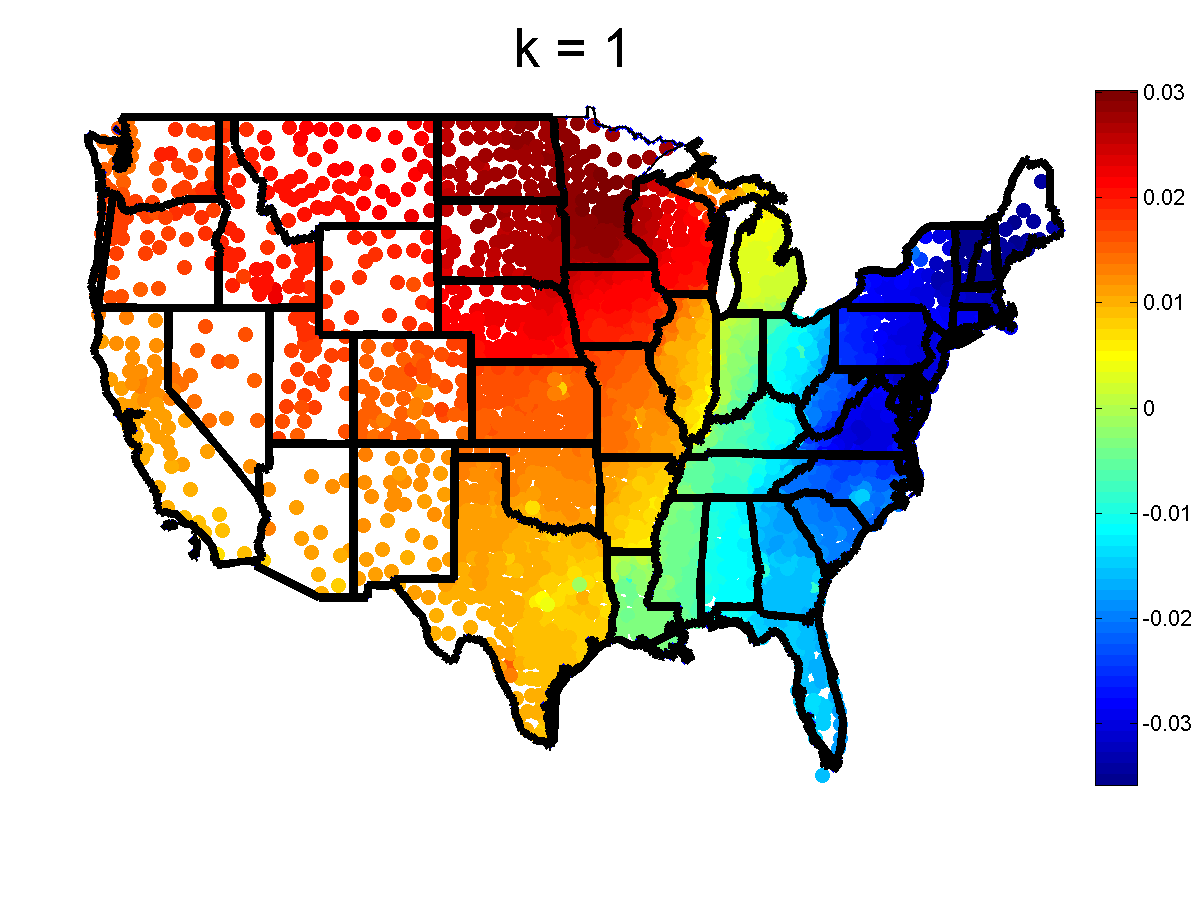}
\includegraphics[width=0.32 \textwidth]{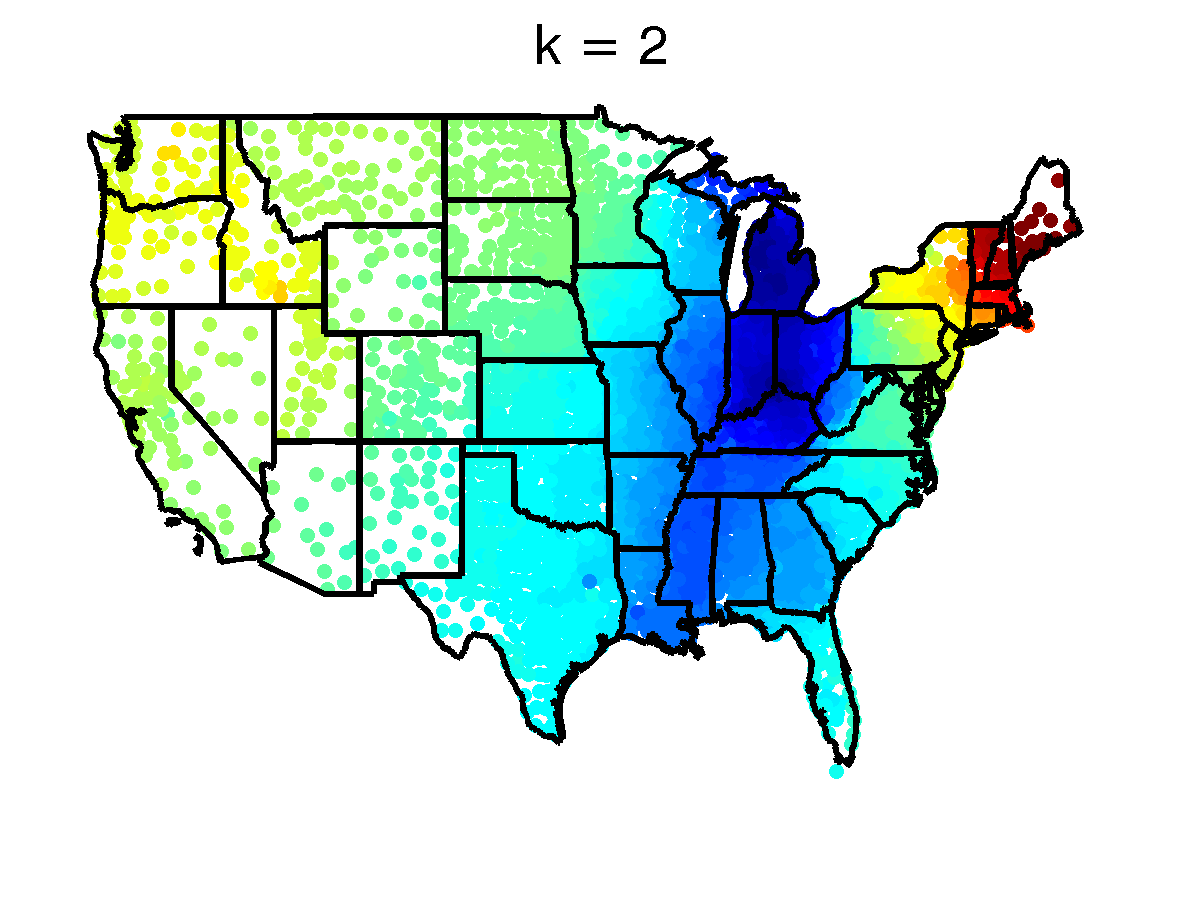}
\includegraphics[width=0.32 \textwidth]{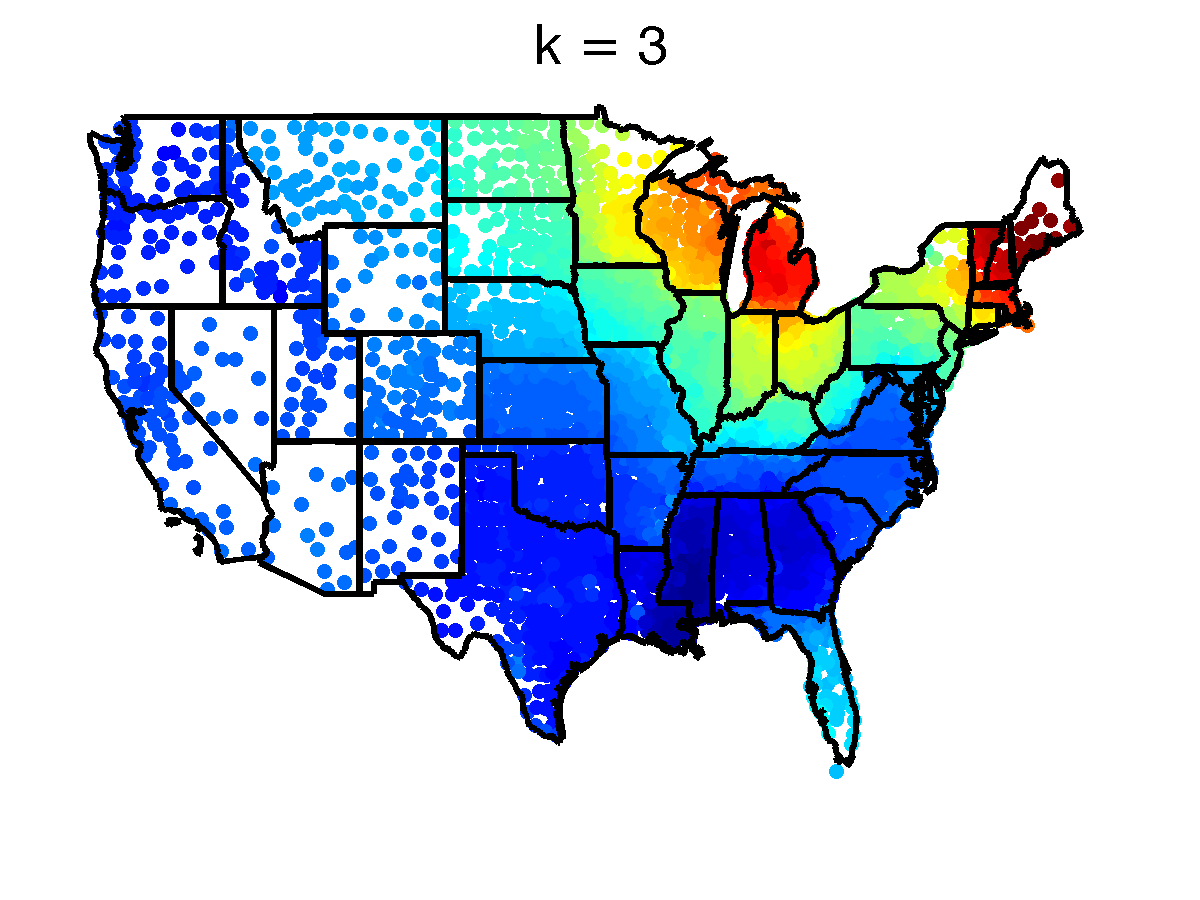}
 \includegraphics[width=0.32 \textwidth]{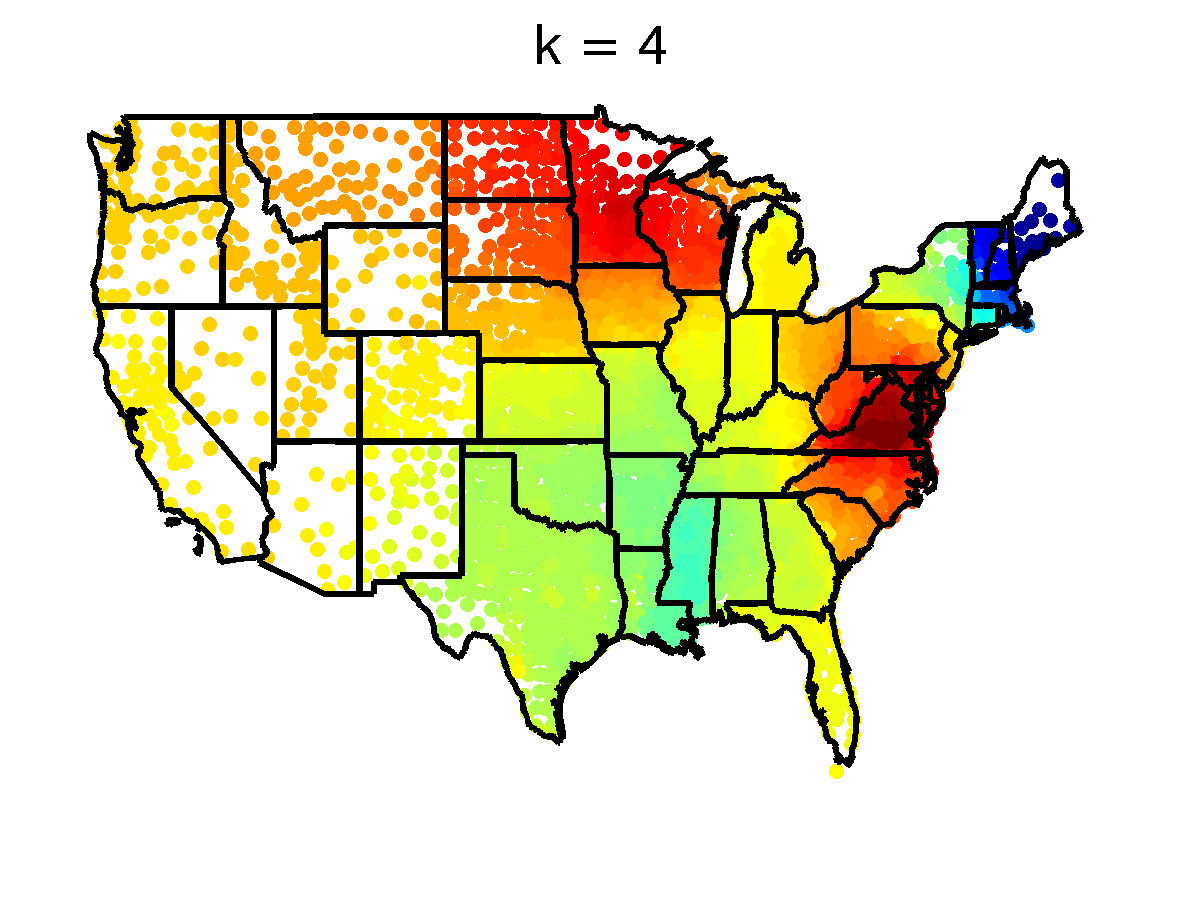}
\includegraphics[width=0.32 \textwidth]{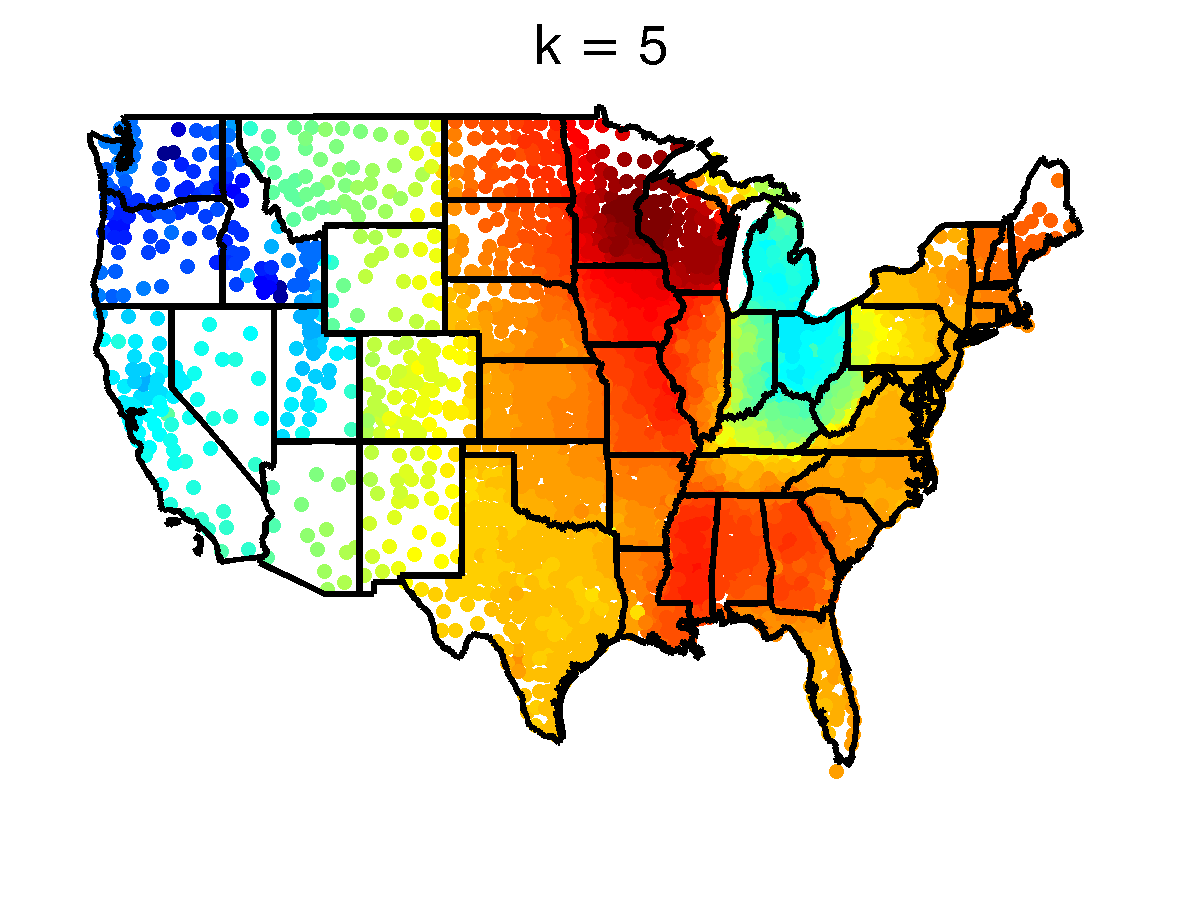}
\includegraphics[width=0.32 \textwidth]{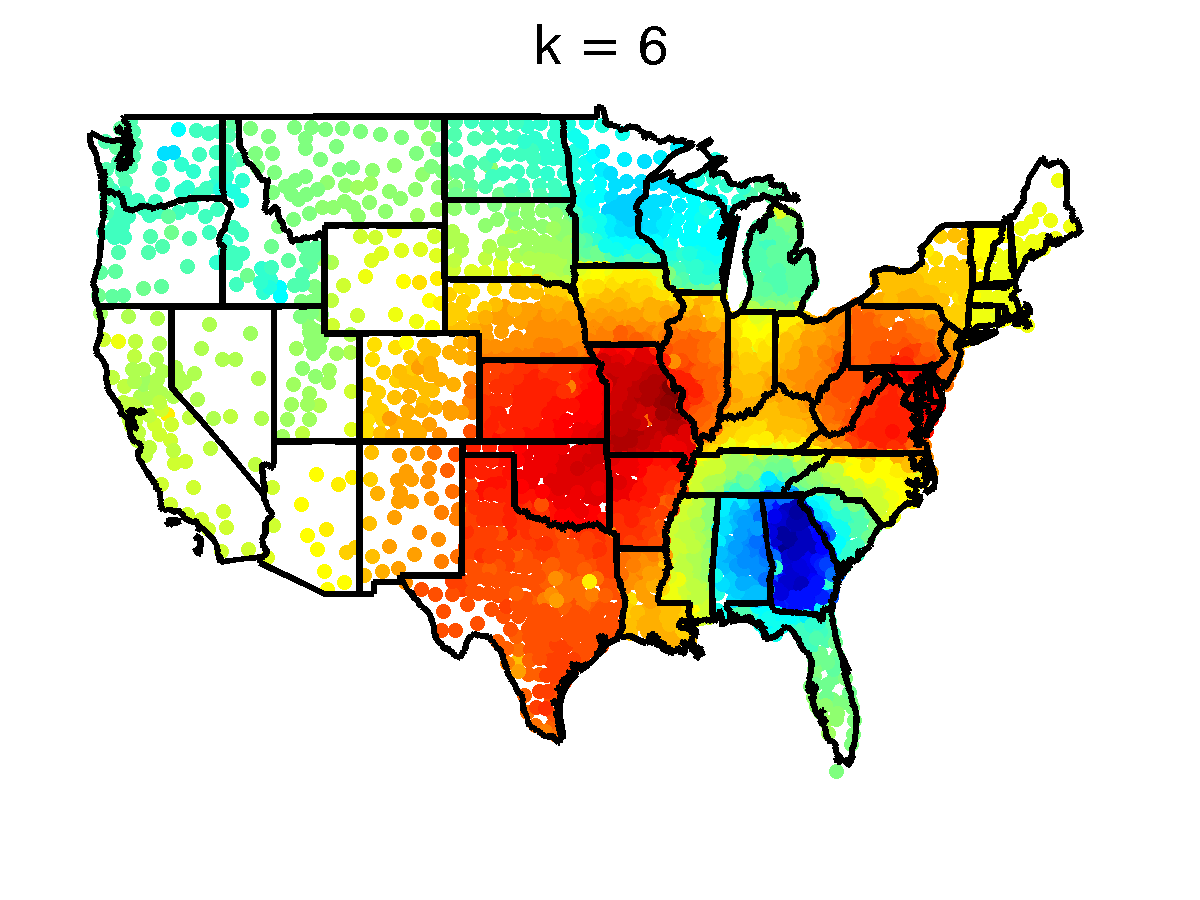}
\includegraphics[width=0.32 \textwidth]{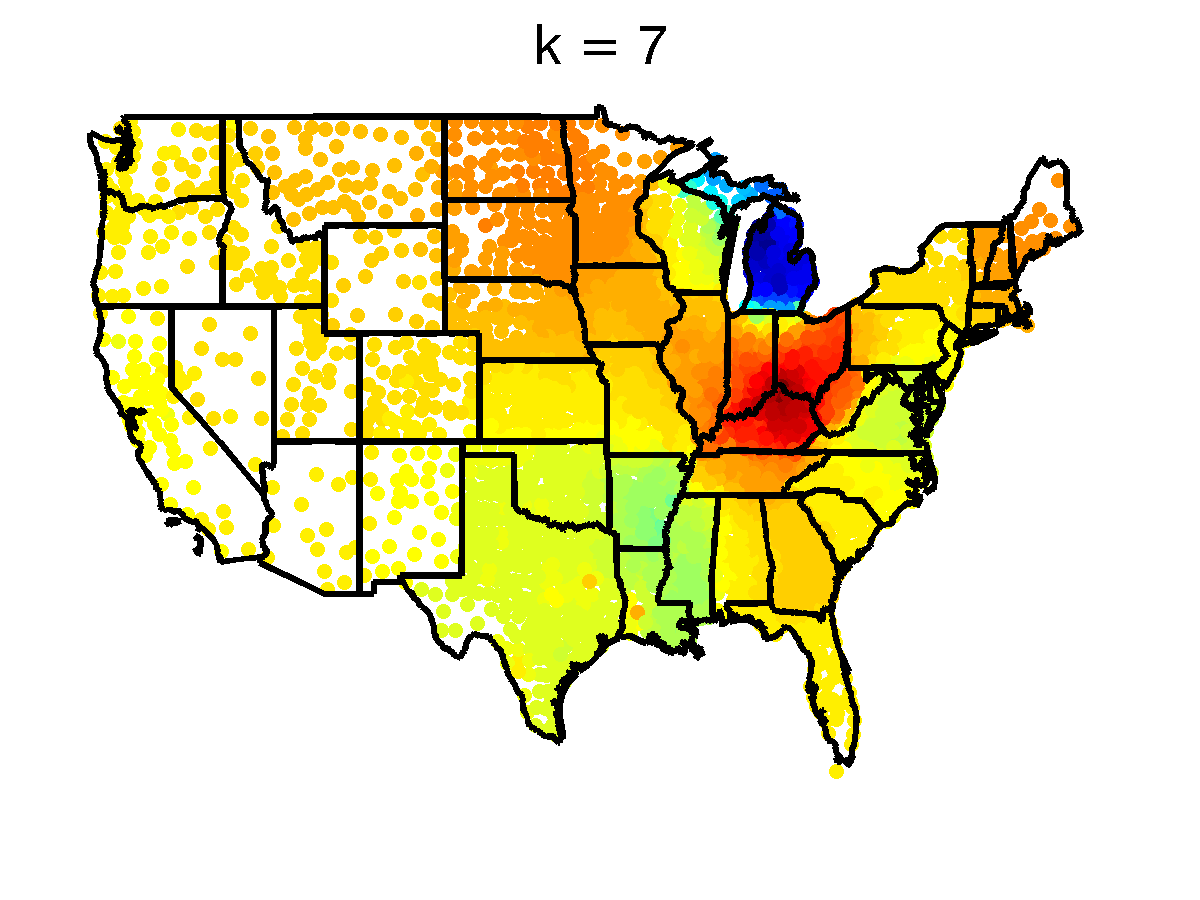}
\includegraphics[width=0.32 \textwidth]{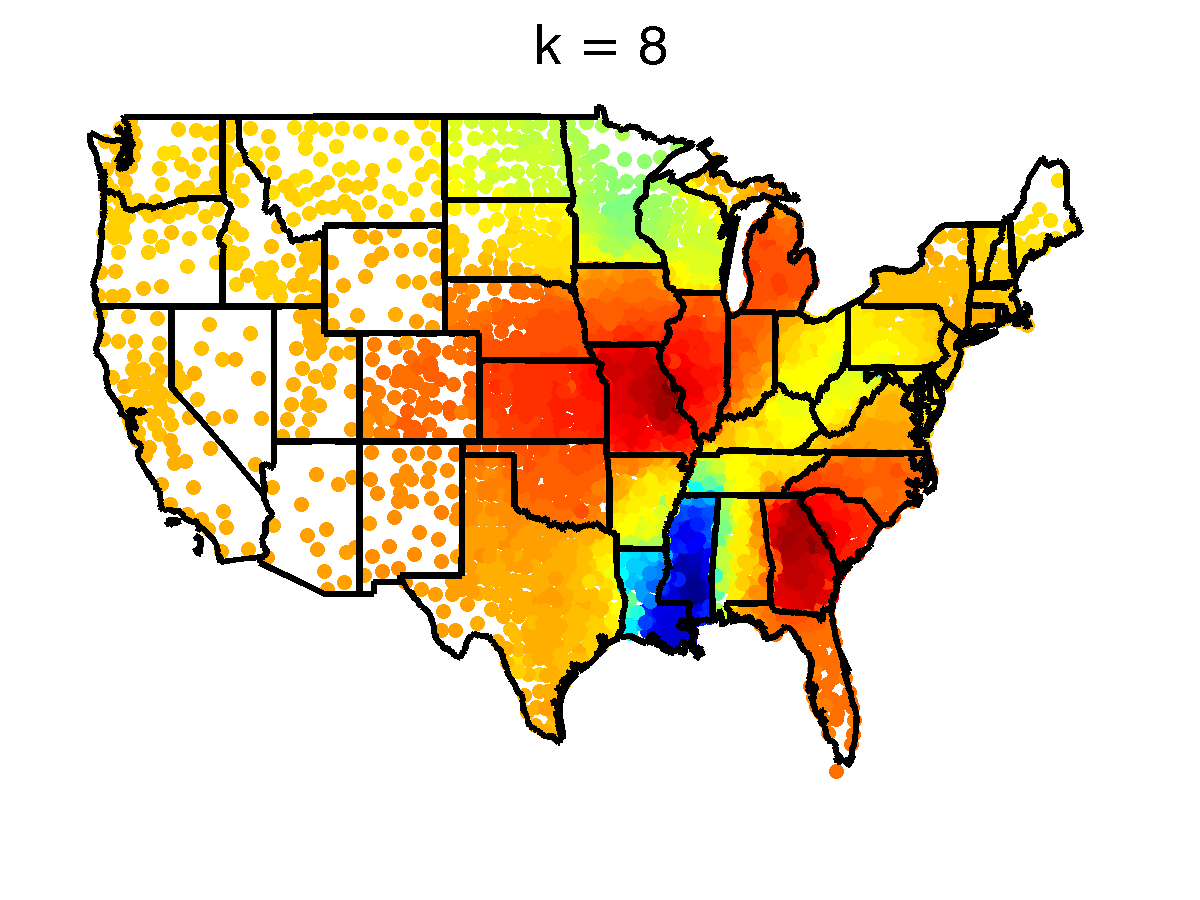}
\includegraphics[width=0.32 \textwidth]{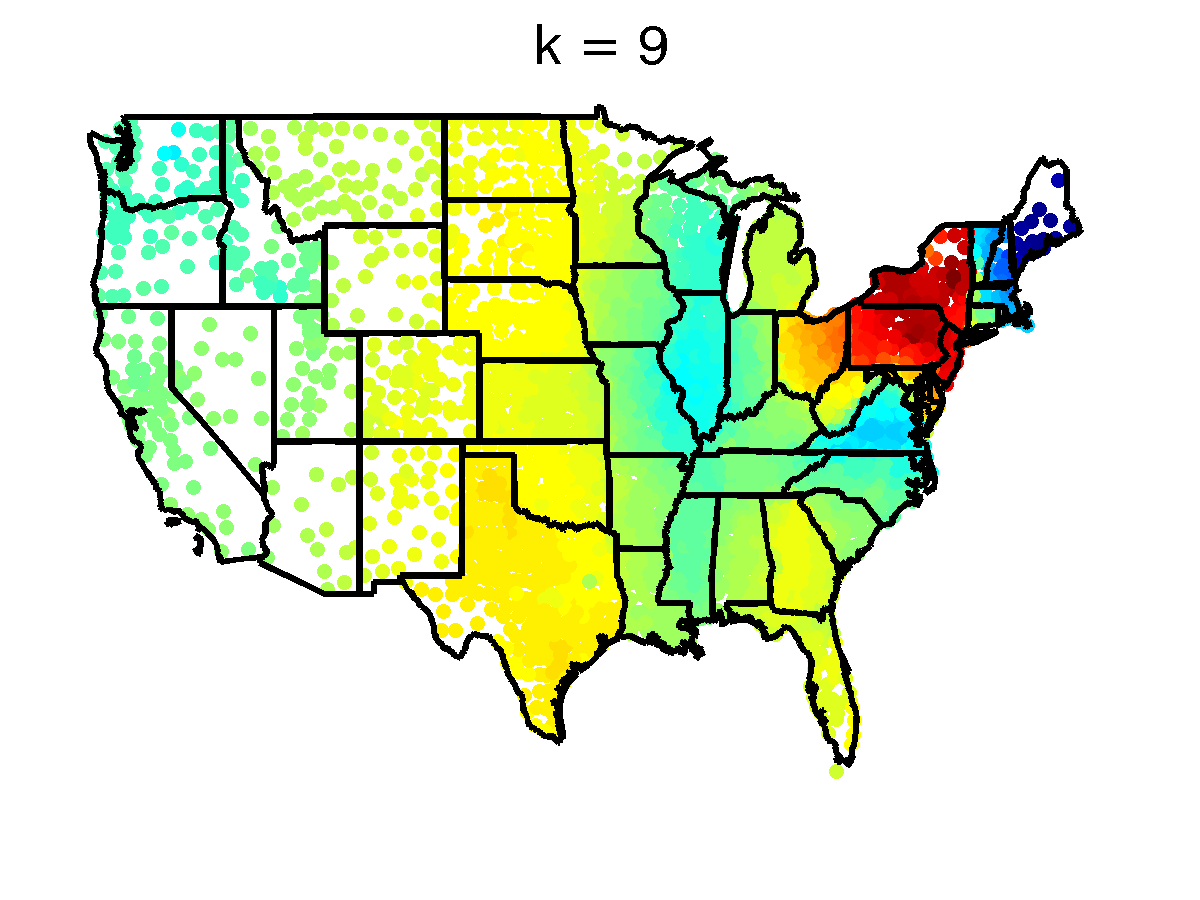}
\includegraphics[width=0.32 \textwidth]{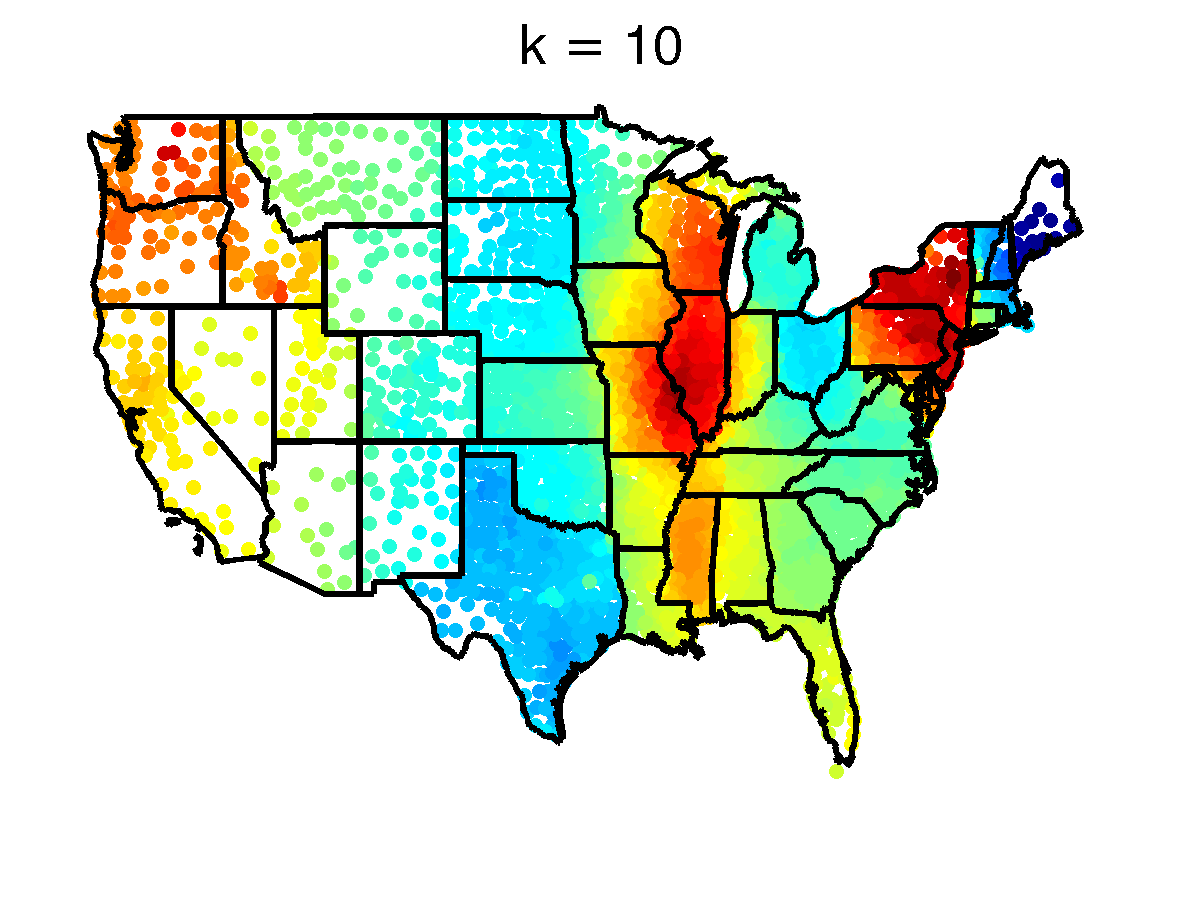}
\includegraphics[width=0.32 \textwidth]{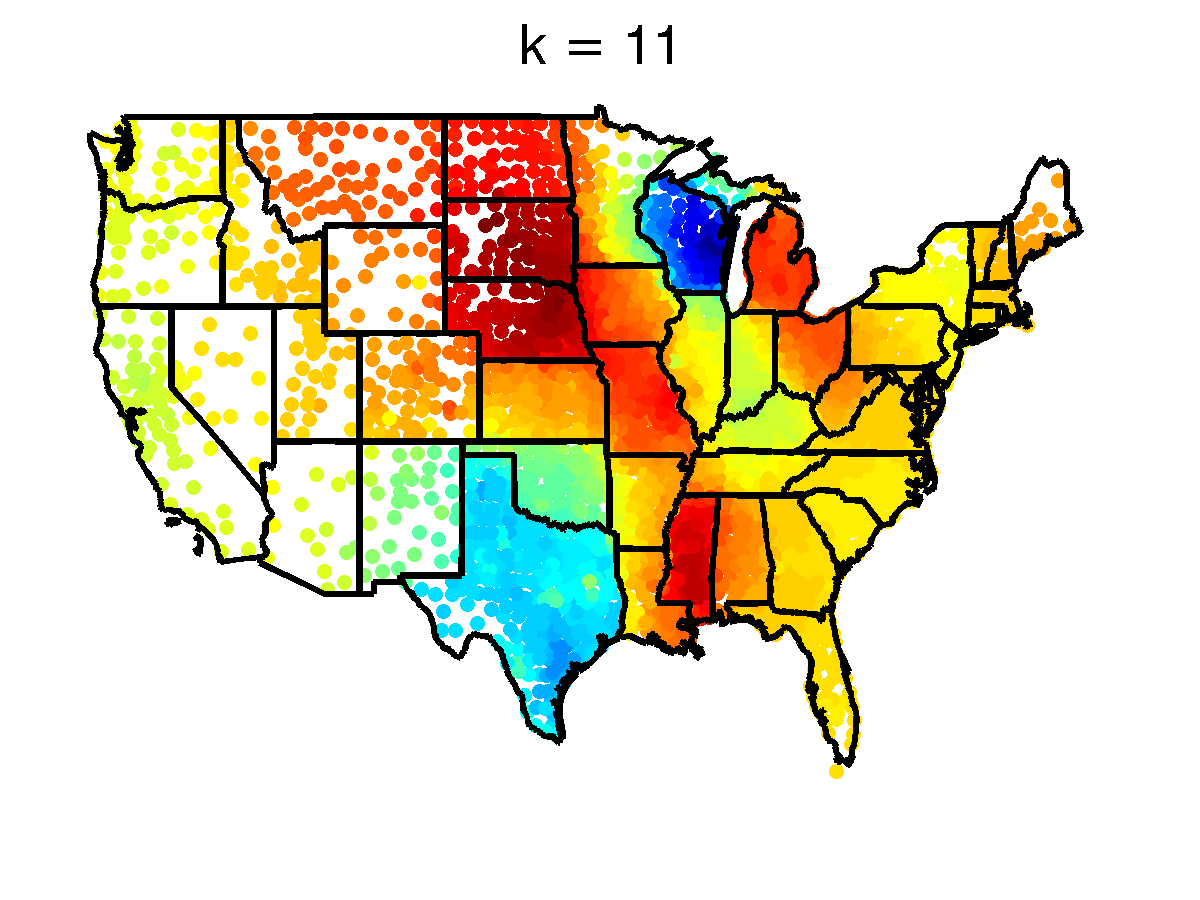}
\includegraphics[width=0.32 \textwidth]{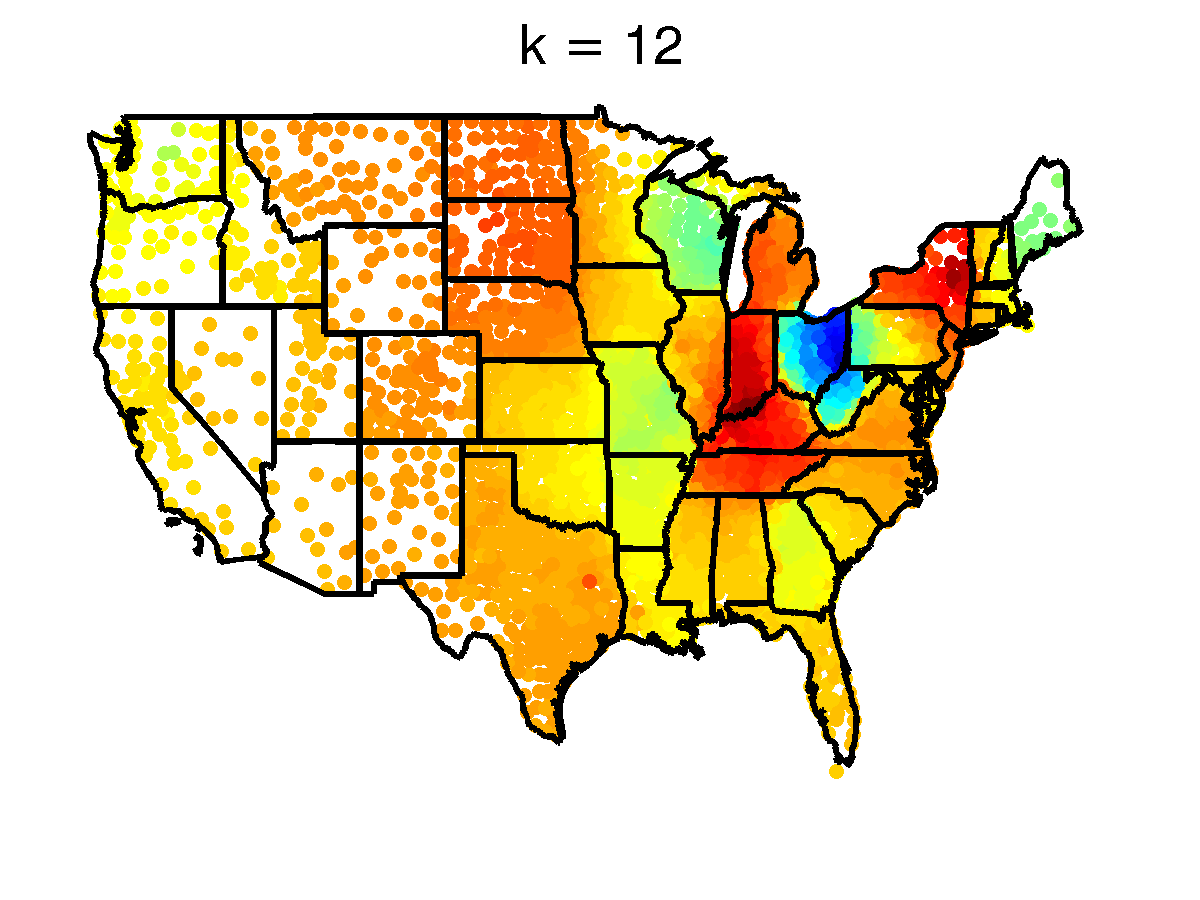}
\includegraphics[width=0.32 \textwidth]{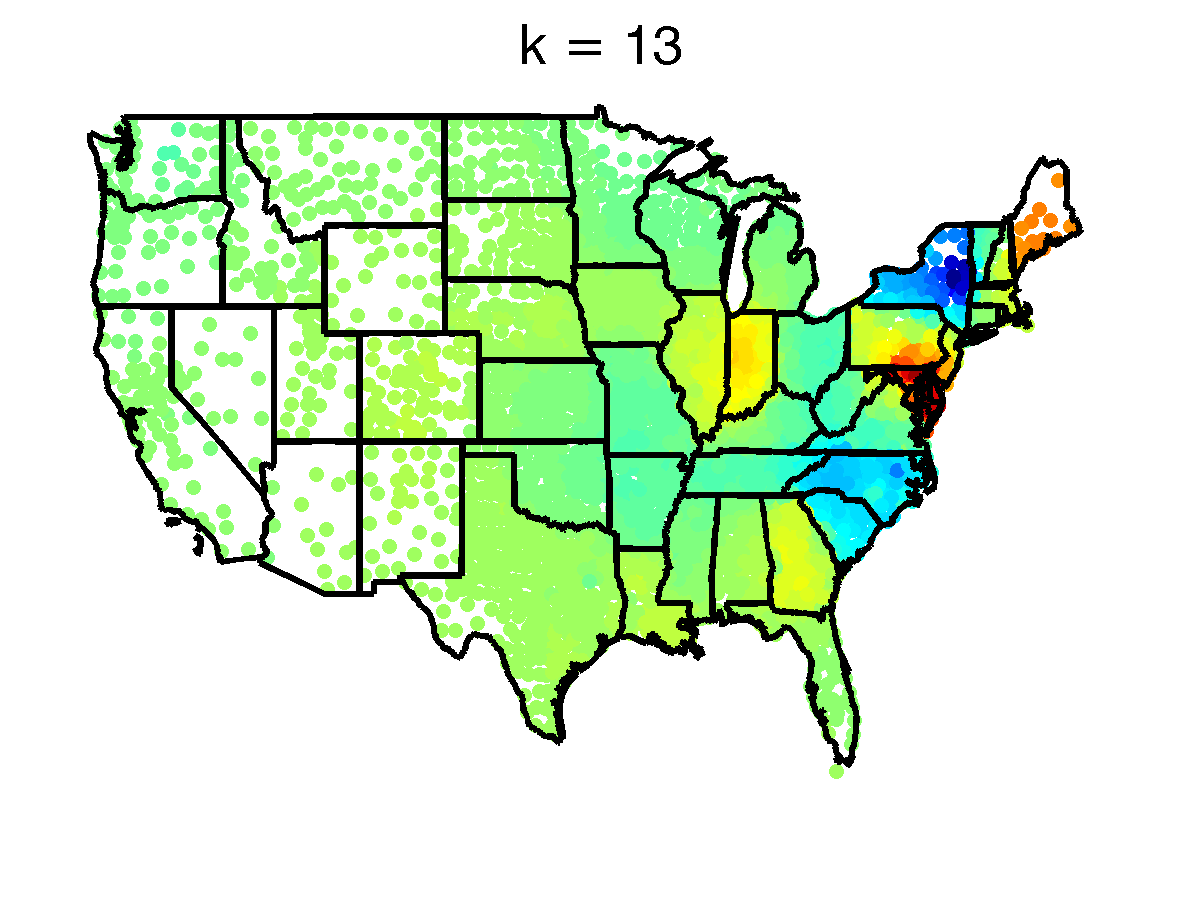}
\includegraphics[width=0.32 \textwidth]{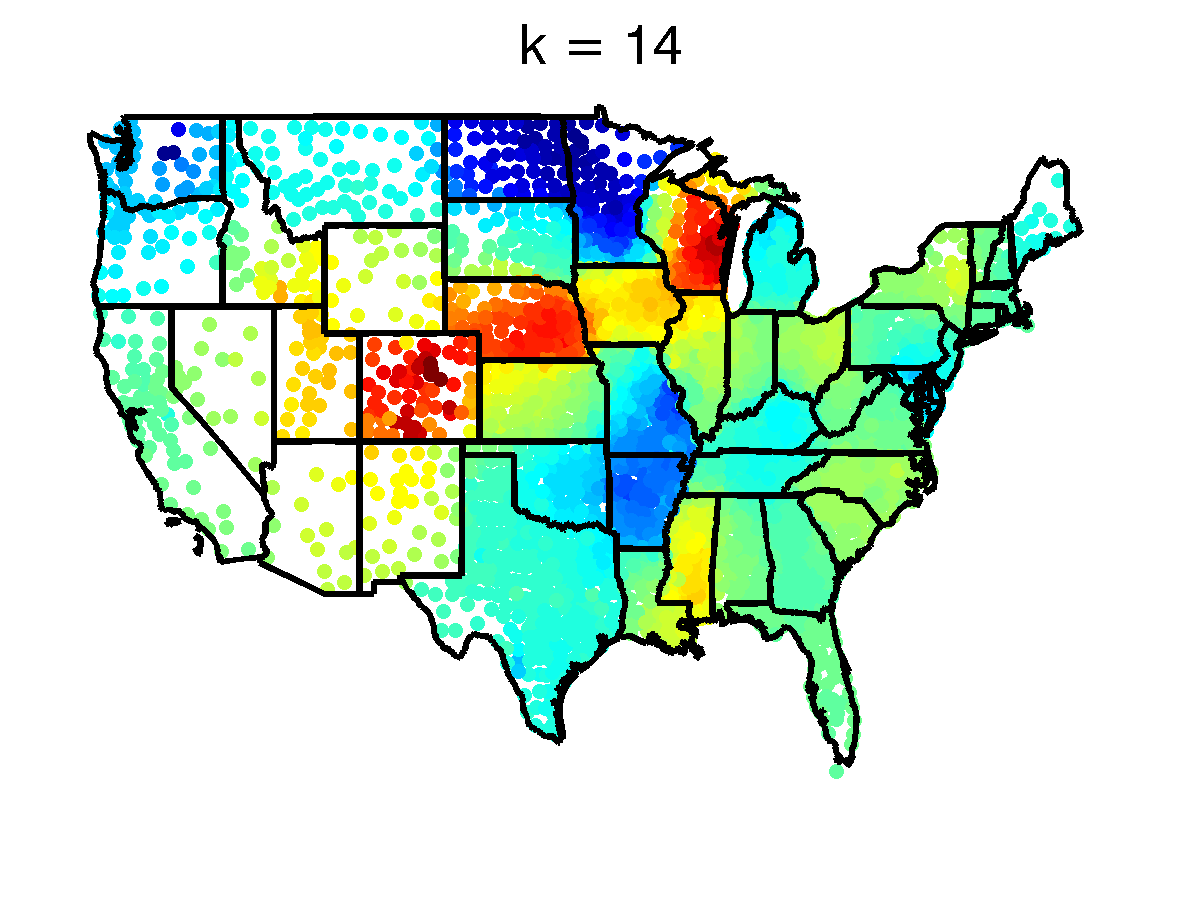}
\includegraphics[width=0.32 \textwidth]{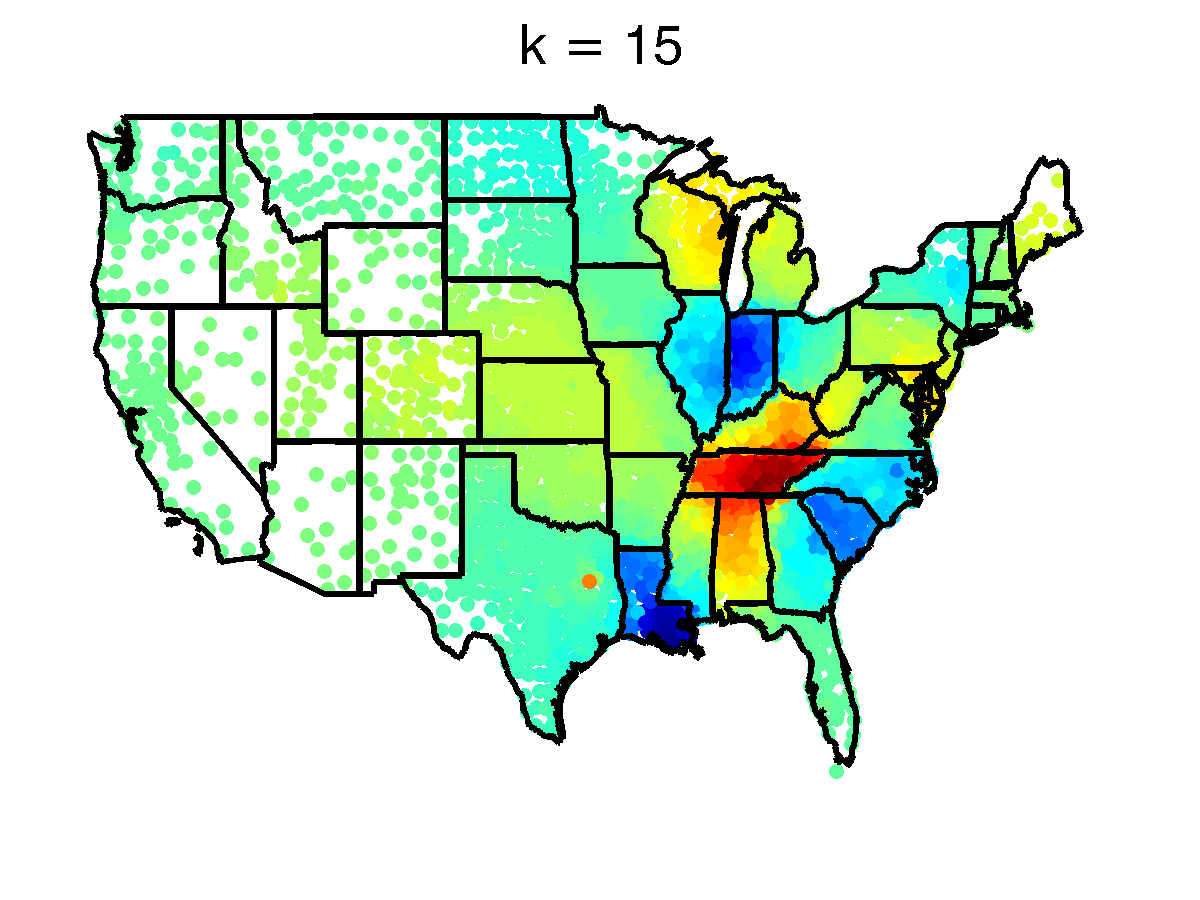}
\includegraphics[width=0.32 \textwidth]{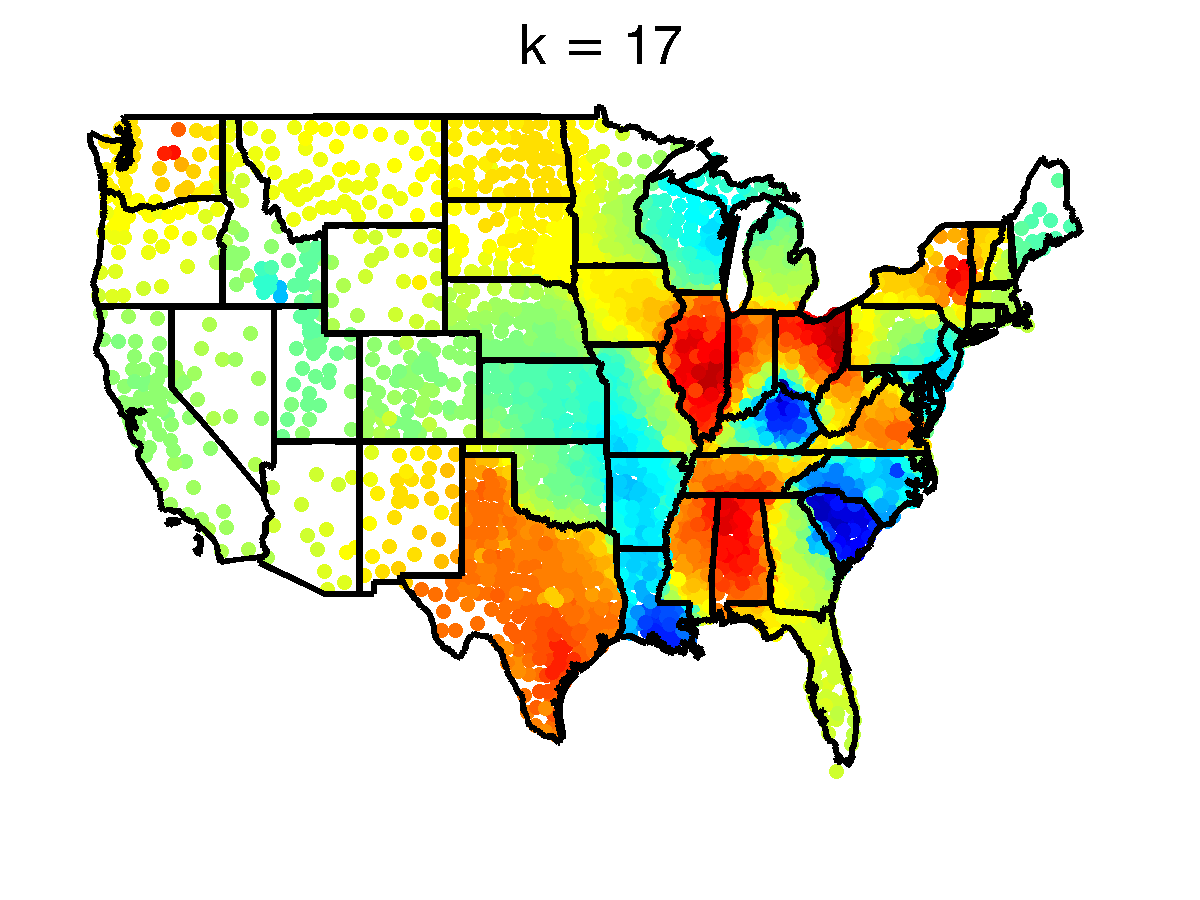}
\includegraphics[width=0.32 \textwidth]{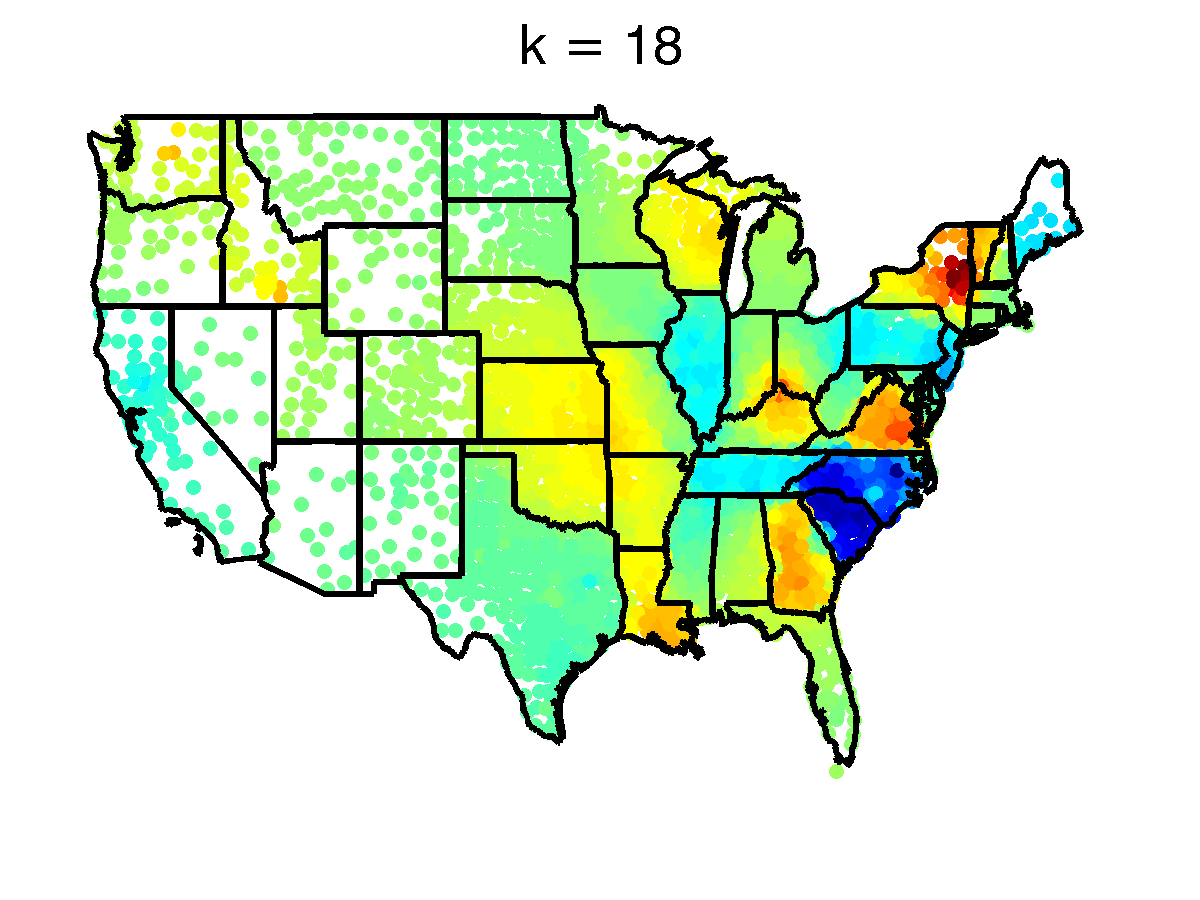}
\includegraphics[width=0.32 \textwidth]{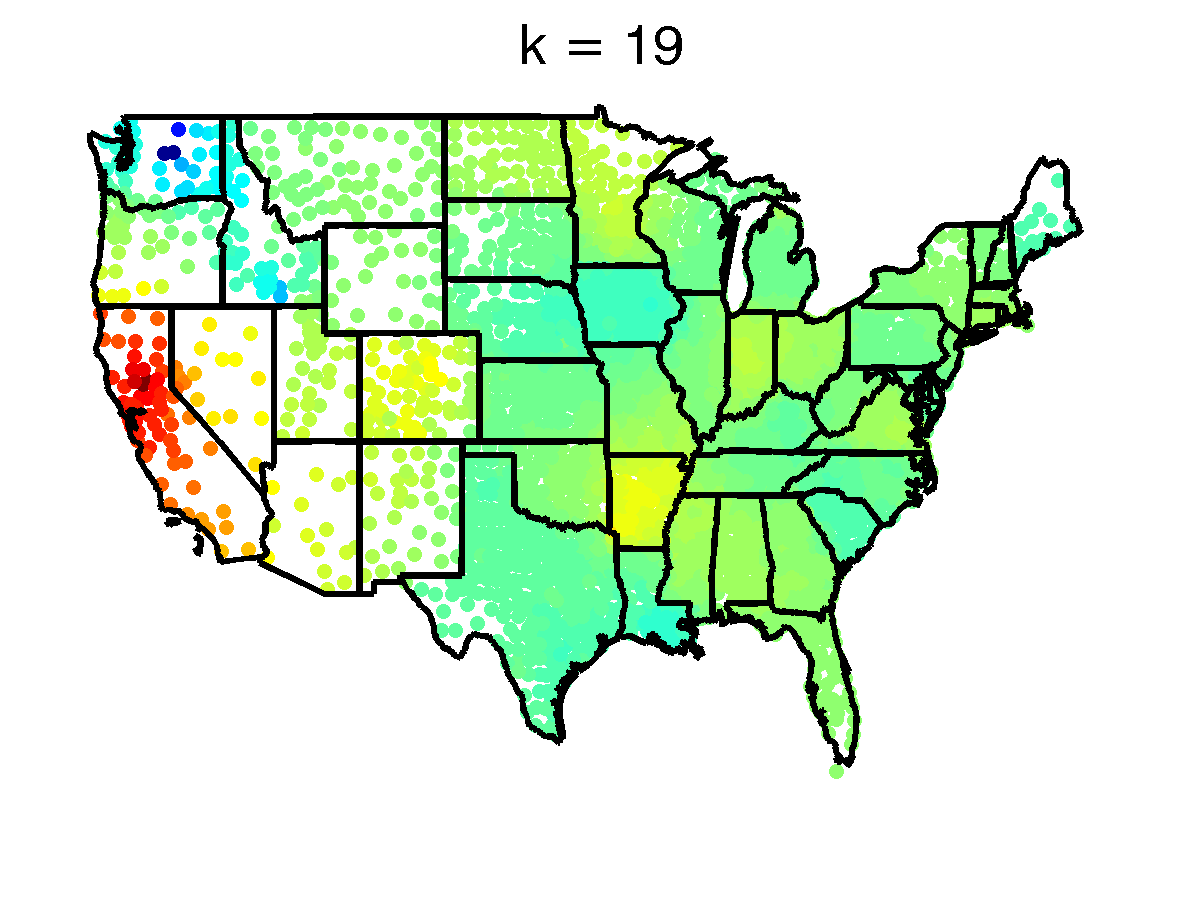}
\end{center}
\caption{ Top eigenvector colorings for the similarity  matrix $W_{ij} = \frac{M_{ij}^2}{P_i P_j}$.}
\label{fig:US_K1_top18}
\end{figure}

\begin{figure}[h!t]
\begin{center}
\includegraphics[width=0.32 \textwidth]{K1/Kernel_1_v_19.png}
\includegraphics[width=0.32 \textwidth]{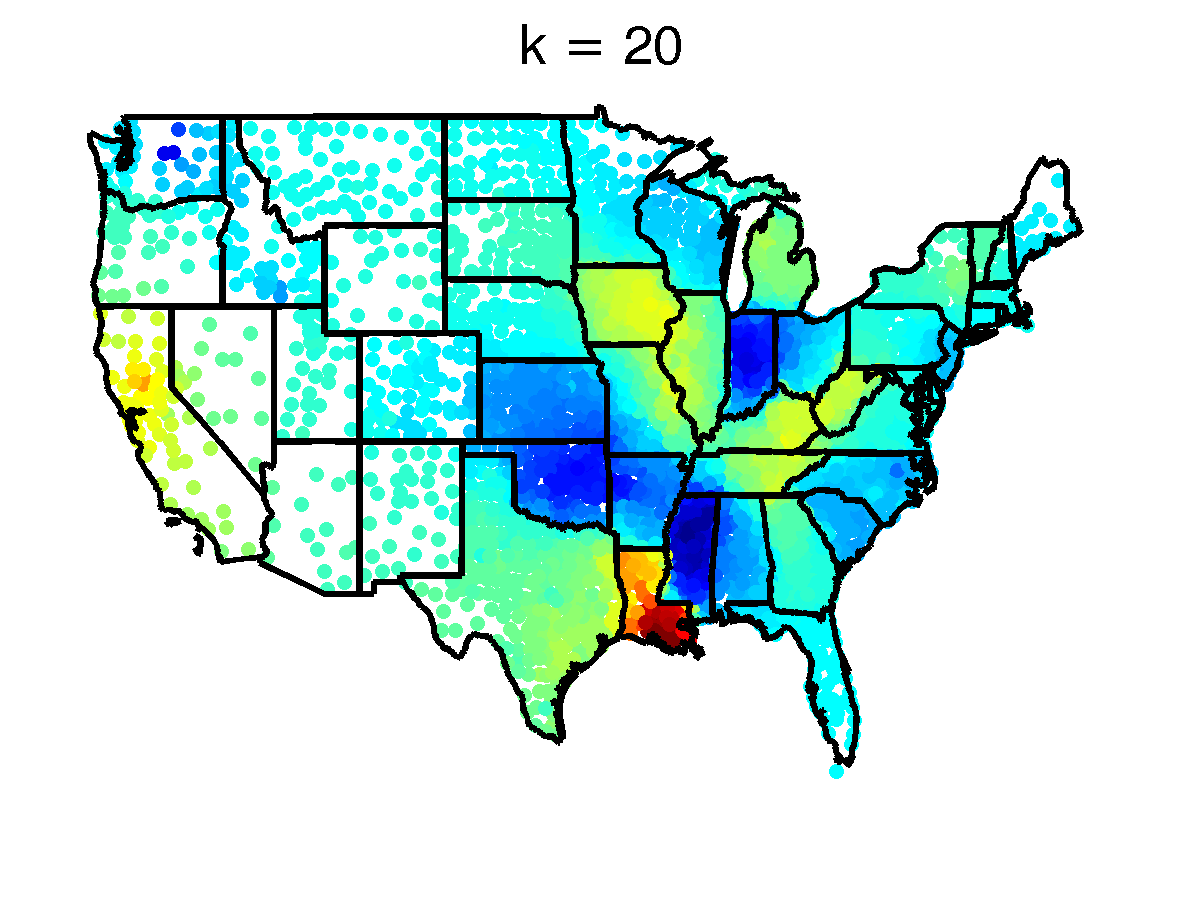}
\includegraphics[width=0.32 \textwidth]{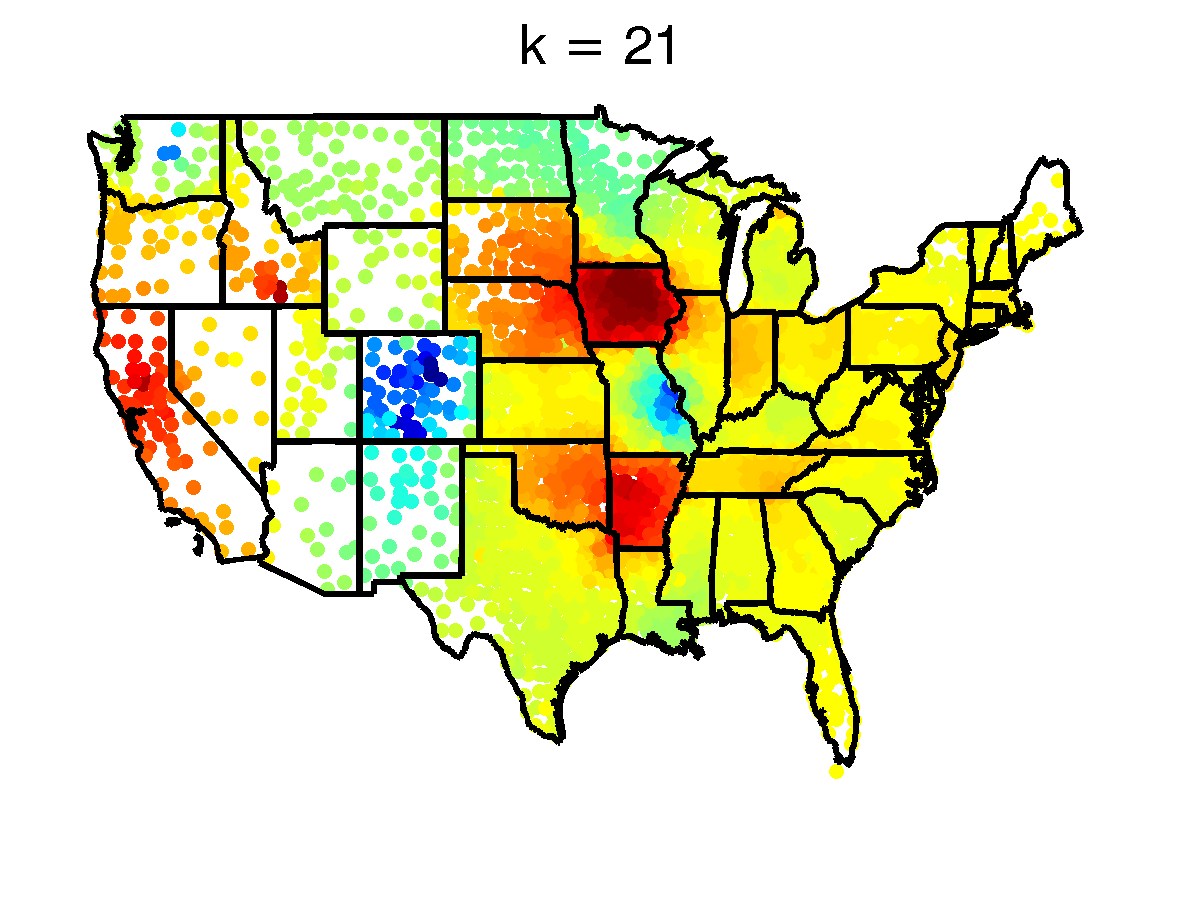}
\includegraphics[width=0.32 \textwidth]{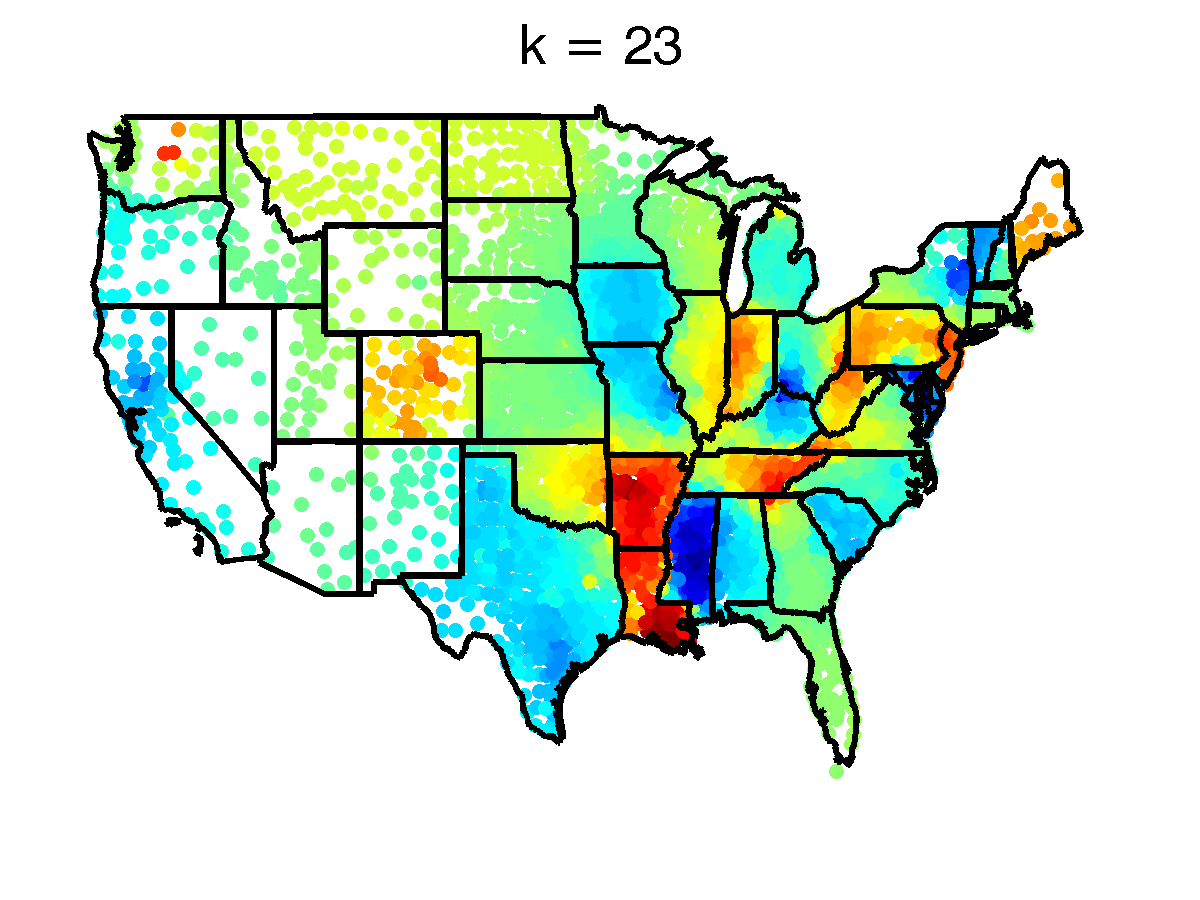}
\includegraphics[width=0.32 \textwidth]{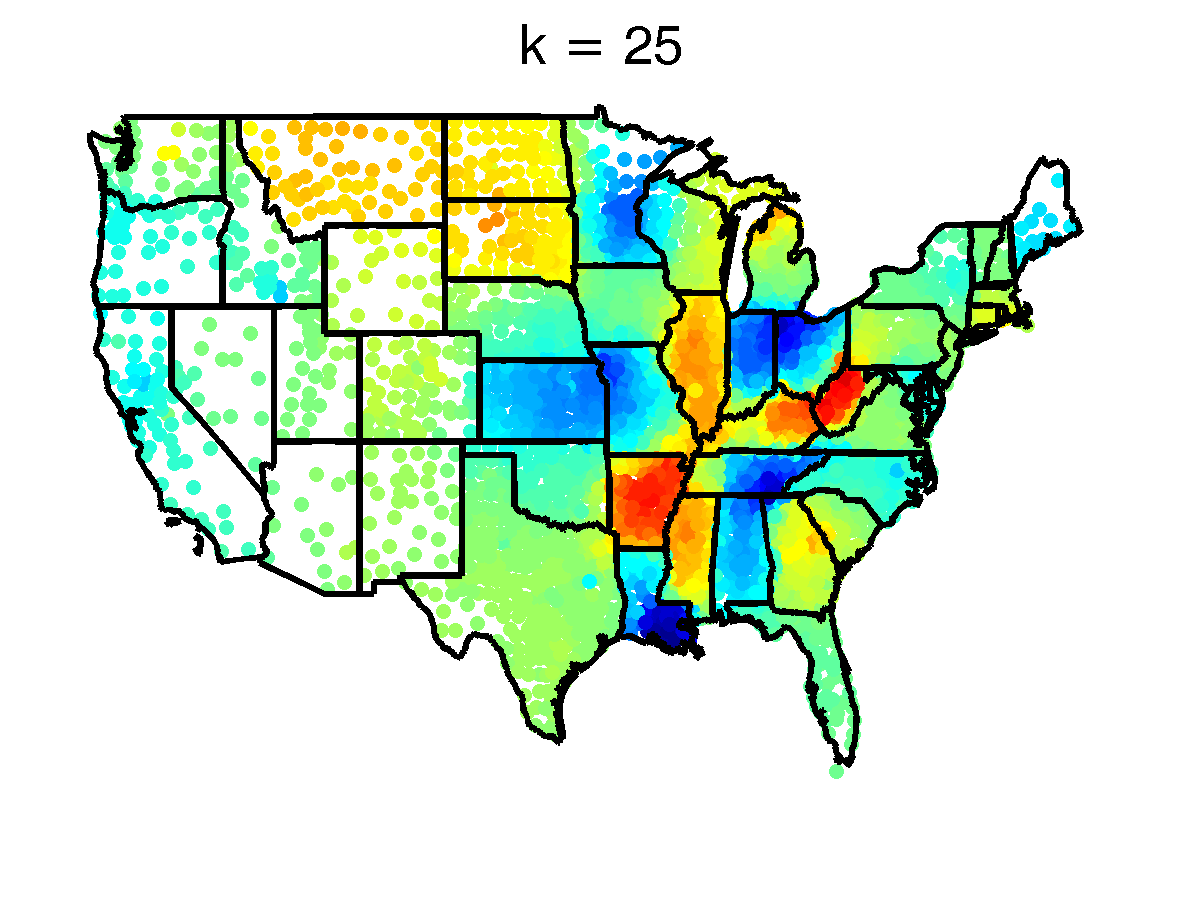}
\includegraphics[width=0.32 \textwidth]{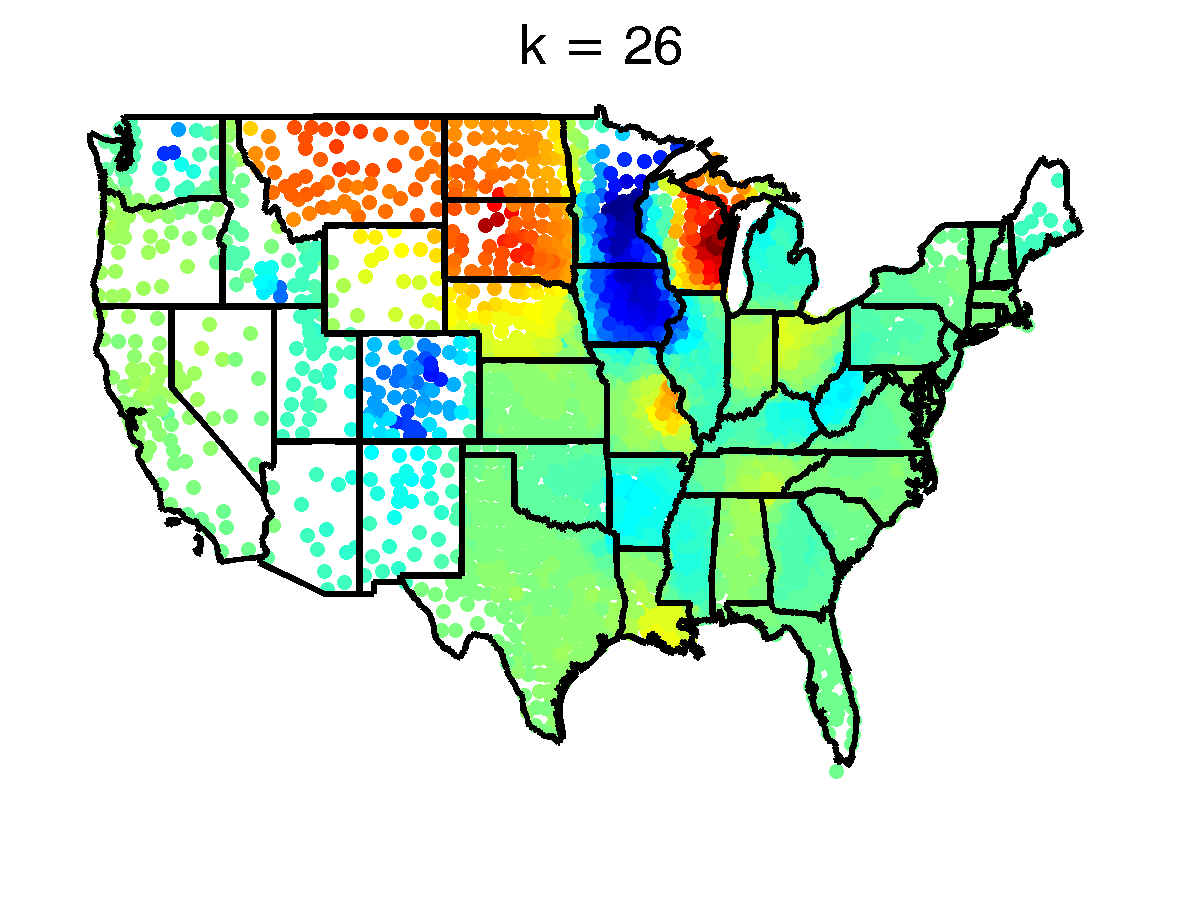}
\includegraphics[width=0.32 \textwidth]{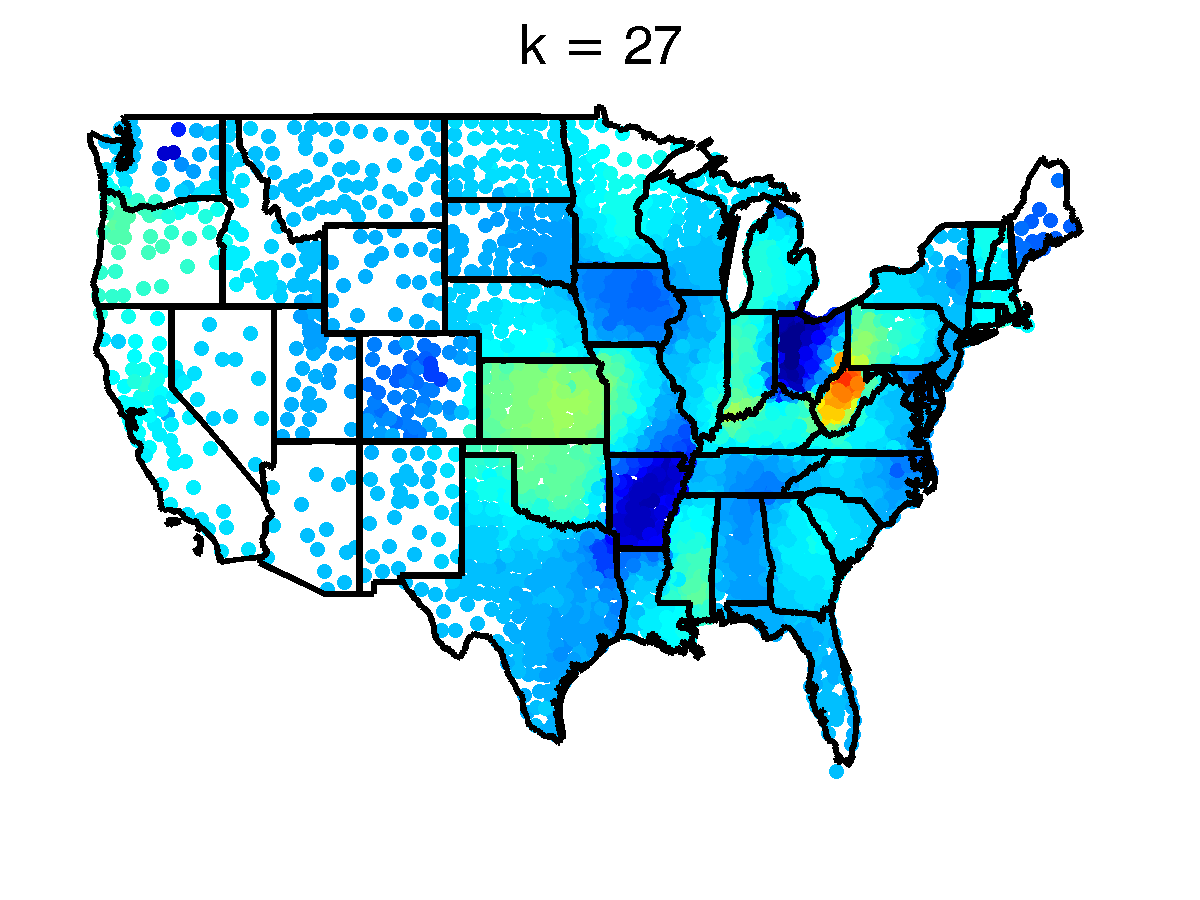}
\includegraphics[width=0.32 \textwidth]{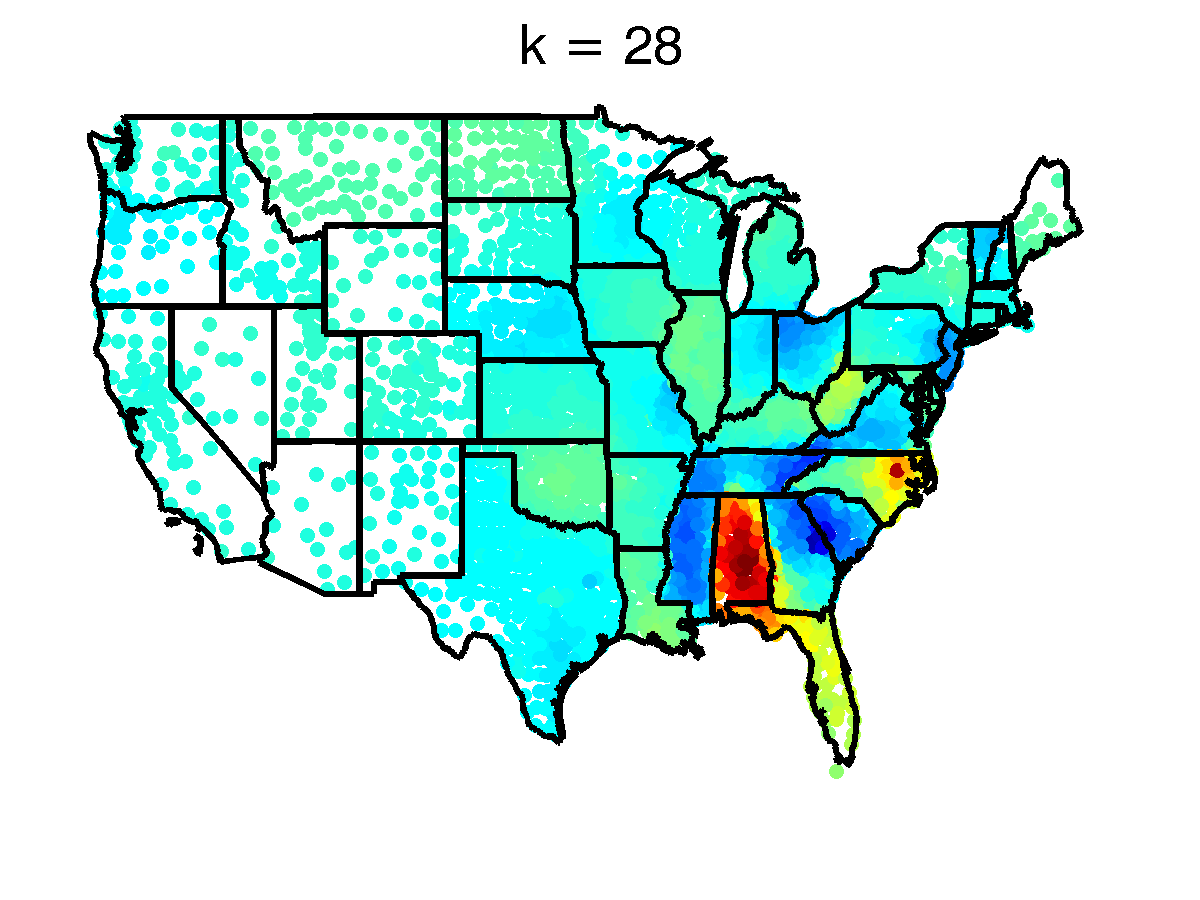}
\includegraphics[width=0.32 \textwidth]{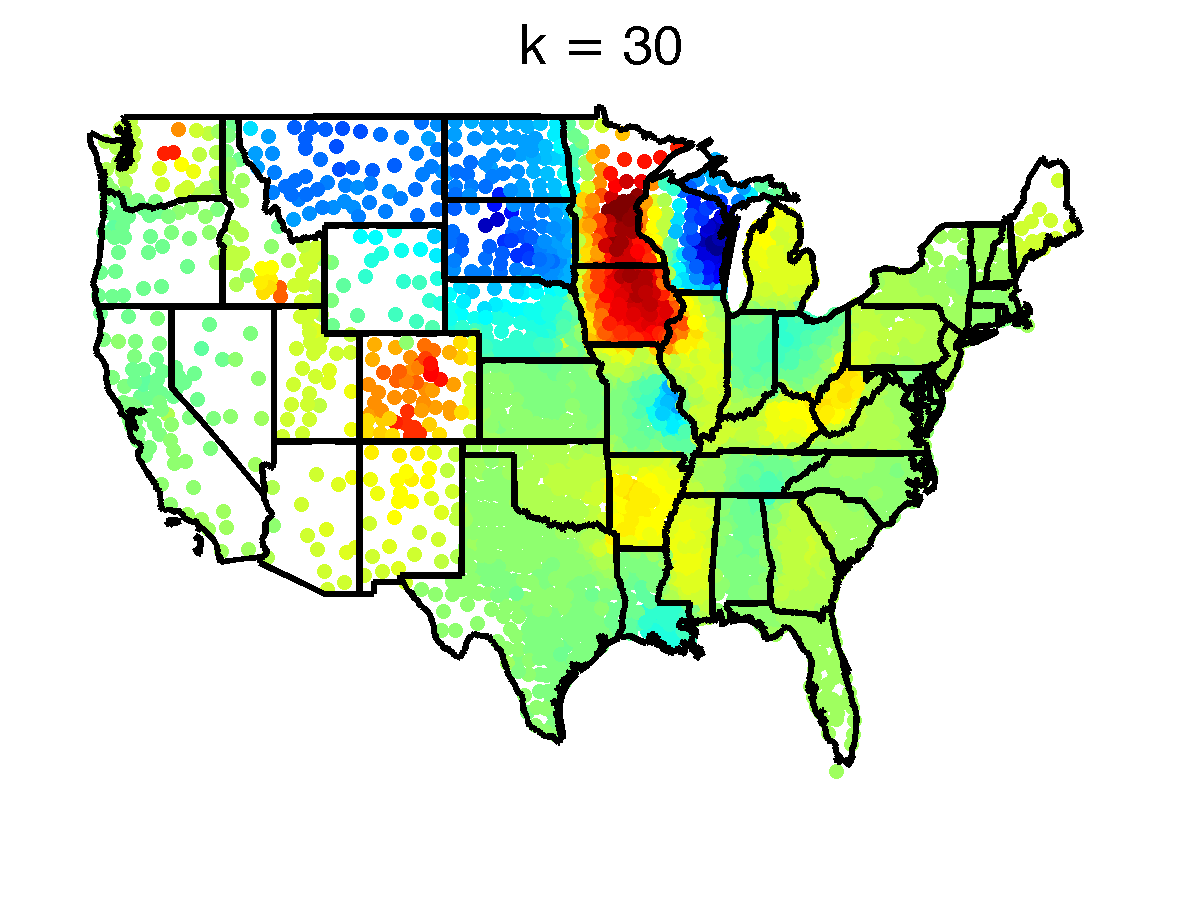}
\includegraphics[width=0.32 \textwidth]{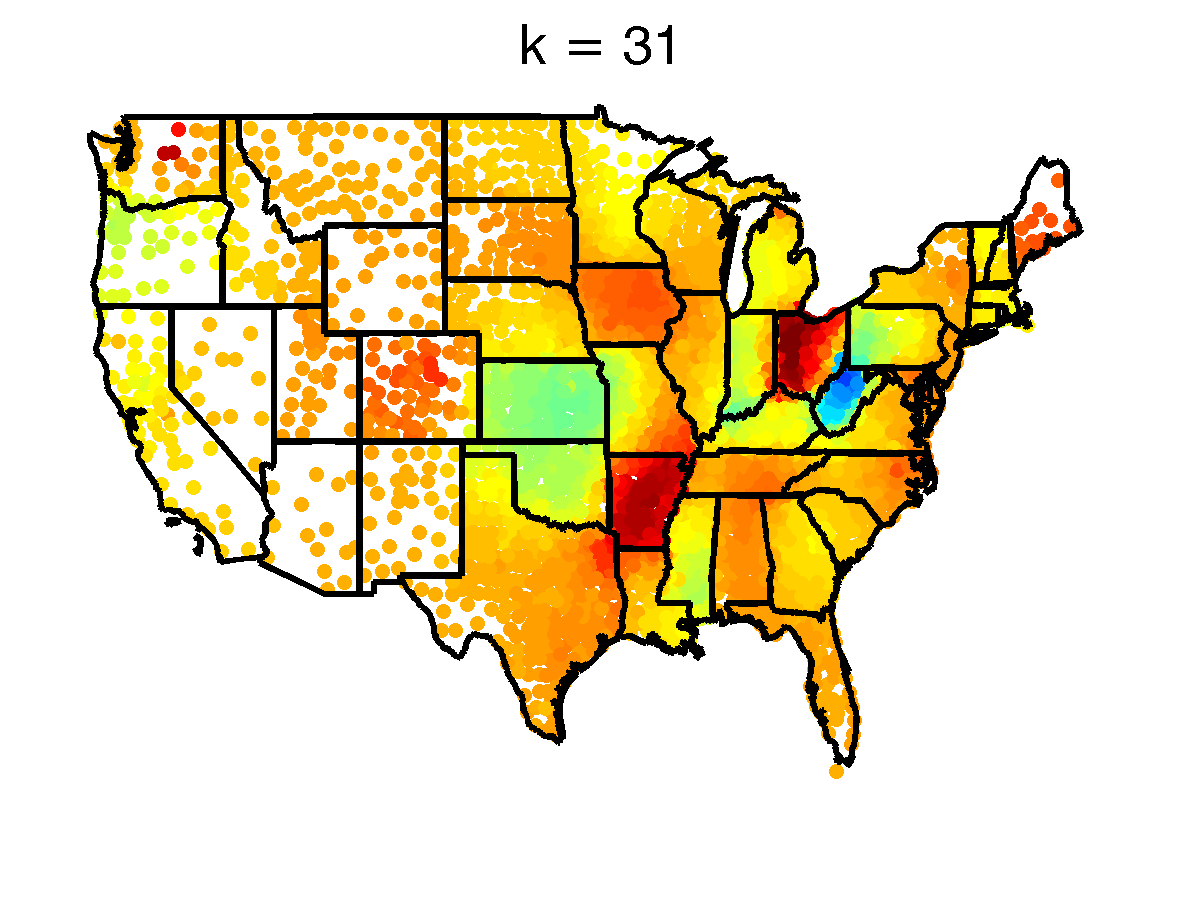}
\includegraphics[width=0.32 \textwidth]{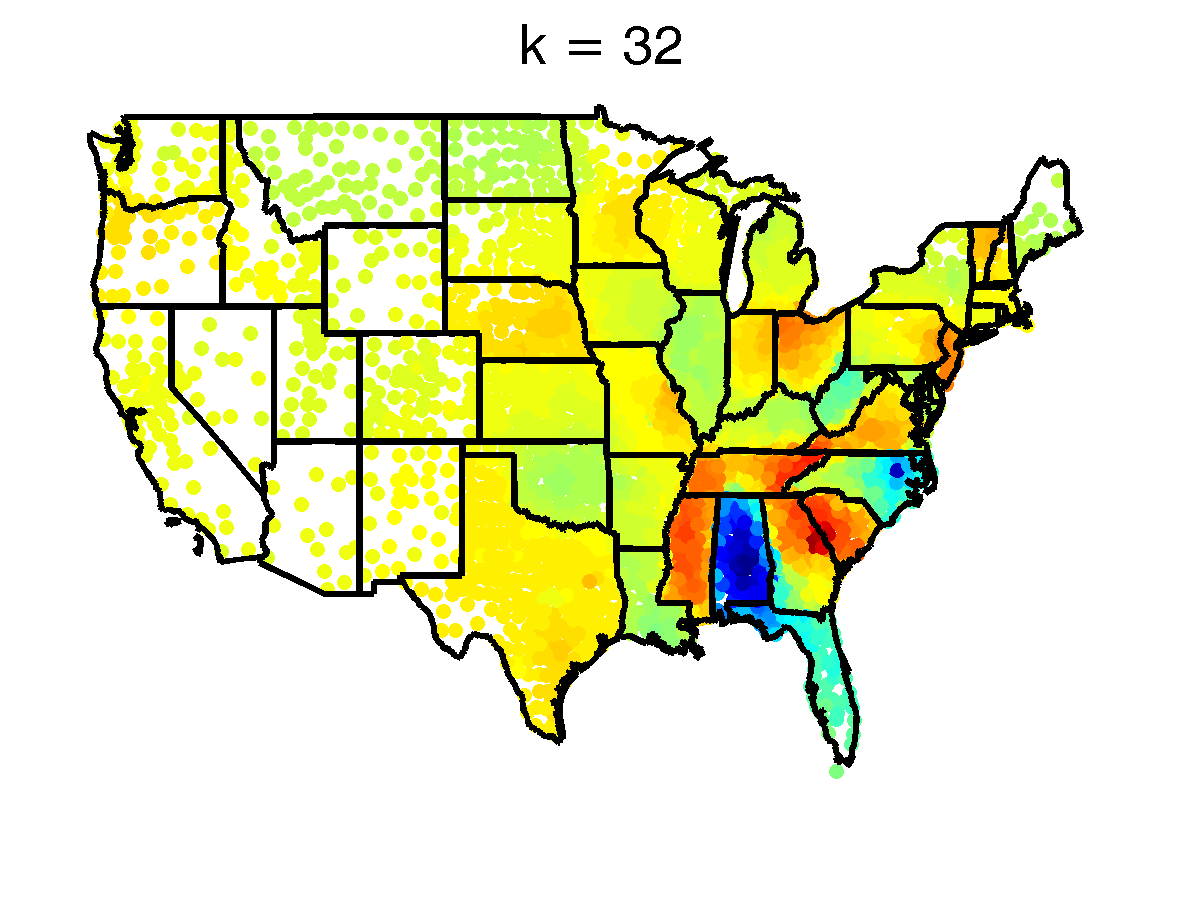}
\includegraphics[width=0.32 \textwidth]{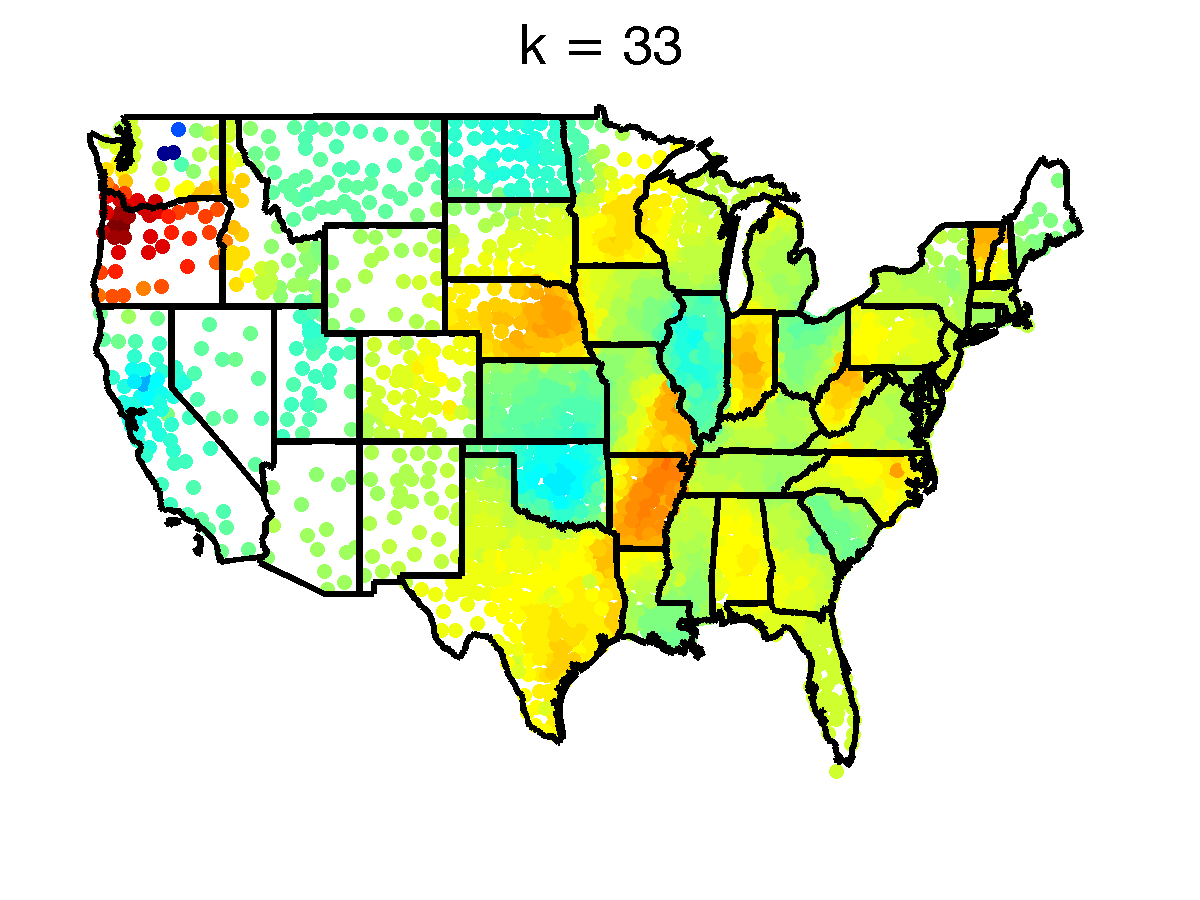}
\includegraphics[width=0.32 \textwidth]{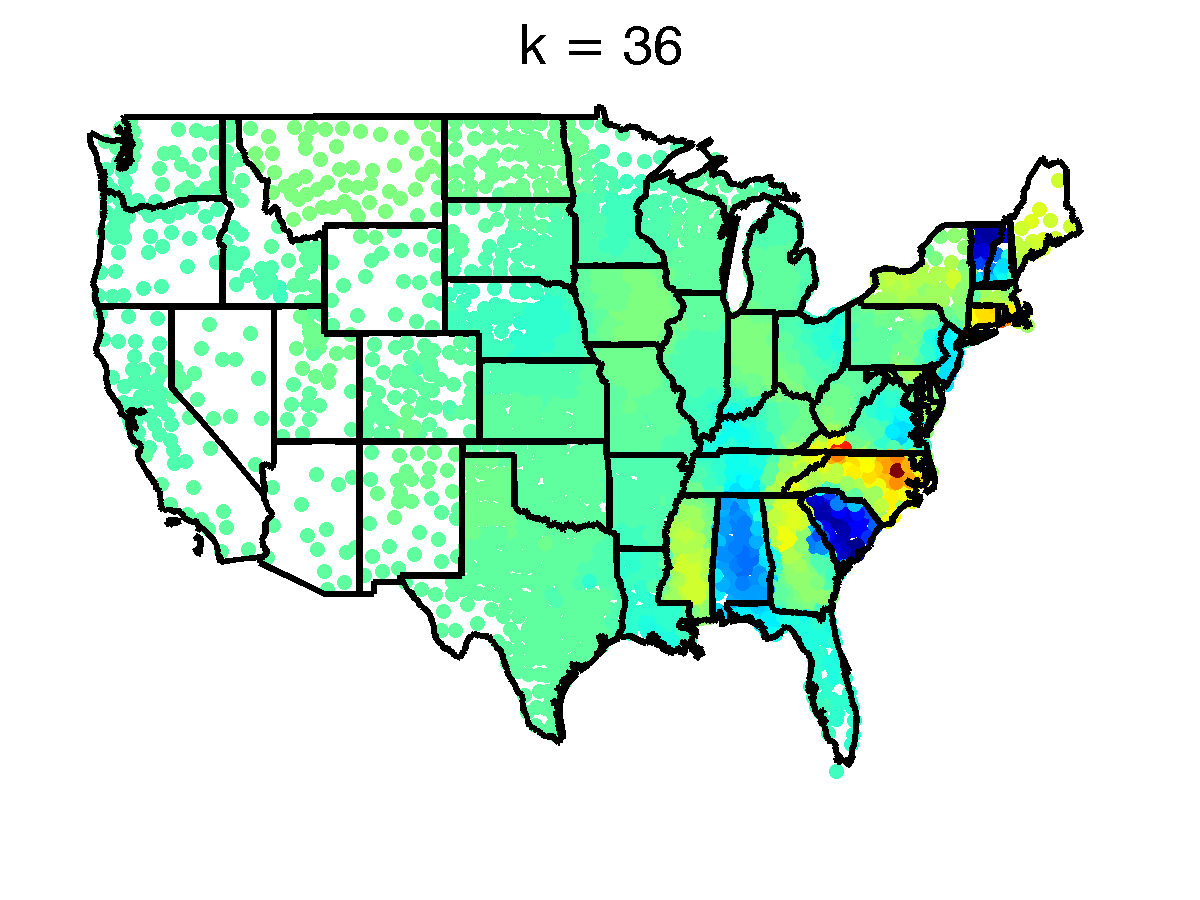}
\includegraphics[width=0.32 \textwidth]{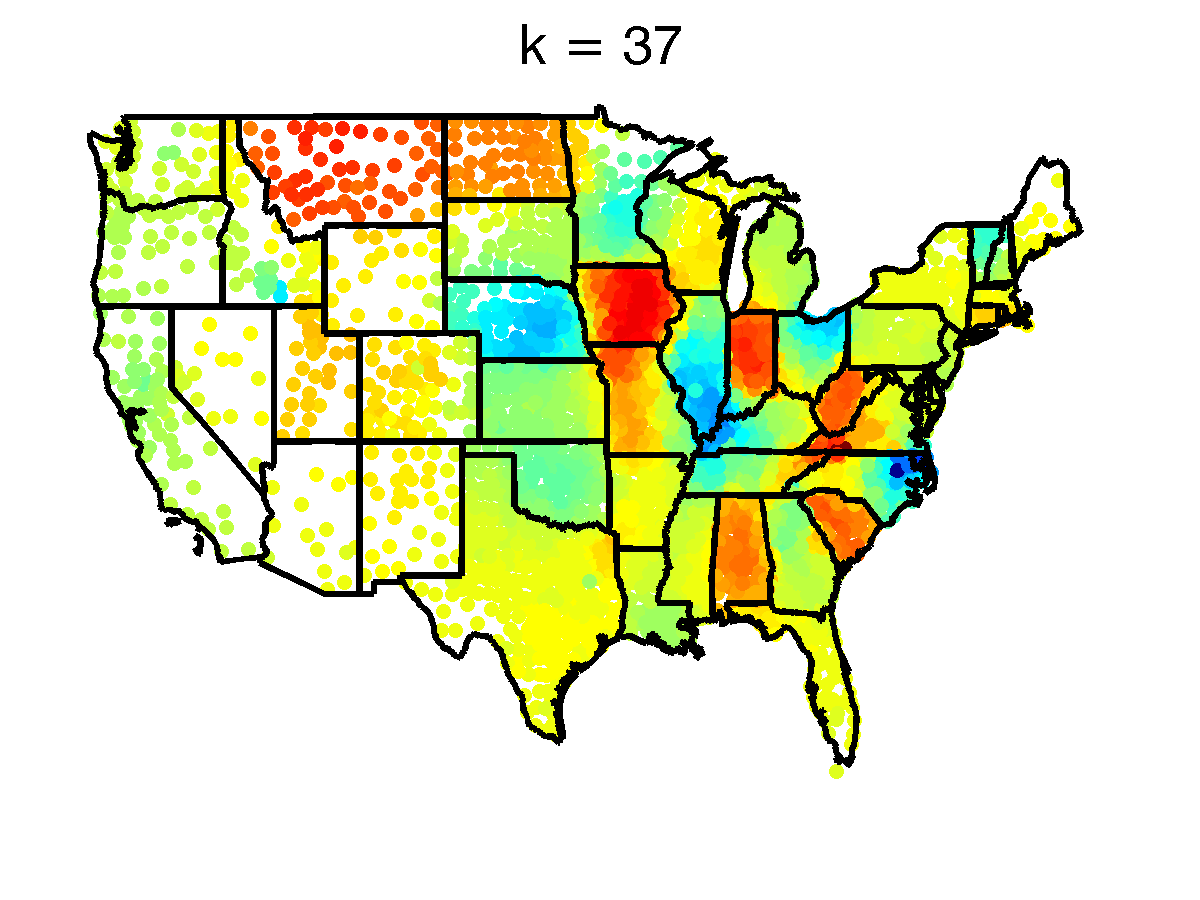}
\includegraphics[width=0.32 \textwidth]{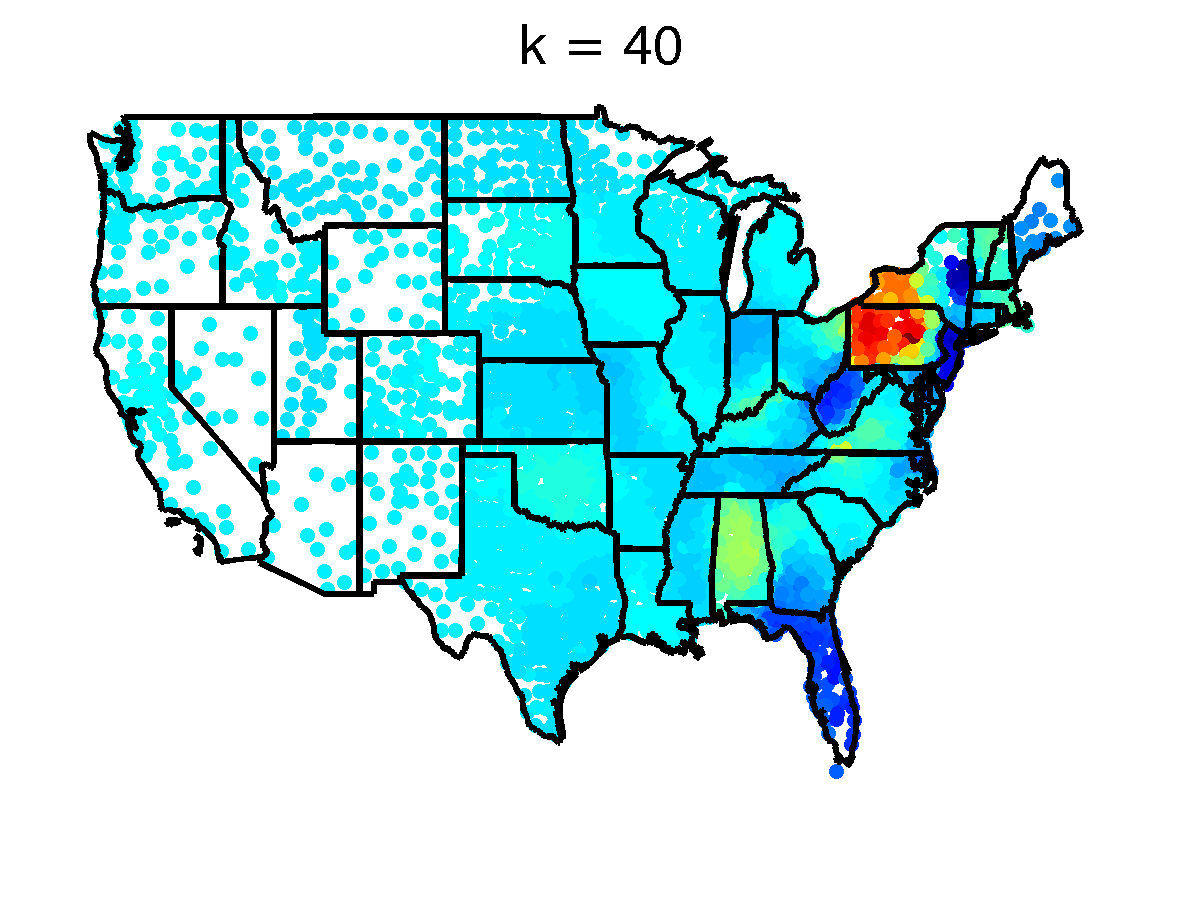}
\includegraphics[width=0.32 \textwidth]{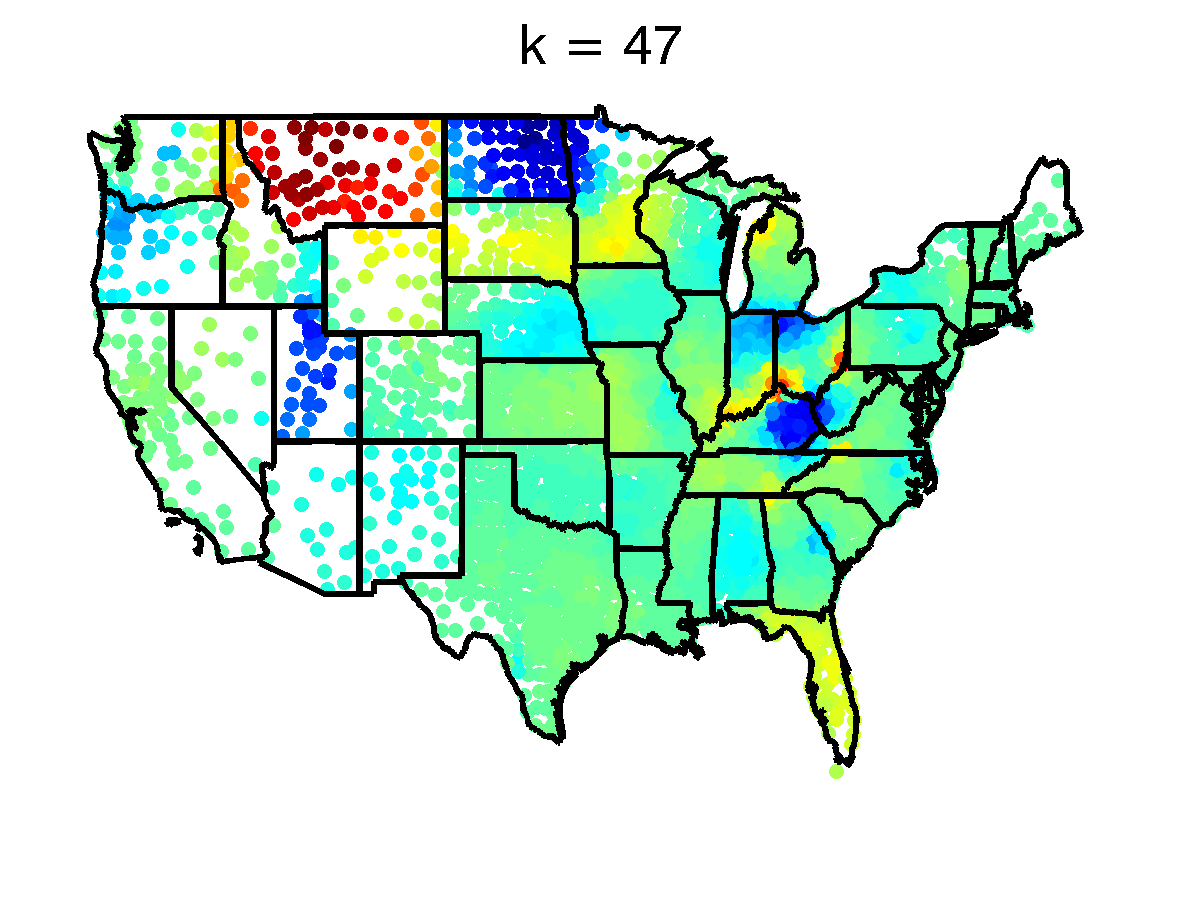}
\includegraphics[width=0.32 \textwidth]{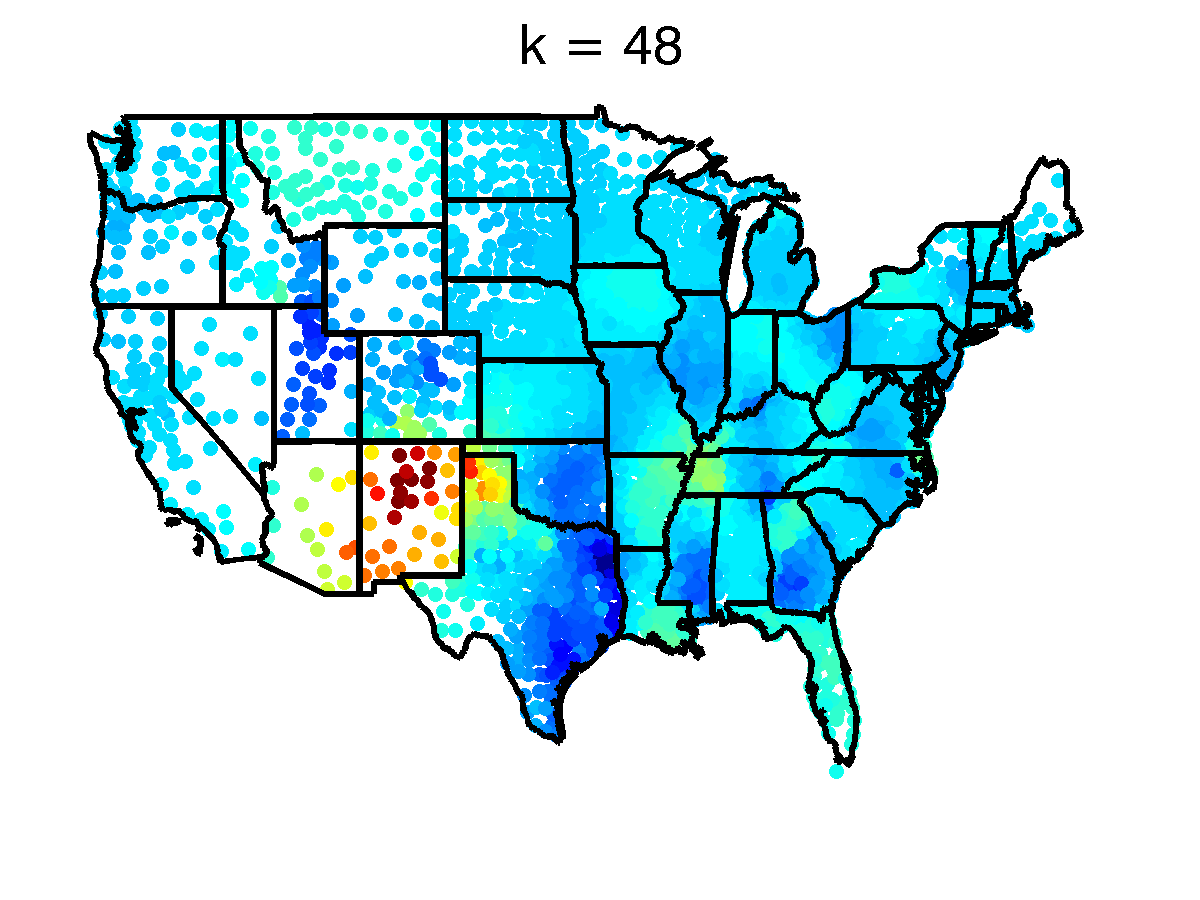}
\includegraphics[width=0.32 \textwidth]{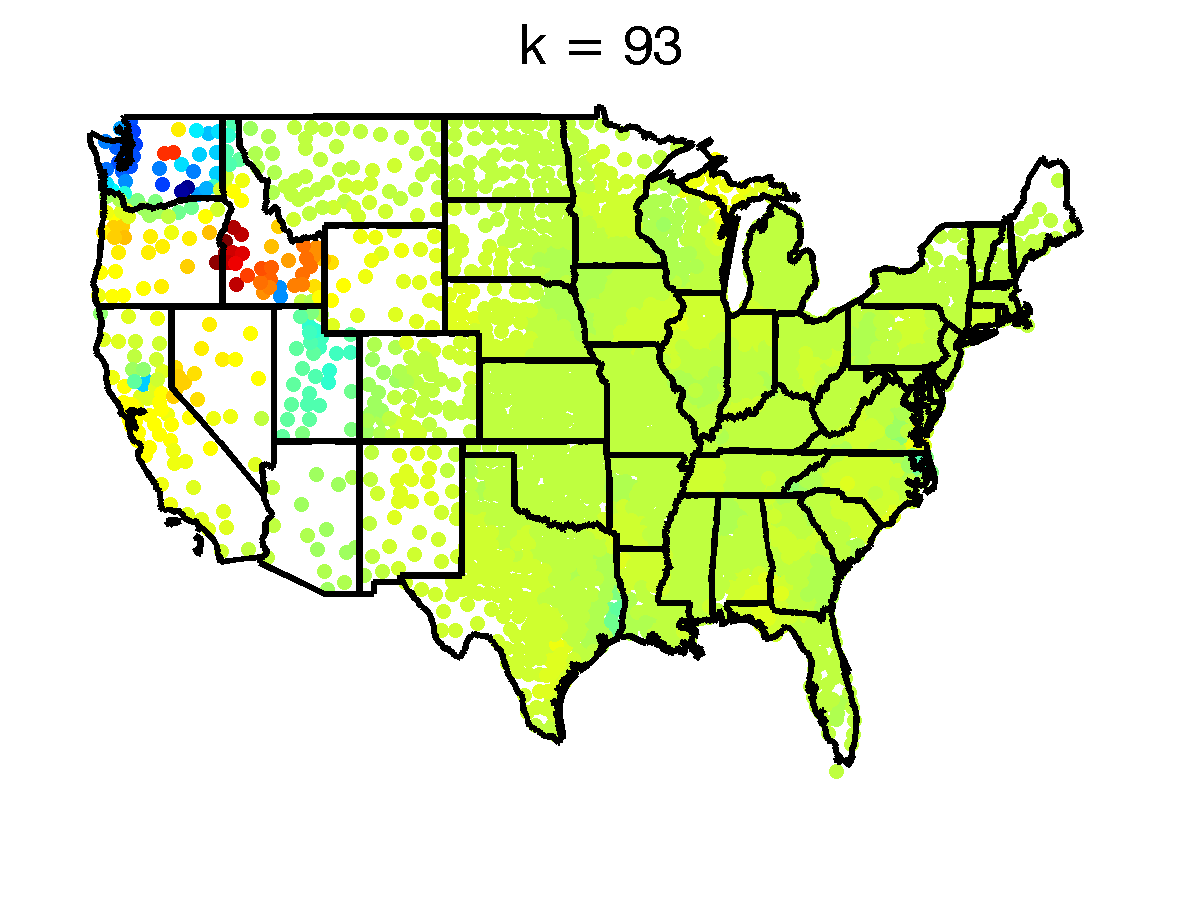}
\end{center}
\caption{Selected colorings by lower order eigenvectors for the similarity matrix $W_{ij} = \frac{M_{ij}^2}{P_i P_j}$.}
\label{fig:US_K1_next18}
\end{figure}

Our kernel of choice for the eigenvector colorings in Figures \ref{fig:US_K1_top18} and \ref{fig:US_K1_next18} was $W^{(1)}$, as it produced more visually appealing results in terms of state boundary detection. For the same reason, we omit the numerical simulations where we used exponential weights to compute the similarity between the nodes. Note also that the spectrum of $ A = D^{-1} W^{(1)}$ in the left of Figure \ref{fig:USspectra} is rather different from the other two spectra, with many more large eigenvalues and without a visible spectral gap. For the rest of this section,
we drop the superscript from matrix $W^{(1)}$ and refer to it as $W$.


\begin{figure}[h!t]
\centering
\subfigure[  $ \phi_{1}$]{
\includegraphics[width=0.23\textwidth]{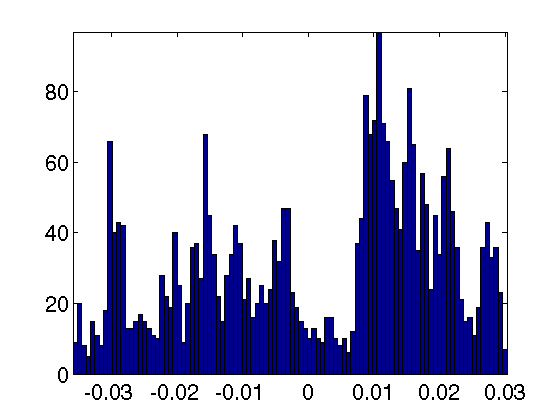}}
\subfigure[  $ \phi_{7}$]{
\includegraphics[width=0.23\textwidth]{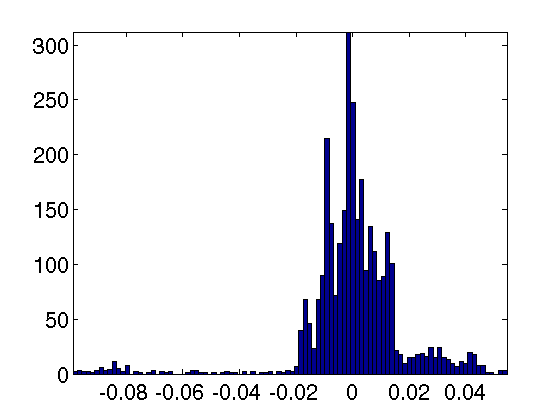}}
\subfigure[  $ \phi_{28}$]{
\includegraphics[width=0.23\textwidth]{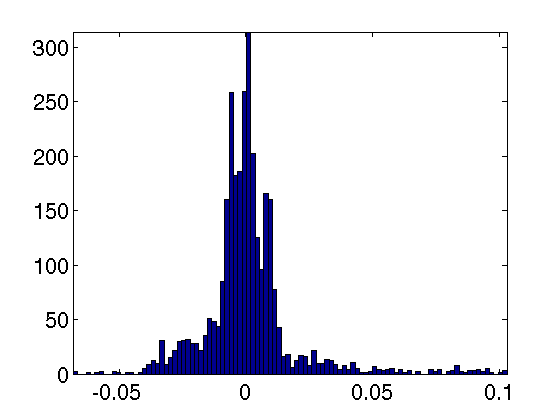}}
\subfigure[  $ \phi_{83} $]{
\includegraphics[width=0.23\textwidth]{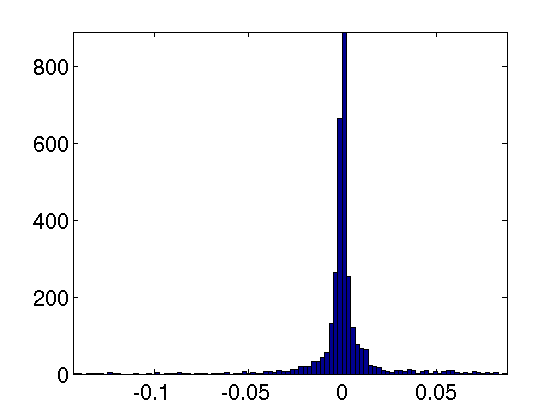}}
\caption[x]{Histogram of the entries in the eigenvectors $\phi_{1},\phi_{7},\phi_{28}$ and $\phi_{83}$ of matrix $ A = D^{-1} W^{(1)}$}
\label{fig:US_eigV_hist}
\end{figure}

In Figure \ref{fig:US_eigV_hist} we plot the histograms of the entries of  several eigenvectors of $A$. Note that the top eigenvector provides a meaningful partitioning that separates the East from the Midwest, and has its entries spread in the interval $[-0.03, 0.03]$ with few entries of zero magnitude. On the other hand, the eigenvectors $ \phi_{7}$, $\phi_{28}$ and $\phi_{83}$  are \textit{localized} in the sense that they have their larger entries localized on a specific subregion of the US map (highlighted in blue or red in the eigenvector colorings), while taking small values in magnitude on the rest of the domain. We explore in Section \ref{localization} the connection with the phenomenon of ``localized eigenfunctions" of the Laplace operator.

We use the rest of this section to provide a possible interpretation of the color coded regions that stand out in the eigenvector colorings in Figures \ref{fig:US_K1_top18} and \ref{fig:US_K1_next18}. By interpreting  the matrix $W$ as a weighted graph, we explore a possible connection of such geographically cohesive colored subgraphs with the graph partitioning problem. In general, the graph partitioning problem seeks to decompose a graph into $K$ disjoint subgraphs (clusters), while minimizing the sum of the weights of the ``cut" edges, i.e., edges with endpoints in different clusters. Given the number of clusters $K$, the Weighted-Min-Cut problem is an optimization problem that computes a partition $\mathcal{P}_1, \ldots ,\mathcal{P}_K$ of the vertex set, by minimizing the weights of the cut edges
\begin{equation}
 \mbox{Weighted Cut}(\mathcal{P}_1, \ldots ,\mathcal{P}_k) = \sum_{i=1}^{k} E_w(\mathcal{P}_i, \overline{\mathcal{P}_i}),
\label{wcut}
\end{equation}
where $E_w(X,Y) = \sum_{i \in X, j \in Y} W_{ij}$, and $\overline{X}$ denotes the complement of $X$.
For an extensive literature survey on spectral clustering algorithms we refer the reader to \cite{luxburg}, and point out the popular spectral relaxation of (\ref{wcut}) introduced by Shi and Malik \cite{shimalik}.

When dividing a graph into two smaller subgraphs, one wishes to minimize the sum of the weights on the edges across two different subgraphs, and simultaneously, maximize the sum of the weights on the edges within the subgraphs. Alternatively, one tries to maximize the ratio between the latter quantity and the former, i.e., between the weights of the inside edges and the weights of the outside edges. To that end, we perform the following experiment, where we regard the US states as the clusters, and investigate the possibility that the
isolated colored regions that emerge correspond to local cuts in the weighted graph.

\begin{figure}[h!t]
\begin{center}
\vspace{-3mm}
\includegraphics[width=1.1 \textwidth]{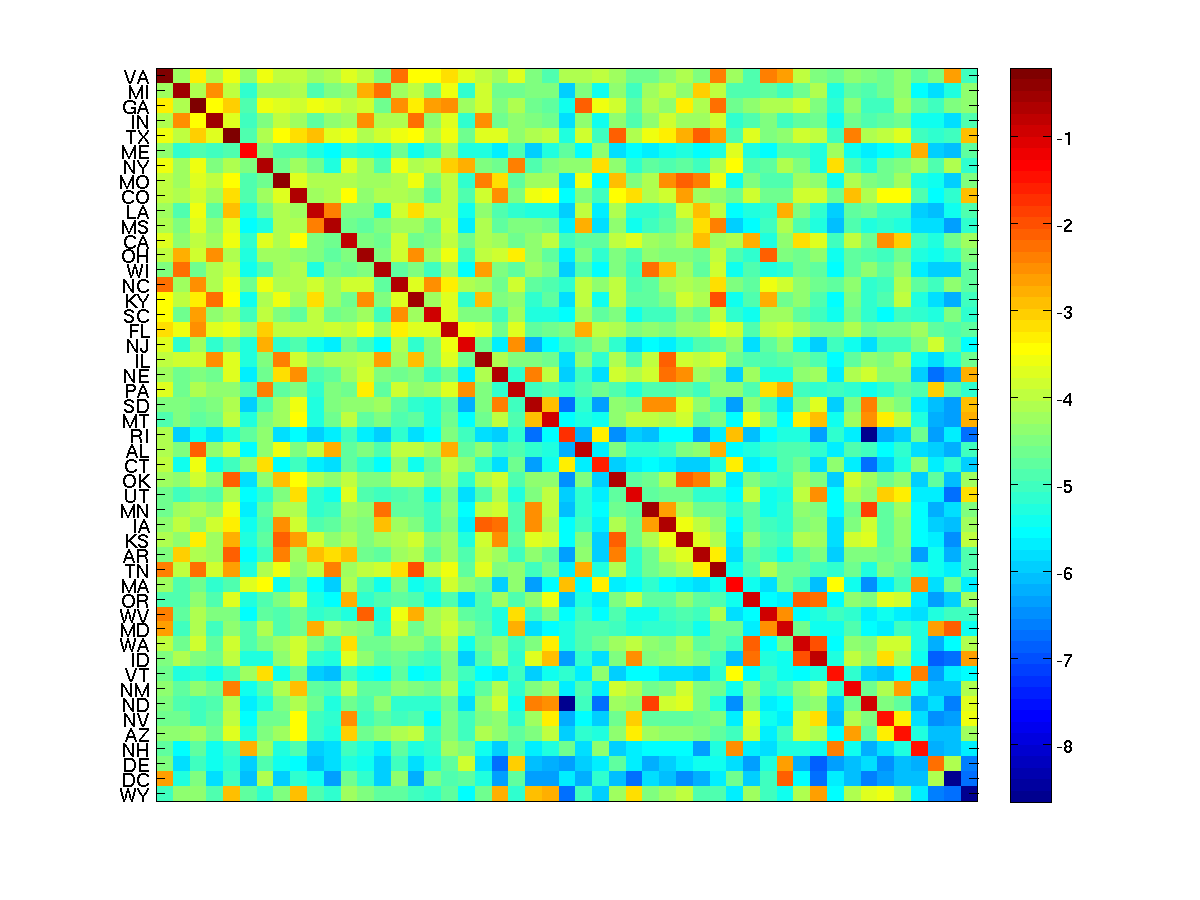}
\vspace{-8mm}
\end{center}
\caption{Heatmap of the inter-state migration flows, where the rows and columns of the matrix are sorted by the ratio degrees of the states. The intensity of entry $(i,j)$ denotes, on a logarithmic scale, the similarity between states $i$ and $j$, i.e., the sum of all entries in the submatrix $\tilde{W}_{i,j}$ defined in \ref{submatrix}. Table \ref{tab:top15states} lists the top 15 states in terms of ratio degree.
}
\label{fig:f3}
\end{figure}

We denote by $S$ the matrix of size $N \times N$ ($N=49$ the number of mainland US states) that aggregates the similarities between counties at the level of states. In particular, if state $i$ has $k$ counties with indices $x_1,\ldots,x_k$, and state $j$  has $l$ counties with indices $y_1,\ldots,y_l$, then we consider the $ k \times l $ submatrix
\begin{equation}
  \tilde{W}_{i,j} = W_{\{x_1,\ldots,x_k\}, \{y_1,\ldots,y_l\} }
\label{submatrix}
\end{equation}
and denote by $S_{ij}$ the sum of the $kl$ entries in  $\tilde{W}_{i,j}$.  In other words, matrix $S$ is a ``state-collapsed" version of the matrix $W$, and gives a measure of similarity between pairs of states.
The heatmap in figure \ref{fig:f3} shows the components of the matrix $S$ on a logarithmic scale, where the intensity of entry $(i,j)$ denotes the aggregated similarity between states $i$ and $j$.

We refer to the diagonal entry $S_{ii}$ as the ``inside degree" of state $i$, $d^{in}_i = S_{ii}$, which measures the internal similarity between the counties of state $i$. We denote by $d^{out}_i = \sum_{u=1, u \neq i}^{N} S_{i,u}$ (i.e., the sum of the non-diagonal elements in row $i$) the ``outside degree" of node $i$, which measures the similarity/migration between the counties of state $i$ and all other counties outside of state $i$. Finally, we denote by $ d_{i}^{ratio}=\frac{d^{in}_i}{d^{out}_i}$, the ``ratio degree" of node $i$ which straddles the boundary between  intra-state and inter-state migration. A large ratio degree is a good indicative that a state is very well connected internally, and has little connectivity with the outside world, and thus is a good candidate for a cluster. In other words, a large ``ratio degree" of a cluster (i.e., state) denotes a high measure of separation between that cluster and its environment, which is something discovered by the localization properties of the low-order eigenvectors. Table \ref{tab:top15states} ranks the top 15 states within the US in terms of their ratio degree.

Next, we examine the top several eigenvector colorings in Figure \ref{fig:US_K1_top18}, and point out the individual states on which the eigenvectors localize, together with its rank in terms of ``ratio degree". Note that the entries of large magnitude are colored in red and blue, while the rest of the spectrum denotes  values of smaller magnitude or very close to zero. The top three eigenvectors correspond to global cuts between various coasts within the US. The only state that stands out individually is Michigan (MI) for $k=3$, which has rank 2. For $k=4$, the largest entries correspond to counties in Virginia (VA) which is also ranked $1^{st}$, and similarly for Wisconsin (WI) for $k=5$, ranked 14. For $k=6$, the states colored in dark red and dark blue are Georgia (GA) with rank 3, and Missouri (MO) of rank 8. When $k=7$, Michigan (MI), of rank 2, stands out as the only dark blue colored state. For $k=8$, we point out Georgia, rank 3, together with Mississippi (MS) of rank 11, and Louisiana (LA) of rank 10. Eigenvector $k=9$ localizes mostly on Maine (ME) of rank 6, and the New York (NY) area with rank 7. Finally, eigenvector $k=10$ localizes on a combination of states we already pointed out. We have thus enumerated nine states that stand out in the top ten eigenvector colorings, and all nine of them appear in Table \ref{tab:top15states} that ranks the top fifteen states in terms of ``ratio degree". Although this experiment does not provide a formal justification for the eigenvector localization phenomenon, we believe it is a first step in providing evidence that the low order eigenvectors point out local cuts in the network.

\begin{table}[h]
\begin{minipage}[b]{0.3\linewidth}
 \begin{center}
\begin{tabular}{l|l|l}
rank & state & ratio degree \\
\hline
1.& VA& 26.7 \\
2.& MI& 20.4\\
3.& GA& 19.9\\
4.& IN&19.7\\
5.& TX& 19.0\\
6.& ME& 18.9\\
7.& NY& 18.7\\
8.& MO&  18.5\\
9.& CO & 17.1\\
10.& LA&  16.6\\
11.& MS&  16.1\\
12.& CA & 15.7\\
13.& OH&  15.6\\
14.& WI&  14.5\\
15.& NC&  14.4\\
\end{tabular}
\label{tab:top15states}
\end{center}
\end{minipage} \;\;\;\;\;\;
\begin{minipage}[b]{0.7\linewidth}
\centering
\includegraphics[width=1 \textwidth]{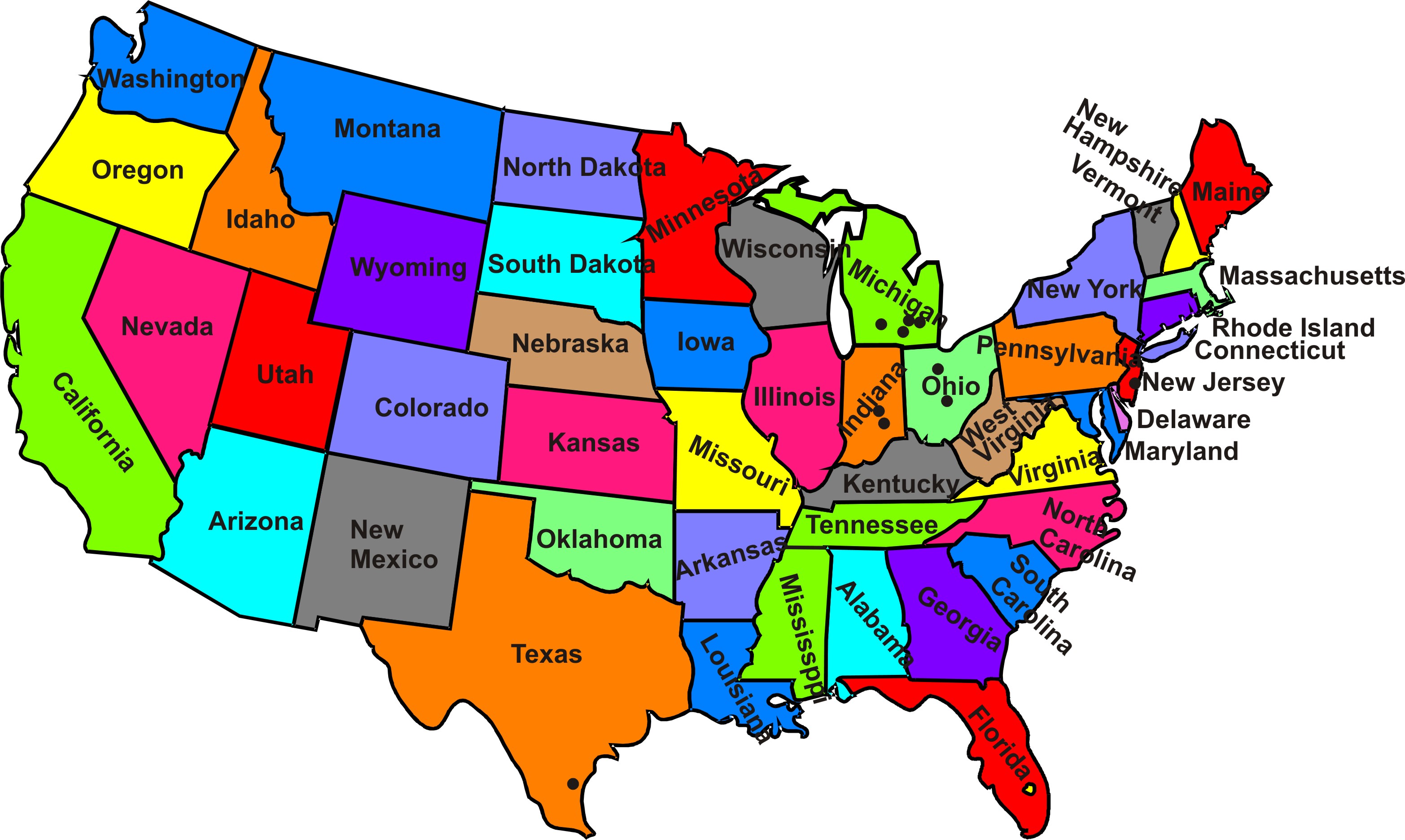}
\vspace{3mm}
\label{tab:ANE_1d3z}
\end{minipage}
\caption{Left: top 15 states within the US, ordered by ratio degree. Right: map of mainland US showing the names of the states.}
\vspace{-5mm}
\end{table}

\section{Belgium Mobile Network}
\label{BelgiumSection}

In a recent work \cite{gravity}, we studied the anonymized mobile phone communication from a Belgian operator and derived a statistical model of interaction between cities, showing that inter-city communication intensity is characterized be a gravity model: the communication intensity between two cities is proportional to the product of their sizes divided by the square of their distance. In this section, we briefly describe the Belgium mobile data set, summarize the results in \cite{gravity}, and apply the diffusion map technique. We refer the reader to \cite{unfolding} for more information on the mobile phone data set.

The data set  contains anonymous communication patterns of 2.5 million mobile phone customers, grouped in 571 cities in Belgium over a period of six months in 2006 (see also  \cite{lamb} for a description of the data set). Every customer is associated with the ZIP code of her/his billing address. Note that calls involving other operators were filtered out, meaning that both the calling and receiving individuals in the data set are customers of the mobile phone company. Also, there is a link between two customers if at least  three calls were made in both directions during the six month interval. After this pre-processing, the network has 2.5 million nodes and 38 million links. For every pair of customers  we associate a communication intensity by computing the total communication time in seconds. After grouping the customers into their corresponding cities, we compute $T_{ij}$, the aggregated communication time in seconds between the customers of city $i$ and $j$, and denote the resulting matrix by $T = (T_{ij})_{1\leq i<j \leq n}$. We denote by $N_{ij}$ the number of phone calls between cities $i$ and $j$, by $R_{ij}=\frac{T_{ij}}{N_{ij}}$ the average duration of a call, and by $P_i$ the number of customers that have the zip code billing address of city $i$ (from now on, we refer to $P_i$ as the population of city $i$). Furthermore, the normalized number of phone calls with respect to the population of the cities is denoted by $\bar{N}_{ij} = \frac{N_{ij}}{P_{i} P_{j}}$, and similarly the normalized communication time by $\bar{T}_{ij} = \frac{T_{ij}}{P_{i} P_{j}}$. Finally, $D=(d_{ij})_{1 \leq i<j \leq n}$ represents the distances between the centroids of the areas of cities $i$ and $j$. 
Using these quantities, we now consider the following three kernels: 
$W^{(1)}_{ij} = e^{- \left( R_{ij} \bar{T}_{ij} \right)^2 / 0.2^2}$, 
$W^{(2)}_{ij} = e^{- \left( \frac{R_{ij}^{0.16}}{\bar{N}_{ij}^{0.26}} \right)^2}$, and 
$W^{(3)}_{ij} = \frac{ \bar{T}_{ij} }{  R_{ij} } =  \frac{N_{ij}}{P_i P_j}$.

\begin{figure}[h!t]
\centering
\subfigure[Map of Belgium, colored by latitude]{
\includegraphics[width=0.47\textwidth]{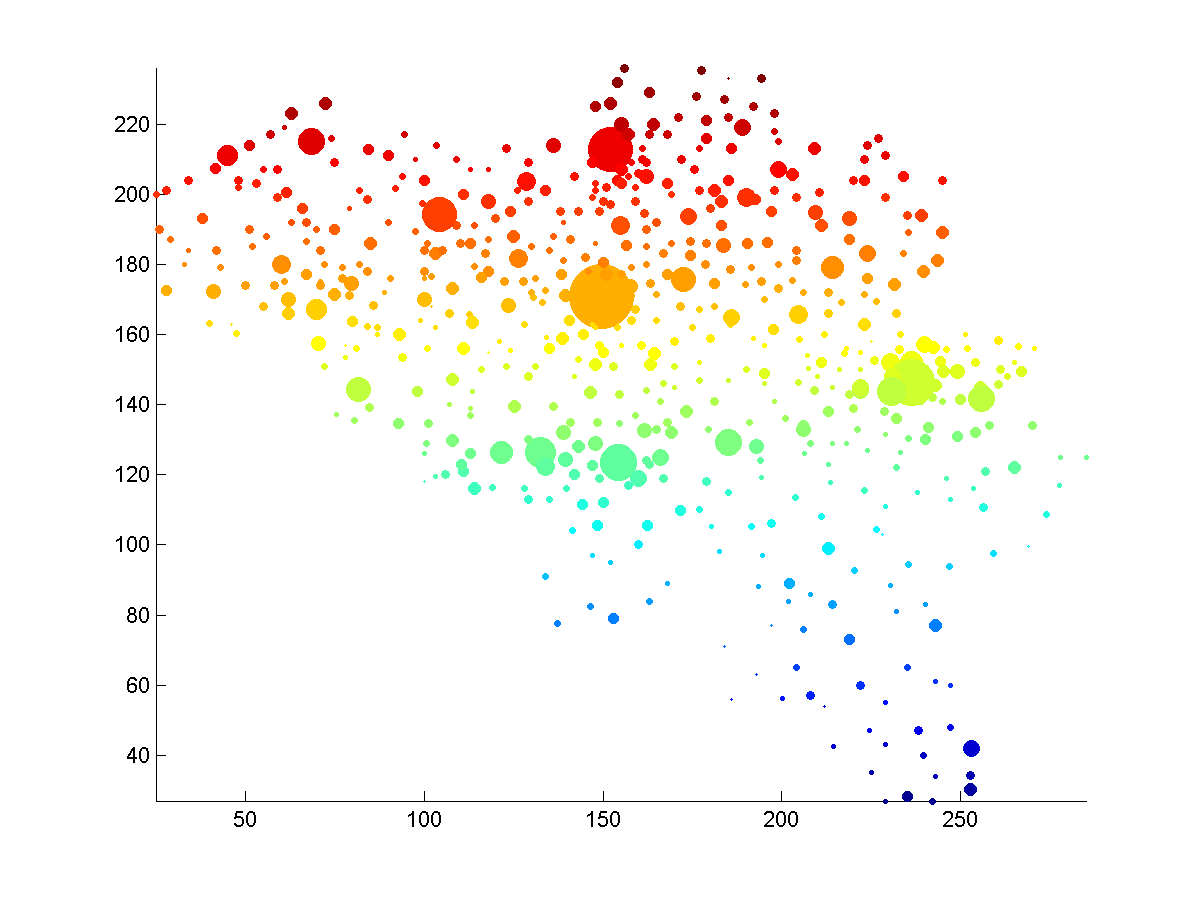}}
\subfigure[ Kernel $W^{(1)}_{ij} = e^{- \left( R_{ij} \bar{T}_{ij} \right)^2 / 0.2^2} $]{  
\includegraphics[width=0.47\textwidth]{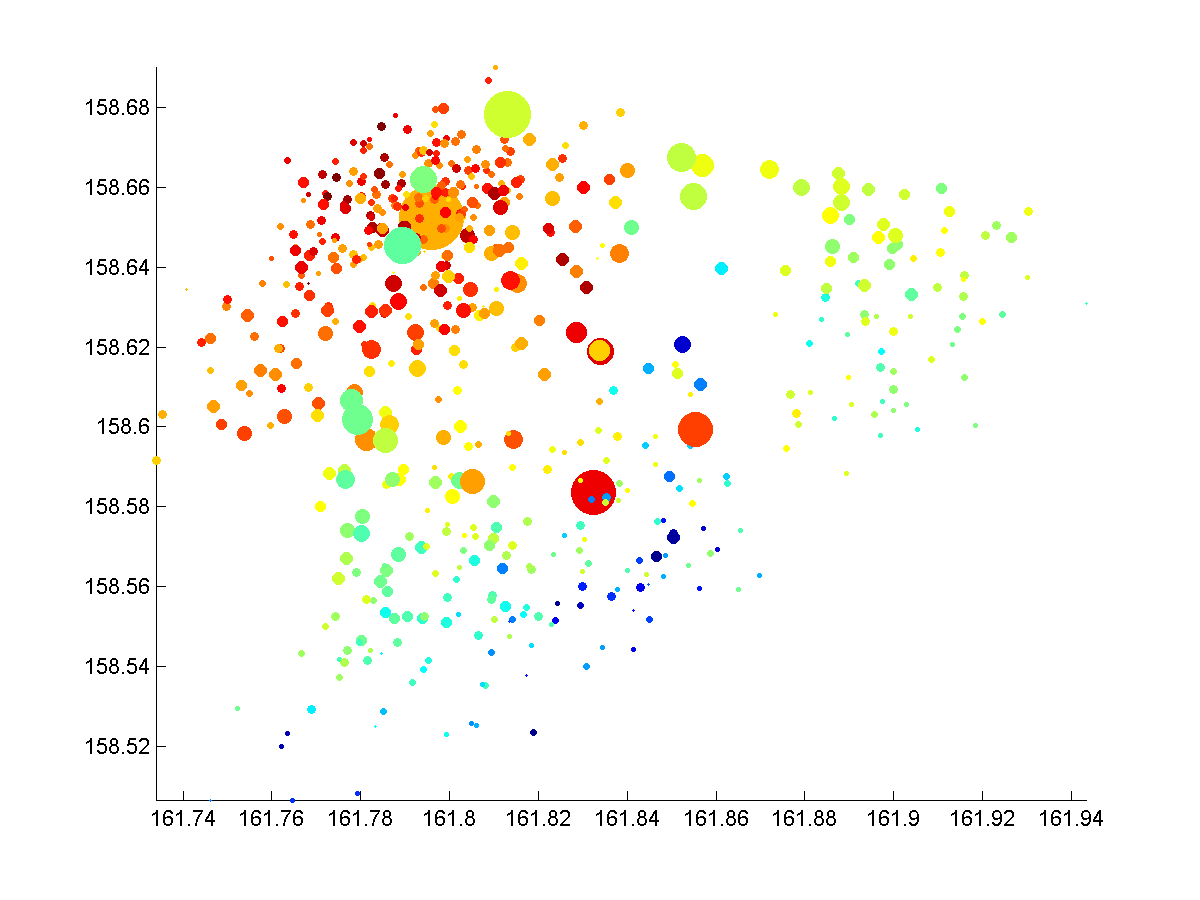}}
\subfigure[ Kernel $W^{(2)}_{ij} = e^{- \left( \frac{R_{ij}^{0.16}}{\bar{N}_{ij}^{0.26}} \right)^2 }$]{ 
\includegraphics[width=0.47\textwidth]{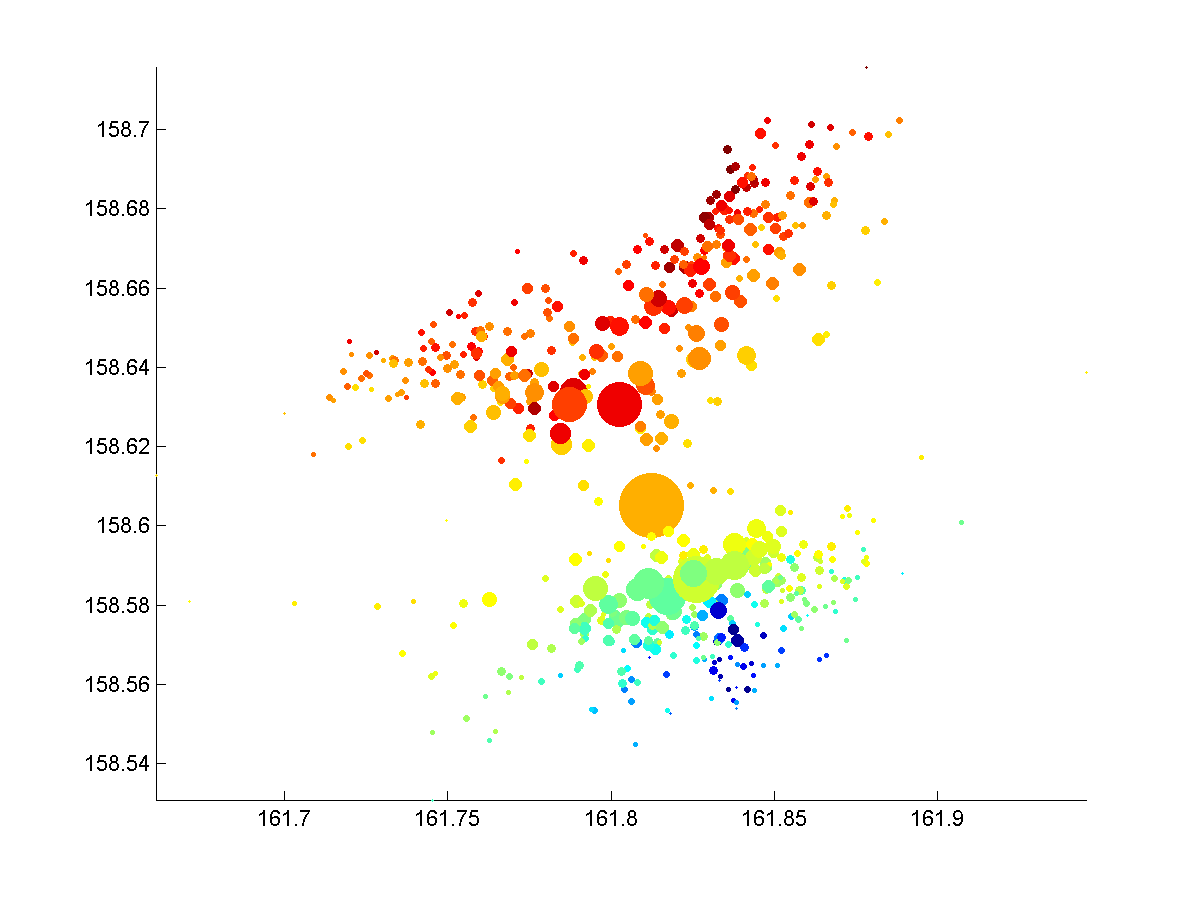}}
\subfigure[ Kernel $W^{(3)}_{ij} = \frac{ \bar{T}_{ij} }{  R_{ij} } =  \frac{N_{ij}}{P_i P_j} $]{
\includegraphics[width=0.47\textwidth]{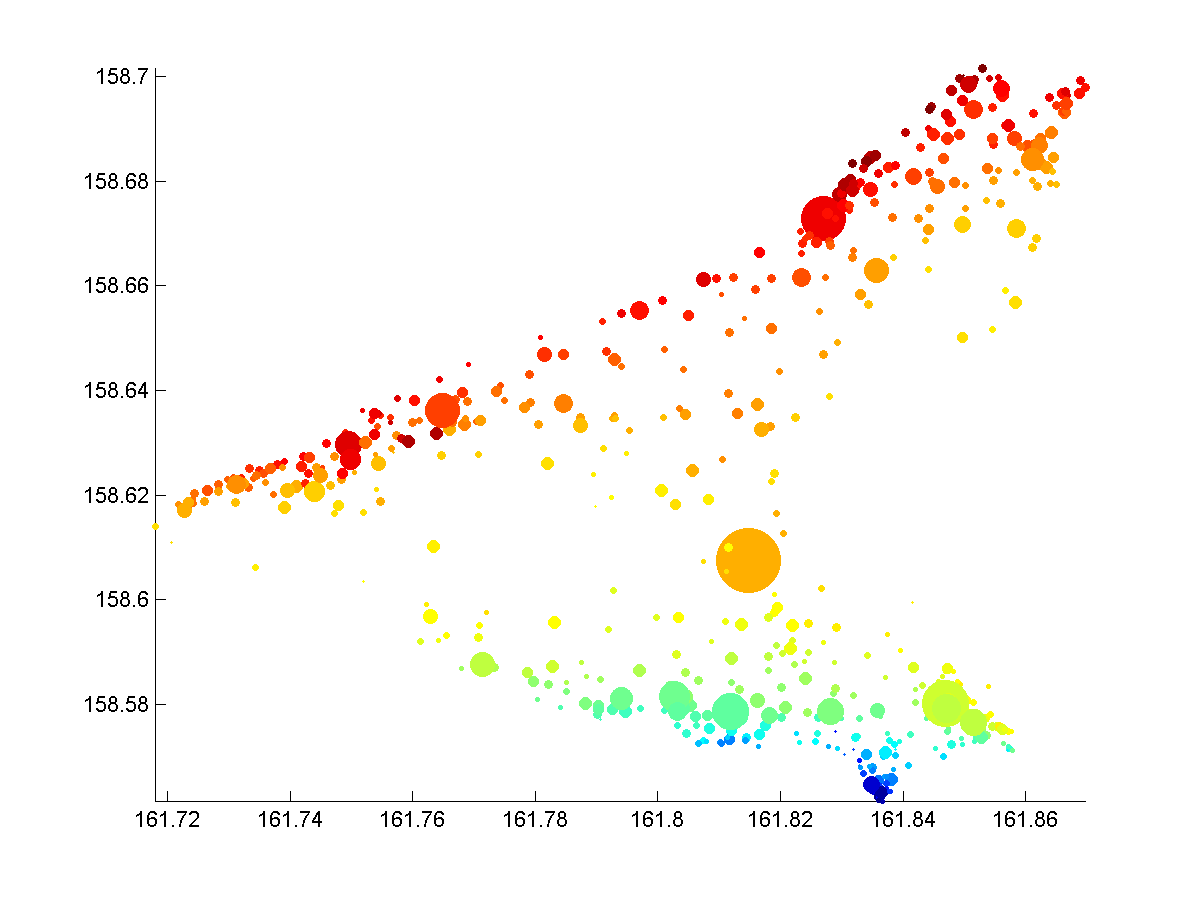}}
\caption[x]{Diffusion map reconstructions from the top two eigenvectors, for various kernels. $T_{ij}$ denotes the aggregated communication time in seconds,  $N_{ij}$ the number of phone calls between cities $i$ and $j$, $R_{ij}=\frac{T_{ij}}{N_{ij}}$ the average duration of a call, and $P_i$ the population of city $i$. We normalize by the population size by defining $\bar{N}_{ij} = \frac{N_{ij}}{P_{i} P_{j}}$  and  $\bar{T}_{ij} = \frac{T_{ij}}{P_{i} P_{j}}$.}
\label{fig:mob_2D_dif}
\end{figure}

\begin{figure}[h!t]
\begin{center}
\includegraphics[width=0.32 \textwidth]{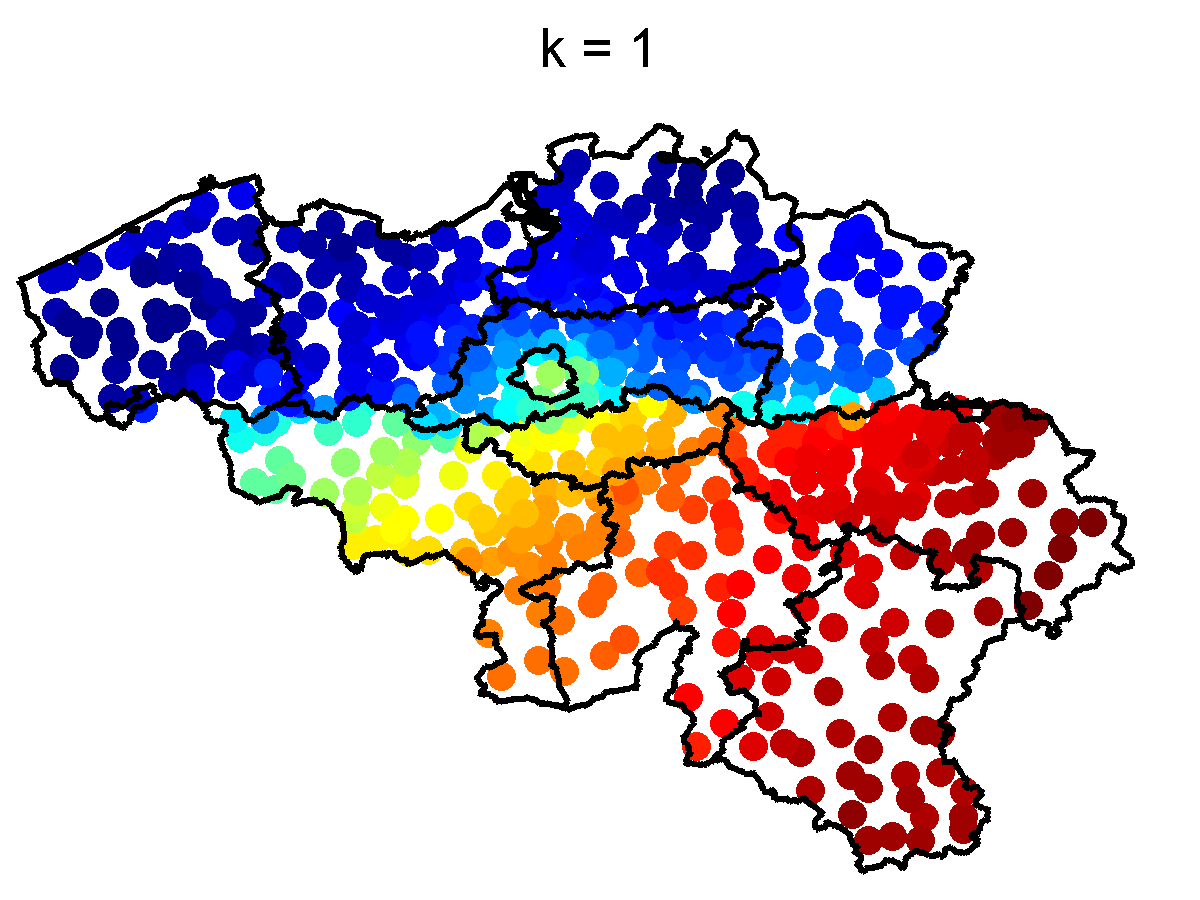}
\includegraphics[width=0.32 \textwidth]{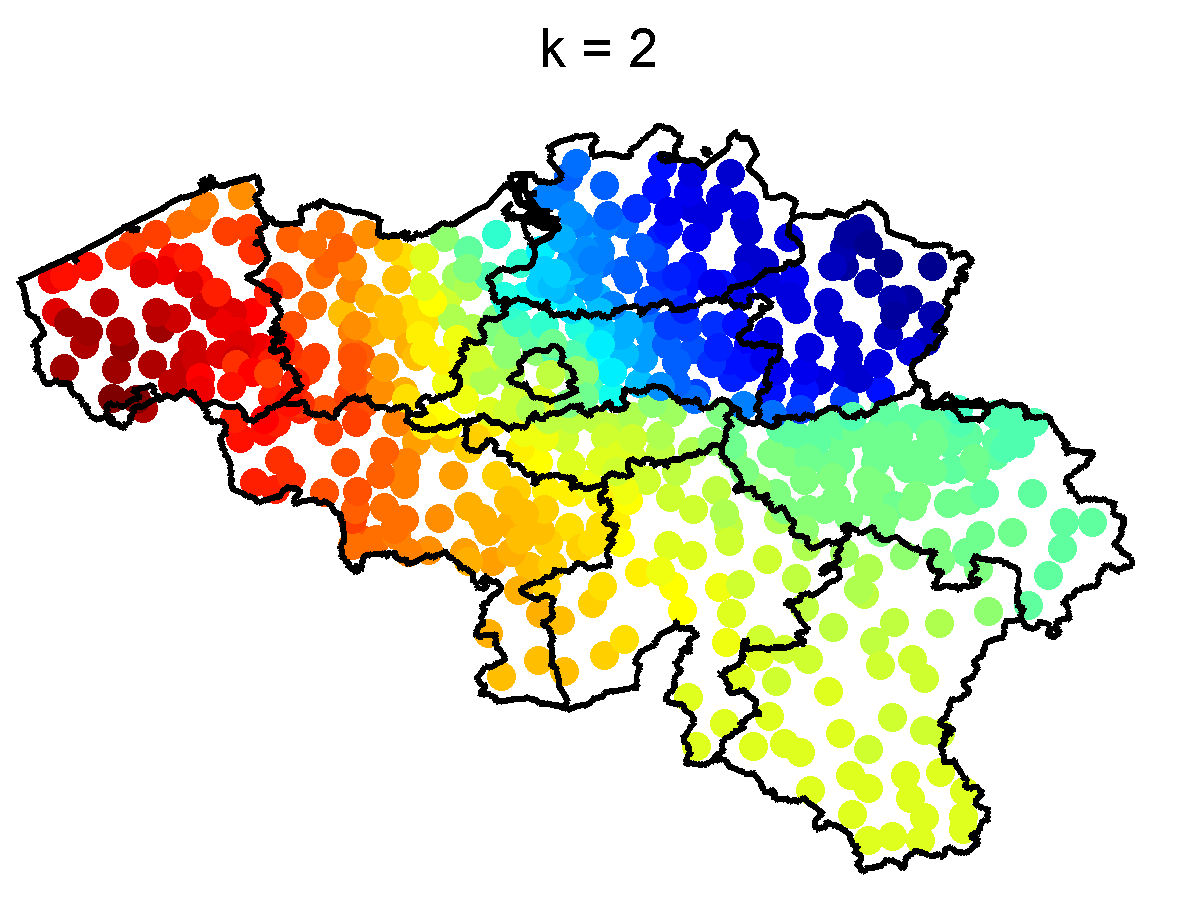}
\includegraphics[width=0.32 \textwidth]{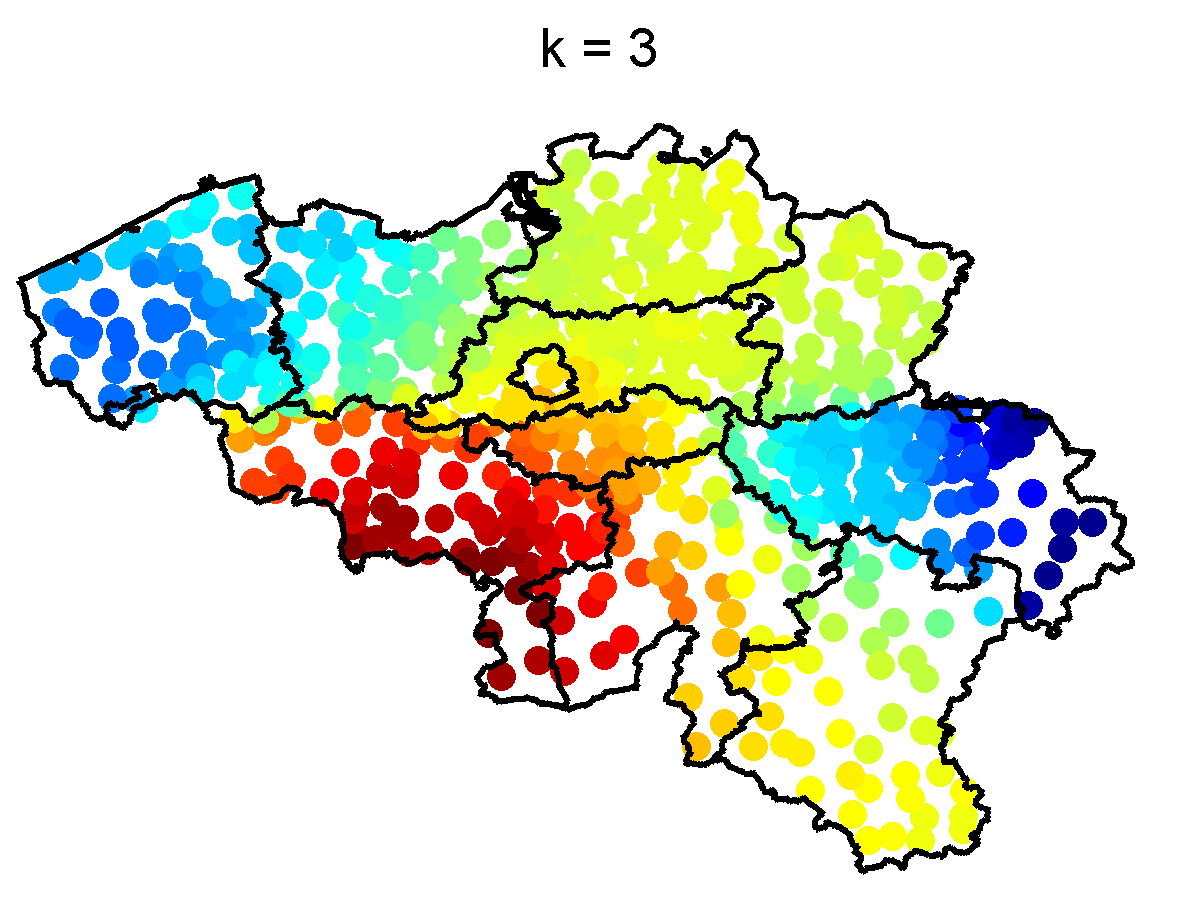}
 \includegraphics[width=0.32 \textwidth]{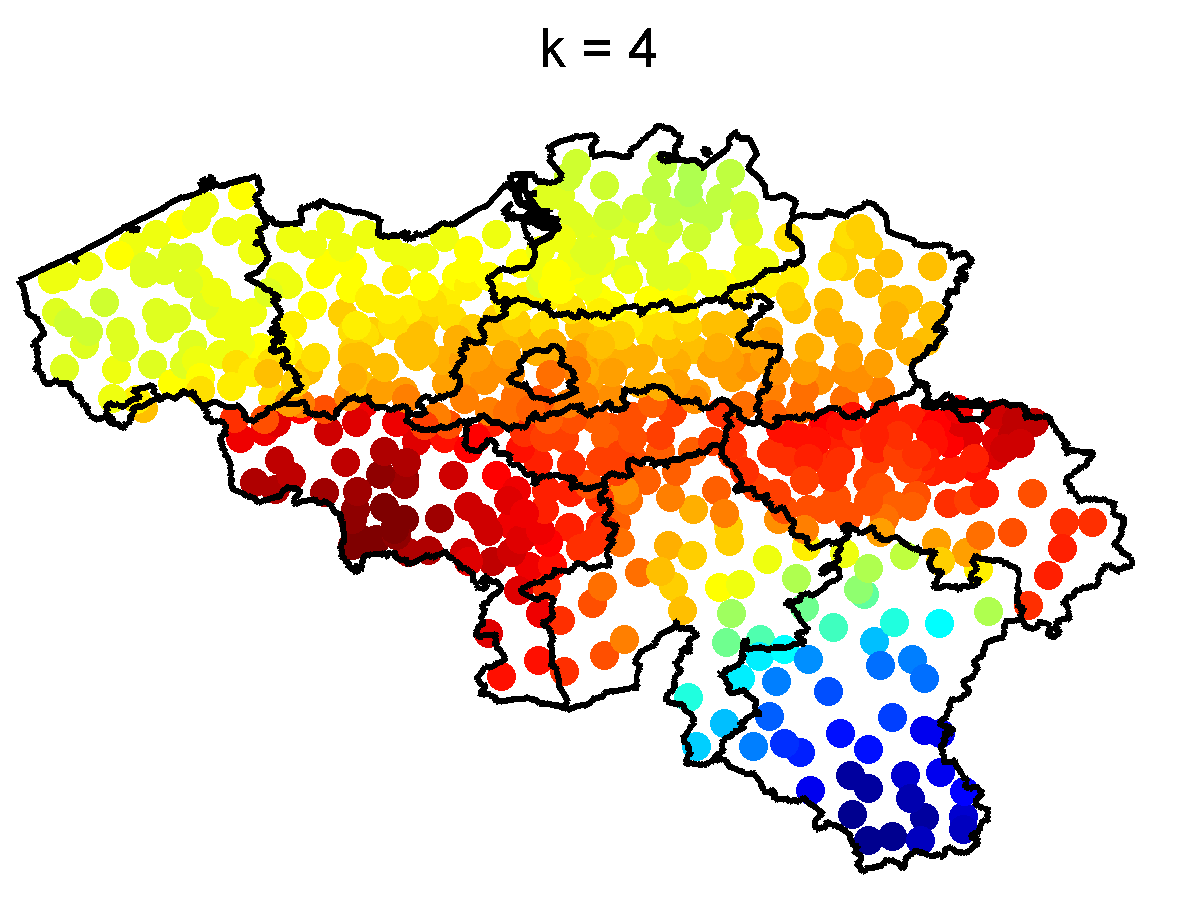}
 \includegraphics[width=0.32 \textwidth]{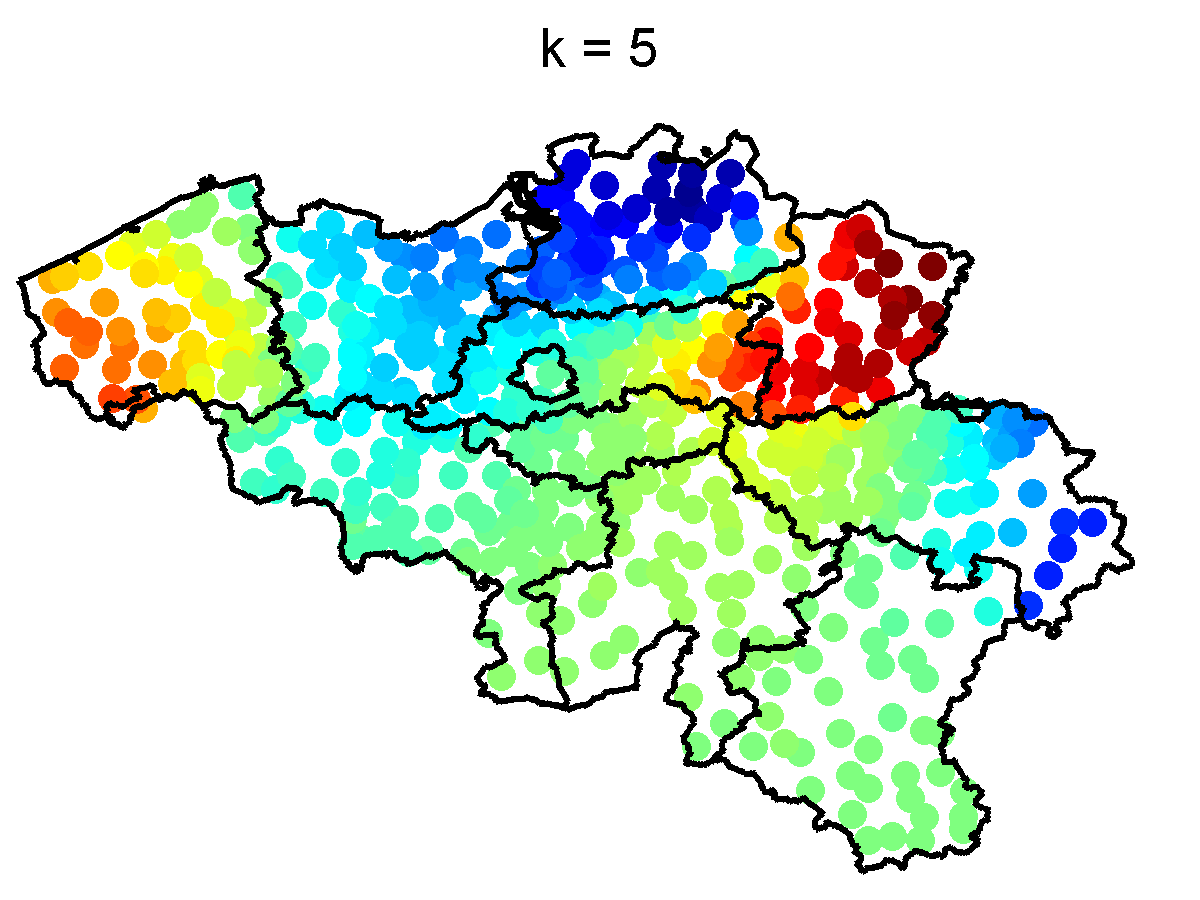}
 \includegraphics[width=0.32 \textwidth]{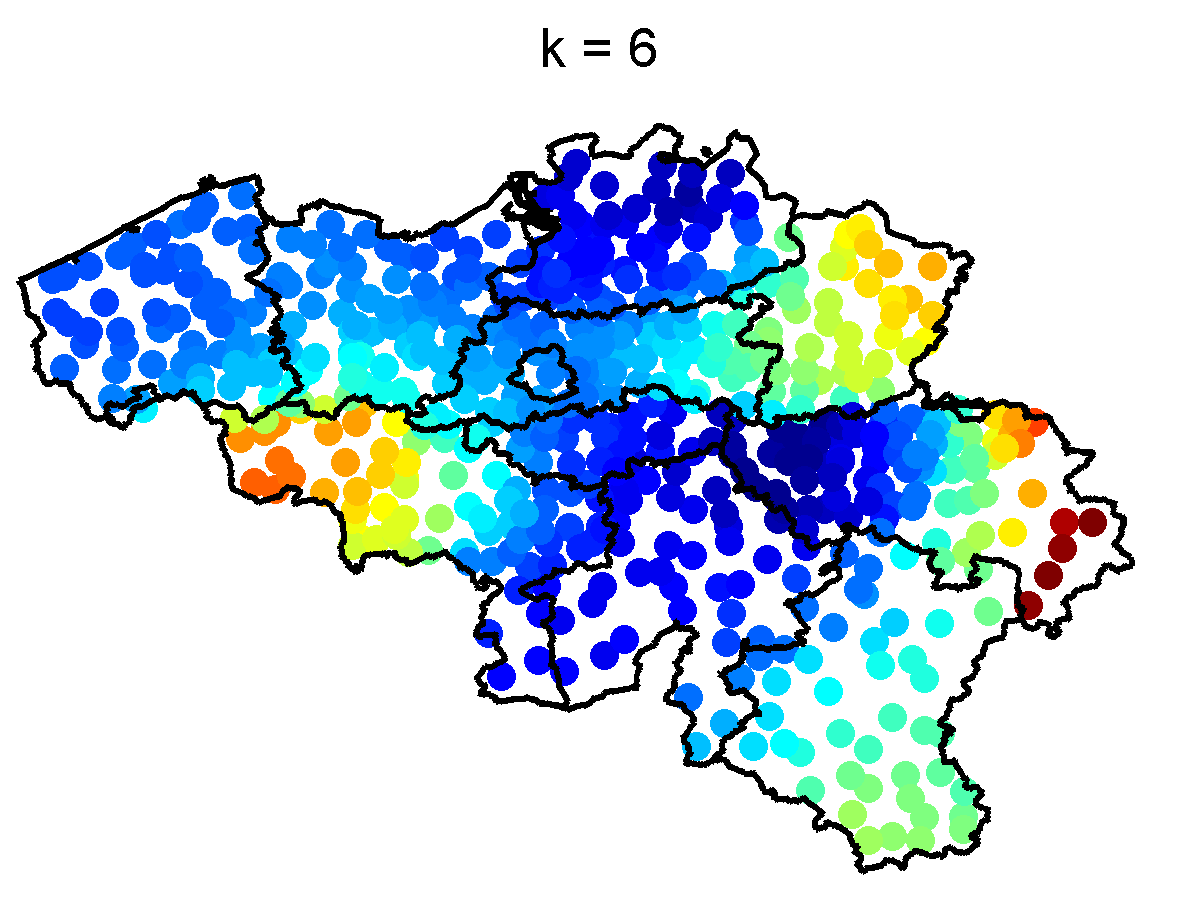}
 \includegraphics[width=0.32 \textwidth]{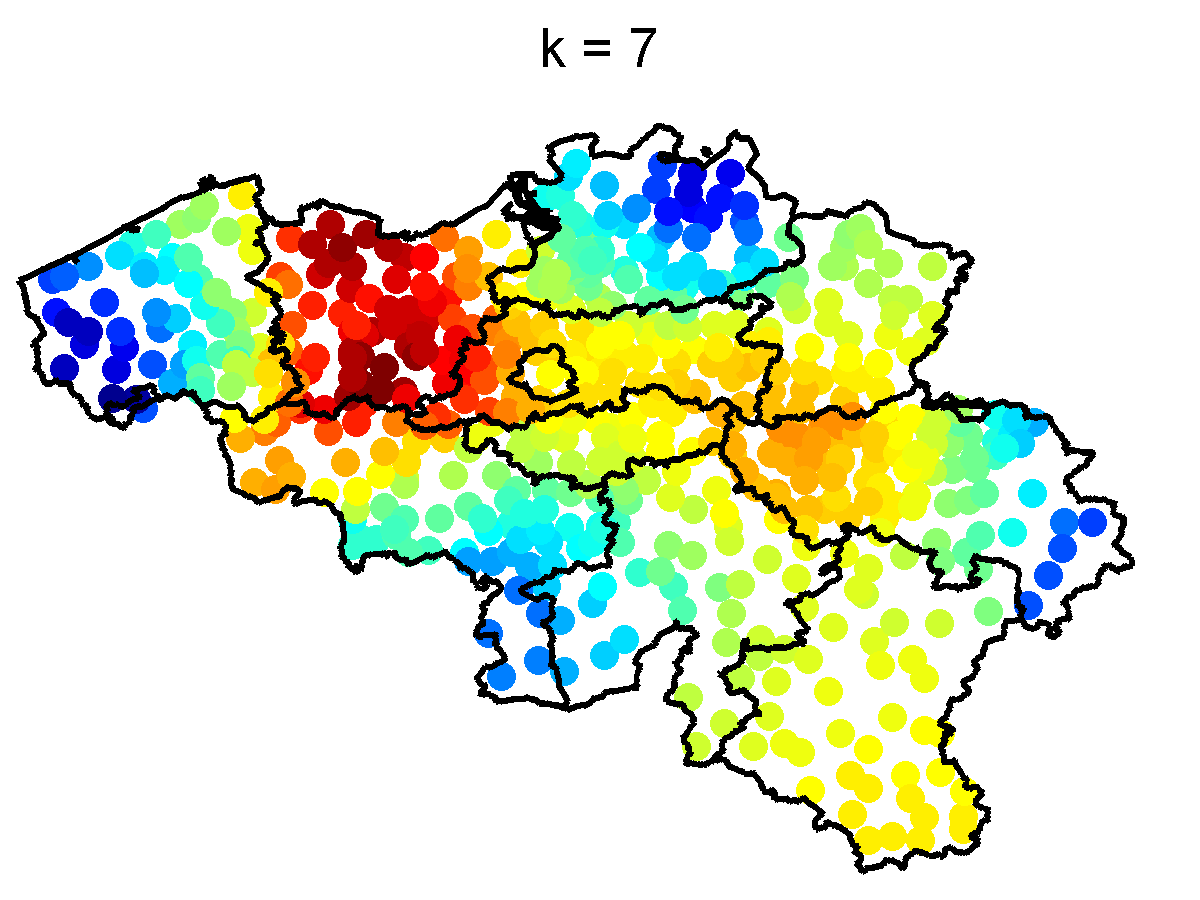}
 \includegraphics[width=0.32 \textwidth]{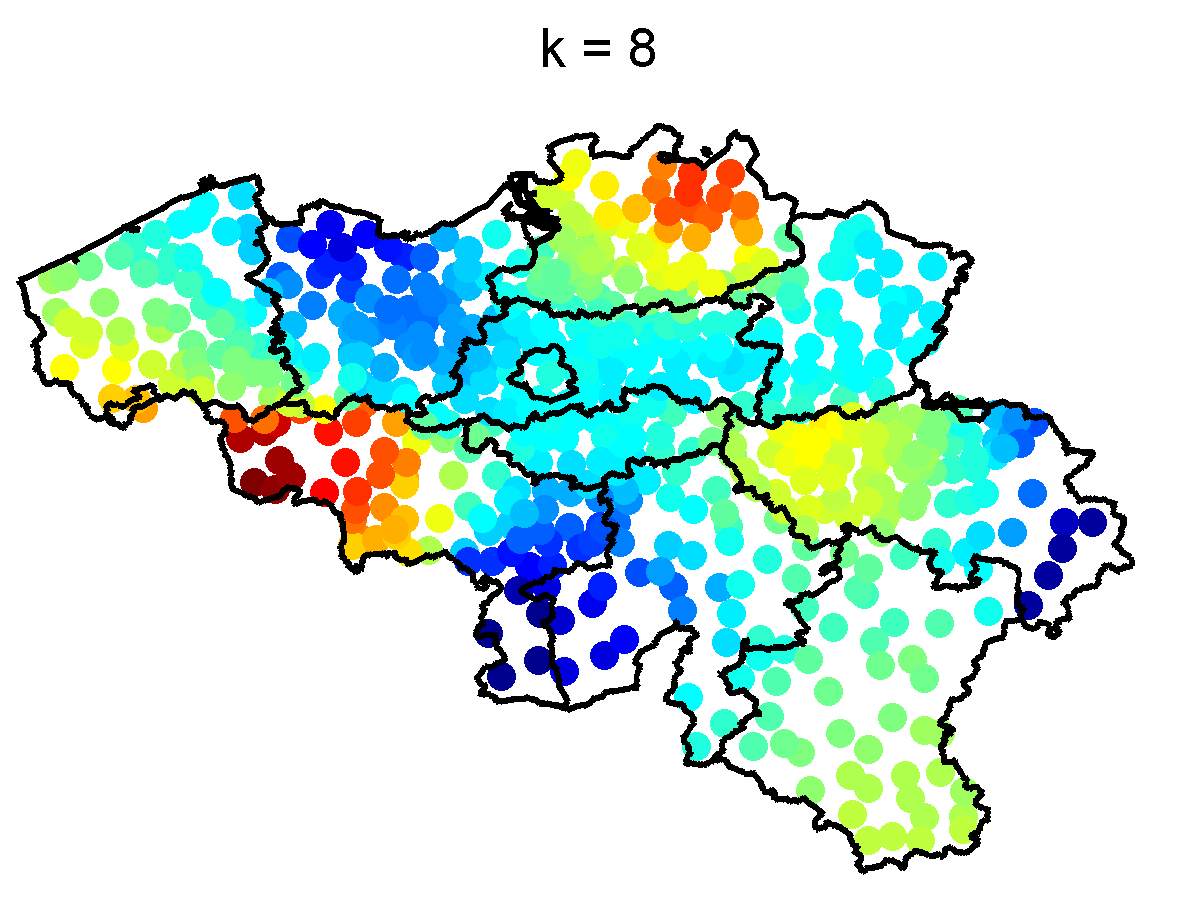}
 \includegraphics[width=0.32 \textwidth]{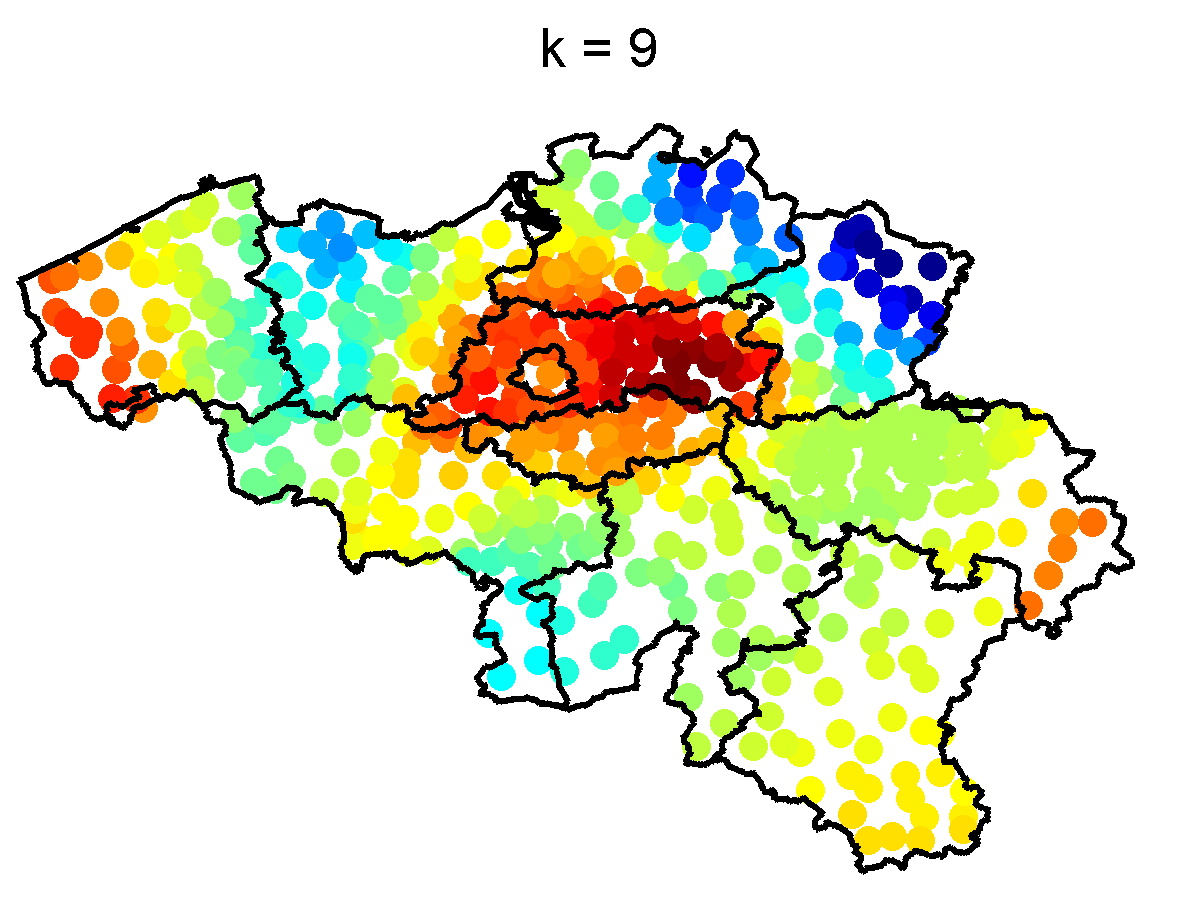}
 \includegraphics[width=0.32 \textwidth]{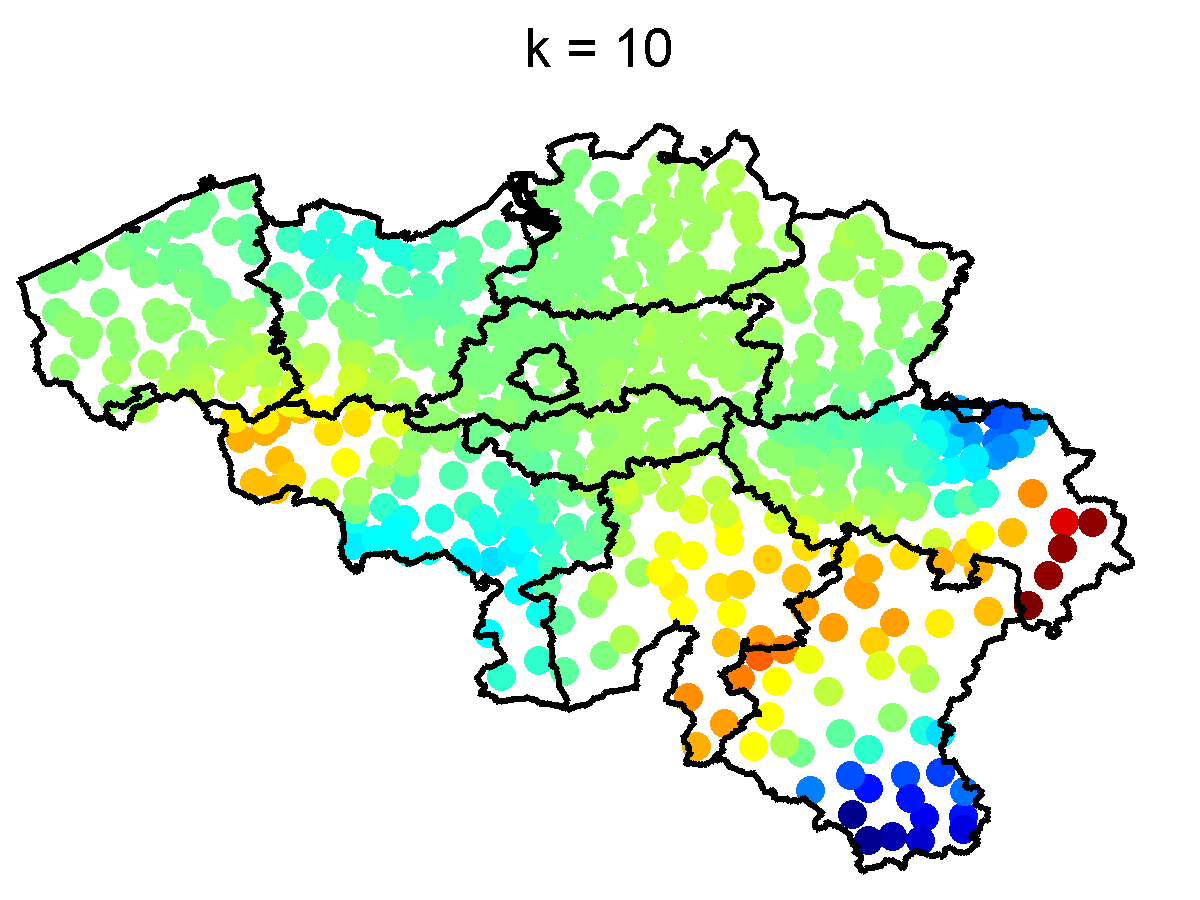}
 \includegraphics[width=0.32 \textwidth]{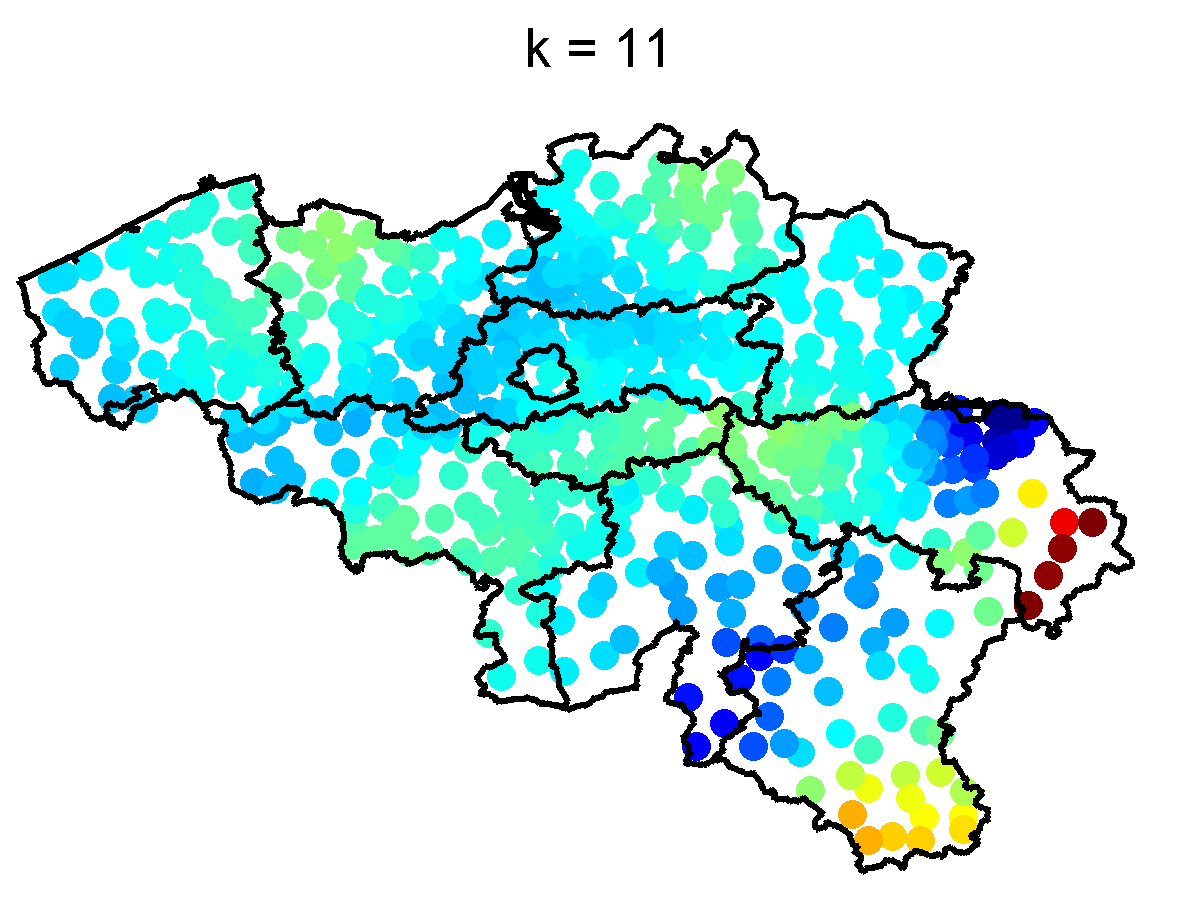}
 \includegraphics[width=0.32 \textwidth]{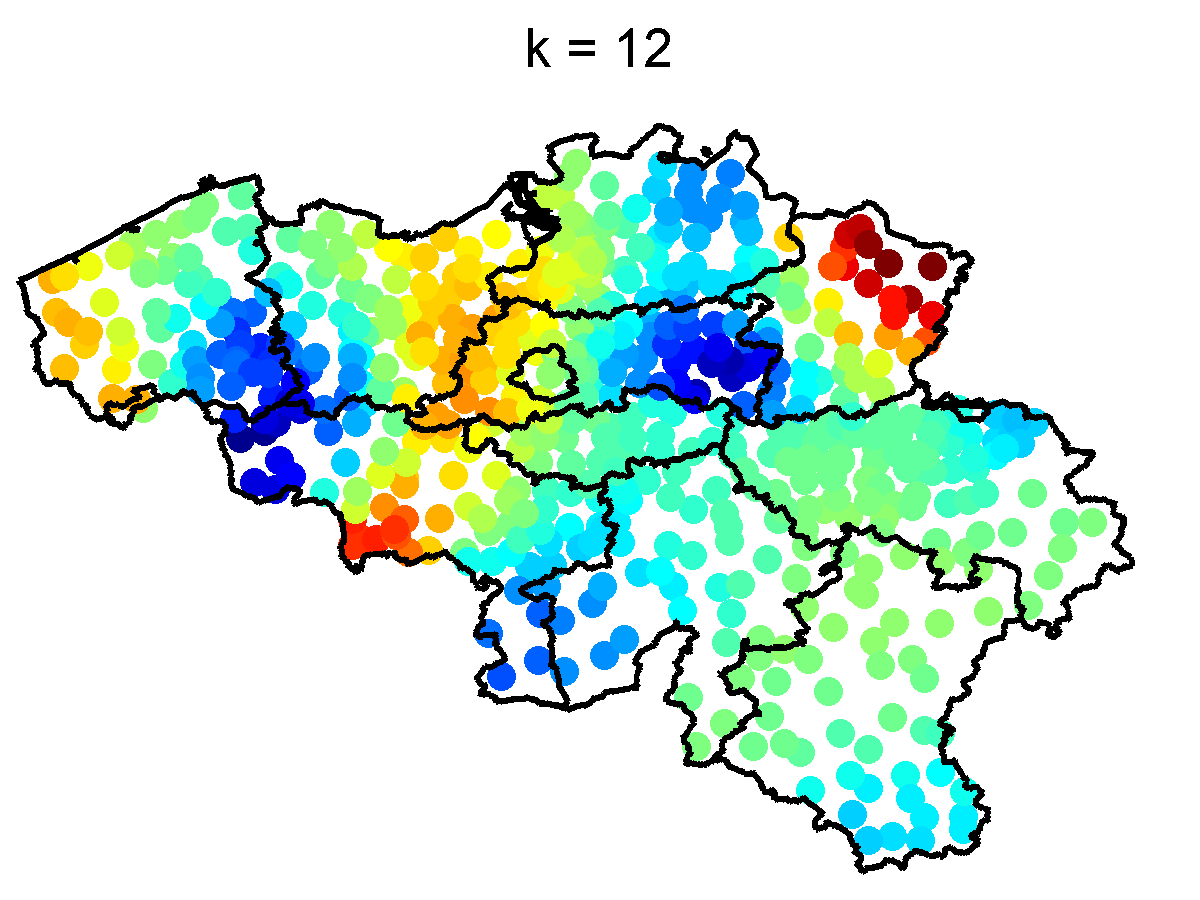}
 \includegraphics[width=0.32 \textwidth]{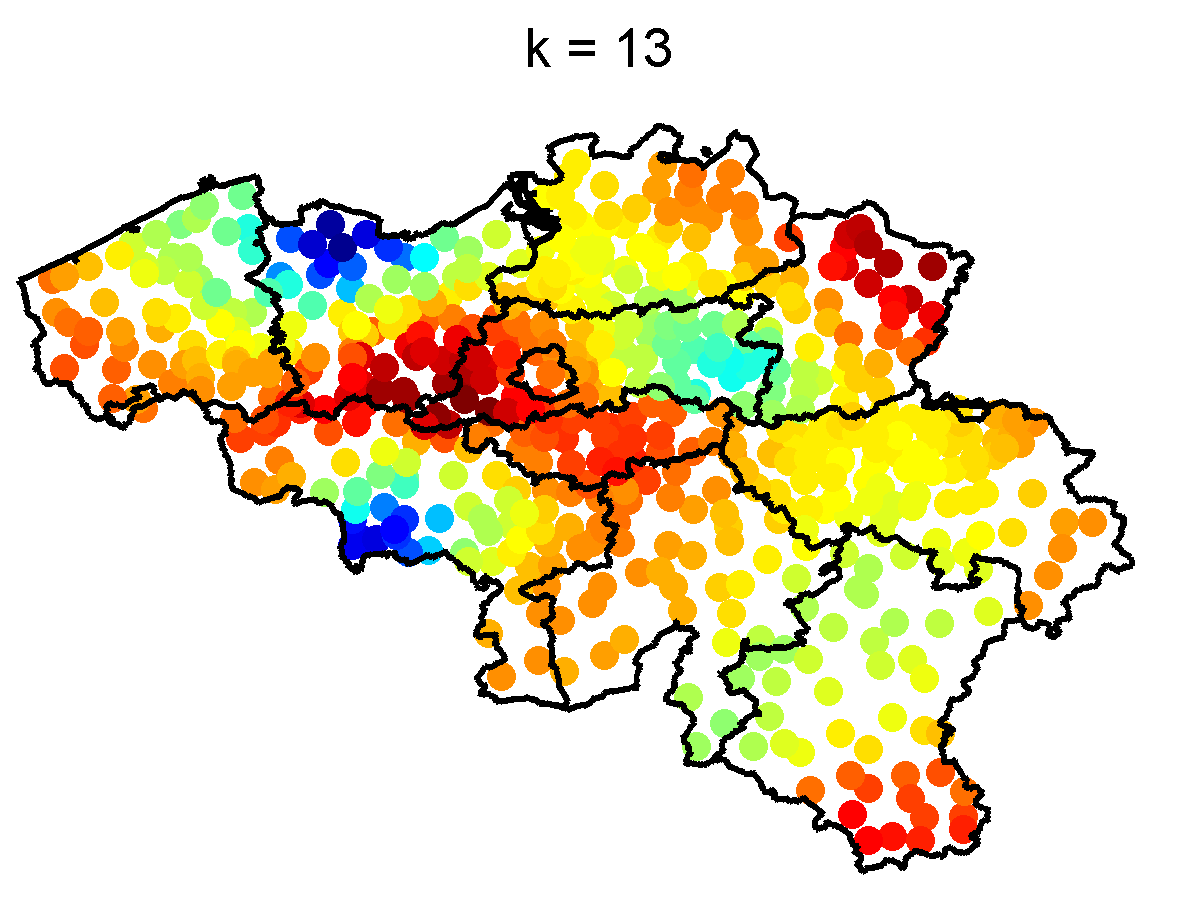}
 \includegraphics[width=0.32 \textwidth]{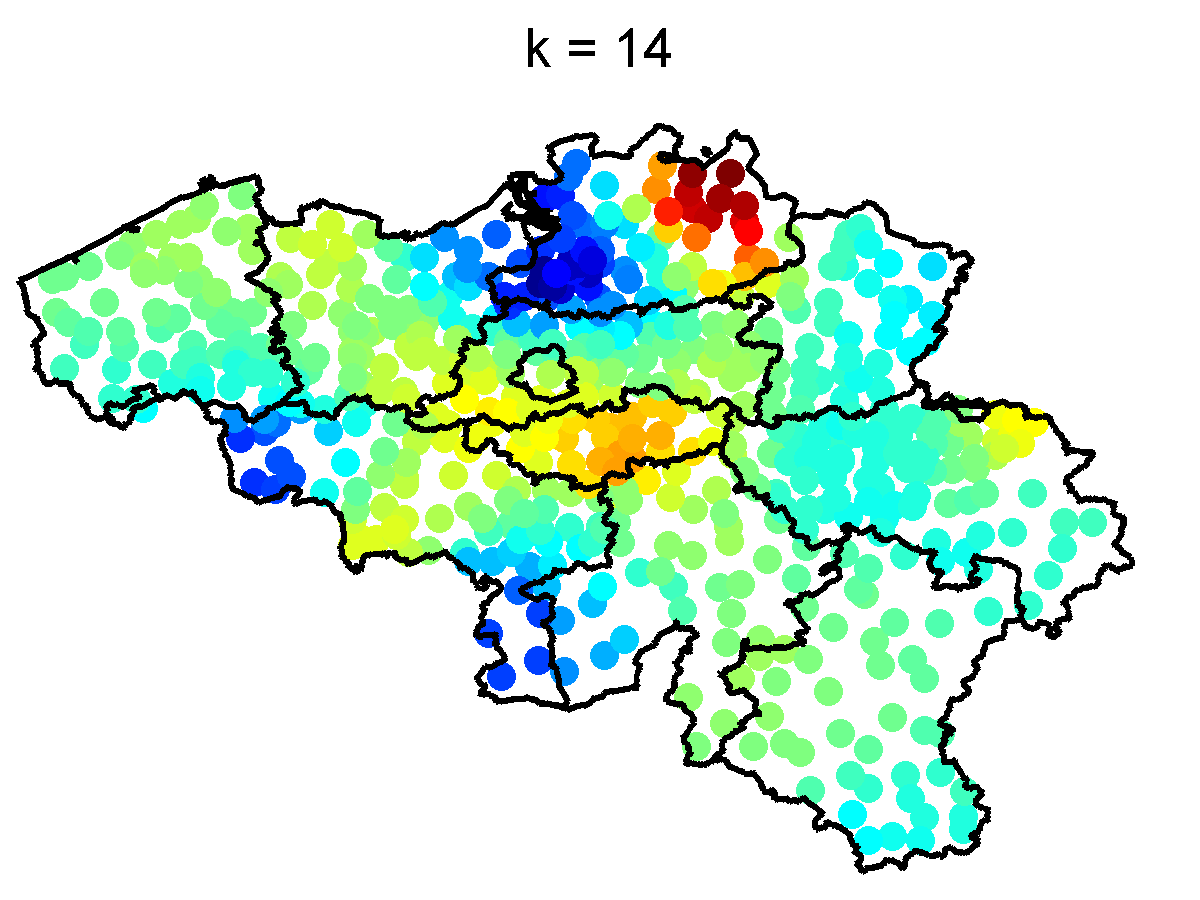}
 \includegraphics[width=0.32 \textwidth]{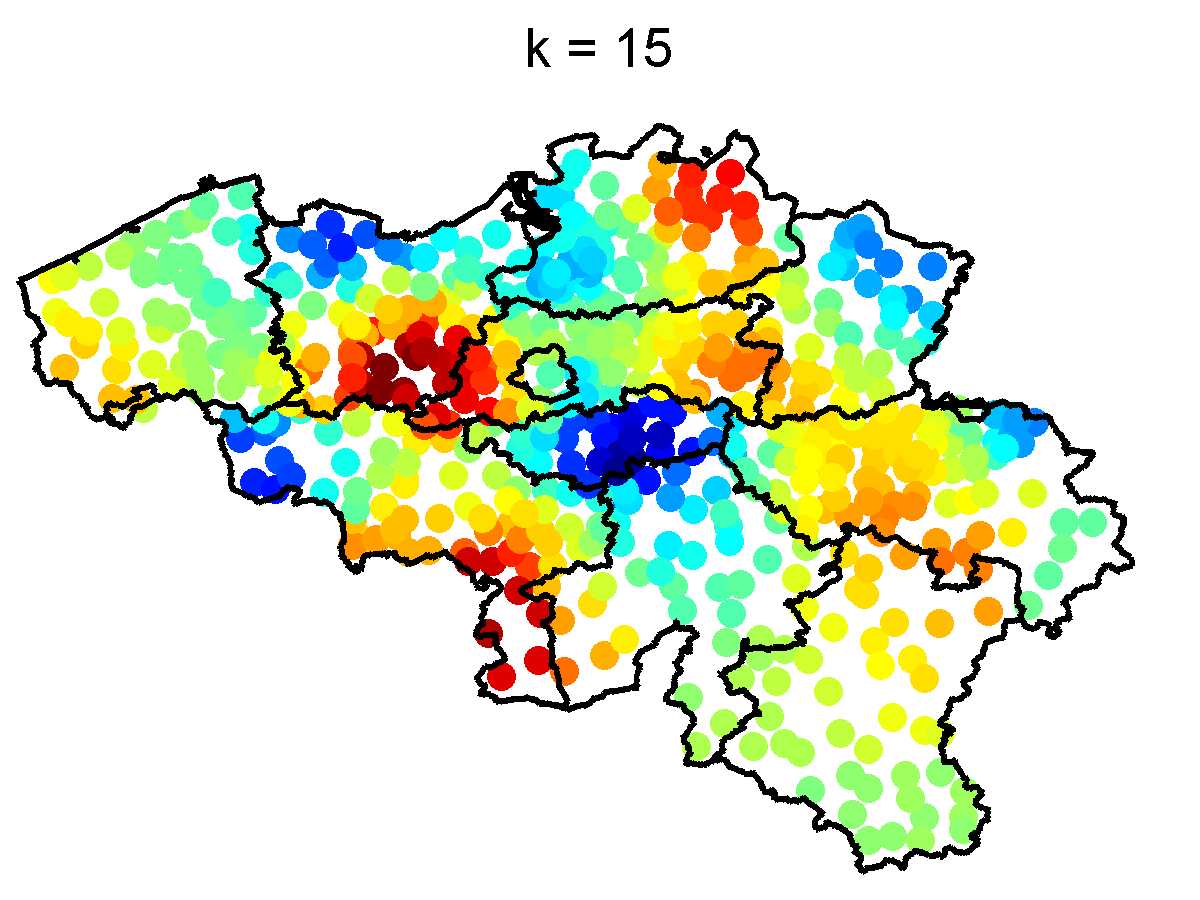}
\end{center}
\caption{Colorings by the top $18$ eigenvectors of $ A = D ^{-1} W^{(3)}$, where $W^{(3)}_{ij} = \frac{ \bar{T}_{ij} }{  R_{ij}  } = \frac{N_{ij}}{P_i P_j}$}
\label{fig:mob_K1_top18}
\end{figure}

Figure \ref{fig:mob_2D_dif} shows the diffusion map reconstructions for various  matrices $W$ that relate cities based on their communication intensities and population sizes. For $W^{(2)}$ and $W^{(3)}$, there is an obvious separation between the north and south parts of Belgium, which stems from the fact that the two regions belong to different linguistic  groups. The same separation is emphasized by the colorings associated to the top eigenvector of matrix $A$, shown in Figure \ref{fig:mob_K1_top18}. The remaining eigenvector colorings in Figure \ref{fig:mob_K1_top18} clearly isolate various subregions in Belgium. For example, eigenvectors $\psi_1$ and $ \psi_{11}$ highlight language communities (French, Dutch and German), while  $\psi_3$ and $ \psi_5$ isolate the regions of Li\a`ege  and Limburg.

\section{Localized eigenfunctions}
\label{localization}

Let us first make more precise what is meant by a localized eigenfunction. This phenomenon of localization occurs when there exist eigenfunctions supported by small regions of the domain, i.e. they are localized in these regions. An eigenfunction localized on a domain $\Omega_1$ has support on $\Omega_1$ significantly larger than on the complement $\Omega \backslash \Omega_1$, and yet it cannot vanish on $\Omega \backslash \Omega_1$ since eigenfunctions of isolated eigenvalues are real analytic functions and cannot vanish on any open set.
This is also observed in the histogram of the entries of the eigenvectors $\phi_{7},\phi_{28}$ and $\phi_{83}$ shown in Figure \ref{fig:US_eigV_hist} and the corresponding colorings in the figures \ref{fig:US_K1_top18} and \ref{fig:US_K1_next18}. In contrast, eigenfunctions that do not localize have their support ``uniformly" distributed across the domain, similar to the case of eigenvector $\phi_{1}$ from Figure \ref{fig:US_eigV_hist}. For example, in the case of the unit interval, the eigenfunctions of the Laplacian are the sine or cosine functions (depending on the boundary conditions) with the larger eigenvalues corresponding to higher oscillations, and they are not localized in the sense that there is no specific subinterval that carries the most (potential) energy of the eigenfunction, and any subinterval supports an amount of energy that is proportional to its length. In other words, the energy of the top eigenfunctions is distributed uniformly across the domain, and similar results are known to hold for the disk and the sphere, where the Laplacian eigenvalues and eigenfunctions are explicitly known.

The behavior observed in the eigenvector colorings from figures \ref{fig:US_K1_top18}, \ref{fig:US_K1_next18} and \ref{fig:mob_K1_top18} is
related to the notion of \textit{localized eigenfunctions}, a phenomenon observed before in the mathematics and physics community. The spectrum of the (continuous) Laplace operator has been extensively studied, and there exists a rich literature on the relationship between the spectrum and the geometry of the domain. As more complicated objects, eigenfunctions are more difficult to analyze than the spectrum, and less is known about them. Most of the literature is focused on high frequency eigenfunctions (associated to larger eigenvalues), such as \cite{BZ,geod,jms}, although recent studies such as \cite{bob} advocated localized eigenfunctions associated to small eigenvalues. In our experiments, we found the bottom eigenvectors uninteresting as they did not contain any meaningful geometric information.
In his work, Sapoval \cite{RSH} studied localized eigenfunctions in different domains and pointed out their importance for physical applications, such as designing efficient noise-protective walls.


Finally, considering that $A$ is a stochastic matrix, one may further explore ideas from the theory of nearly completely decomposable matrices developed in 1961 by Nobel laureate Herbert Simon and his collaborator Albert Ando  to describe and identify the short, medium and long-term behaviors of a dynamical system \cite{simon_ando1}. Very recent work  \cite{simon_ando2} explores this idea in the context of stochastic data clustering, and proposes a technique that uses  the evolution of the system to infer information on the initial structure.


\section{Summary and Discussion}
\label{summary}

We have shown how the diffusion map technique can be used to obtain informative visualizations and capture natural subdivisions within two different real networks. We find surprising that some low order eigenvectors  localize very well and seem to reveal small geographically cohesive regions; it is natural to ask for an explanation for our observation.

In looking at figures \ref{fig:US_K1_top18} and \ref{fig:US_K1_next18} many more questions come to mind. Are the state boundaries a consequence of people migrating within the same state or not? In other words, do states emerge as communities because of people migrating from one county to the other within the state, or because of similar migration patterns directed outside the state? Preliminary analysis on the migration data set in the context of local clustering on graphs supports the idea that the localized low-order eigenvectors highlight local cuts in the network. This is perhaps counter-intuitive since such low-order eigenvectors must satisfy the global requirement of exact orthogonality with respect to all of the earlier delocalized eigenvectors, and they must do so while keeping most of their components zero or close to zero.
Another question to consider is whether, besides the state boundary detection, the eigenvector colorings reveal any extra information on the intensity of the migration from one region to the other. Furthermore, inter-county migration is most common among young adults and declines as people age, and one may ask how the age composition (or income level) of individual US counties impacts the migration pattern.

In answering these questions, one needs to complement the mathematical description of diffusion maps and clustering by eigenvectors with a socio-demographic behavioral interpretation of migration trends, as considered for example in \cite{Lee, intmig}. A more recent paper by Slater \cite{hubsUS} is of particular interest since it analyzes migration patterns in the US Census data from 1965-1970 and 1995-2000. Amongst others, it highlights cosmopolitan or hub like regions, as well as isolated regions that emerge when there is a high measure of separation between a cluster and its environment.

Another interesting direction worth exploring is seeing how the diffusion map reconstructions and colorings change when the matrices used are no longer symmetric. In the case of the US migration data, it may be the case that there are many states for which the most common migration destination is the major city/capital of that state (although there might be other destinations spread across the US that attract people migrating out from that state). It is therefore natural to expect that major cities will stand out in the colorings, however this is not the case in our simulations since we symmetrize the migration matrix and take into account both the in- and out-migration from a given state.





\section{Acknowledgments}
We would like to thank Amit Singer for introducing M.C. to diffusion maps and suggesting this as a possible approach for analyzing the Belgium mobile data set; Etienne Huens and Gautier Krings for useful discussions and help with processing the same data set; Michael Mahoney, Mauro Maggioni, Peter Mucha, Mark Newman and Mason Porter for useful discussions on community detection and references to the literature on localized eigenfunctions; Thomas Espenshade, Kevin O'Neil, John Palmer, and Ashton Verdery for references and useful discussions on demographic trends in US migration.

\bibliographystyle{siam}
\bibliography{Belgium}

\begin{thebibliography}{10}

\bibitem{dif5}
{\sc M.~Belkin and P.~Niyogi}, {\em Laplacian eigenmaps for dimensionality
  reduction and data representation}, Neural Computation, 6 (2003),
  pp.~1373--1396.

\bibitem{unfolding}
{\sc V.~D. Blondel, J.-L. Guillaume, R.~Lambiotte, and E.~Lefebvre}, {\em Fast
  unfolding of communites in large networks}, Journal of Statistical Mechanics:
  Theory and Experiment, 1742-5468 (2008).
\newblock P10008.

\bibitem{BZ}
{\sc N.~Burq and M.~Zworski}, {\em Bouncing ball modes and quantum chaos}, SIAM
  Rev., 47 (2005), pp.~43--49.

\bibitem{geod}
{\sc V.~M.~B. c and V.~F. Lazutkin}, {\em The eigenfunctions which are
  concentrated near a closed geodesic}, Problems of Mathematical Physics,
  Spectral Theory, Diffraction Problems (Russian), 2 (1967), pp.~15--25.

\bibitem{census}
{\sc U.~S. {Census Bureau}}, 2002.
\newblock \url{www.census.gov/population/www.cen2000/ctytoctyflow/index.html}.

\bibitem{dif4}
{\sc R.~R. Coifman, I.~G. Kevrekidis, S.~Lafon, M.~Maggioni, and B.~Nadler},
  {\em Diffusion maps, reduction coordinates and low dimensional representation
  of stochastic systems}, SIAM Multiscale modeling and simulation, 7 (2008),
  pp.~842--864.

\bibitem{dif1}
{\sc R.~R. Coifman and S.~Lafon}, {\em Diffusion maps}, Appl Comput Harmonic
  Anal, 21 (2006), pp.~5--30.

\bibitem{dif3}
{\sc R.~R. Coifman, S.~Lafon, A.~B. Lee, M.~Maggioni, B.~Nadler, F.~Warner, and
  S.~W. Zucker}, {\em Geometric diffusions as a tool for harmonic analysis and
  structure definition of data: {D}iffusion maps}, PNAS, 102 (2005),
  pp.~7426--7431.

\bibitem{expert}
{\sc P.~Expert, T.~Evans, V.~Blondel, and R.~Lambiotte}, {\em Uncovering
  space-independent communities in spatial networks}, PNAS (Proceedings of the
  National Academy of Sciences), 108 (2011), pp.~7663--7668.

\bibitem{fortunato}
{\sc S.~Fortunato}, {\em Community detection in graphs}, Physics Reports, 486
  (2010), pp.~75--174.

\bibitem{bob}
{\sc S.~M. Heilman and R.~S. Strichartz}, {\em Localized eigenfunctions: Here
  you see them, there you don't}, Notices Amer. Math. Soc., 57 (2010),
  pp.~624--629.

\bibitem{jms}
{\sc P.~W. Jones, M.~Maggioni, and R.~Schul}, {\em Manifold parameterizations
  by eigenfunctions of the laplacian and heat kernels}, Proc. Natl. Acad. Sci.
  USA, 6 (2008), pp.~1803--1808.

\bibitem{gravity}
{\sc G.~Krings, F.~Calabrese, C.~Ratti, and V.~D. Blondel}, {\em Urban gravity:
  a model for inter-city telecommunication flows}, Journal of Statistical
  Mechanics: Theory and Experiment, L07003 (2009).

\bibitem{dif2}
{\sc S.~Lafon}, {\em Diffusion maps and geometric harmonics}, Ph.D. Thesis,
  (2004).

\bibitem{lamb}
{\sc R.~Lambiotte, V.~D. Blondel, C.~de~Kerchove, E.~Huens, C.~Prieur,
  Z.~Smoreda, and P.~V. Dooren}, {\em Geographical dispersal of mobile
  communication networks}, Physica A: Statistical Mechanics and its
  Applications, 387 (2008), pp.~5317--5325.

\bibitem{Lee}
{\sc E.~S. Lee}, {\em A theory of migration}, Population Association of
  America, Demography, 3 (1966), pp.~47--57.

\bibitem{intmig}
{\sc D.~S. Massey, J.~A.~G. Hugo, A.~Kouaouci, A.~Pellegrino, and J.~E.
  Taylor}, {\em Theories of international migration: A review and appraisal},
  Population and Development Review, 19 (1993), pp.~431--466.

\bibitem{simon_ando2}
{\sc C.~D. Meyer and C.~D. Wessell}, {\em Stochastic data clustering},  (2010).
\newblock submitted, arXiv:1008.1758.

\bibitem{dif6}
{\sc B.~Nadler, S.~Lafon, R.~Coifman, and I.~Kevrekidis}, {\em Diffusion maps,
  spectral clustering and eigenfunctions of {F}okker-{P}lanck operators},
  vol.~18, 2005.

\bibitem{Newman06}
{\sc M.~E.~J. Newman}, {\em Finding community structure in networks using the
  eigenvectors of matrices}, Physical Review E, 74 (2006).
\newblock 036104.

\bibitem{onnela}
{\sc J.~Onnela, S.~Arbesman, M.~C. Gonzalez, A.~L. Barabasi, and N.~A.
  Christakis}, {\em Geographic constraints on social network groups}, PLoS one,
  6 (2011).
\newblock e16939.

\bibitem{census_rep}
{\sc M.~J. Perry}, {\em State-to-{S}tate {M}igration {F}lows: 1995 to 2000},
  Census 2000 Special Reports,  (2003).

\bibitem{britain}
{\sc C.~Ratti, S.~Sobolevsky, F.~Calabrese, C.~Andris, J.~Reades, M.~Martino,
  R.~Claxton, and S.~H. Strogatz}, {\em Redrawing the map of {G}reat {B}ritain
  from a network of human interactions}, PLoS ONE, 5 (2010).

\bibitem{eigenplaces}
{\sc J.~Reades, F.~Calabrese, and C.~Ratti}, {\em Eigenplaces: analyzing cities
  using the space-time structure of the mobile phone network}, Environment and
  Planning B, 36 (2009), pp.~824--836.

\bibitem{tripart}
{\sc T.~Richardson, P.~J. Mucha, and M.~A. Porter}, {\em Spectral
  tripartitioning of networks}, Physical Review E, 80 (2009).
\newblock 036111.

\bibitem{RSH}
{\sc S.~Russ, B.~Sapoval, and O.~Haeberl\'e}, {\em Irregular and fractal
  resonators with {N}eumann boundary conditions: {D}ensity of states and
  localization}, Phys. Rev. E, 55 (1997), pp.~1413--1421.

\bibitem{shimalik}
{\sc J.~Shi and J.~Malik}, {\em Normalized cuts and image segmentation}, IEEE
  Transactions on Pattern Analysis and Machine Intelligence, 22 (2000),
  pp.~888--905.

\bibitem{simon_ando1}
{\sc H.~A. Simon and A.~Ando}, {\em Aggregation of variables in dynamic
  systems}, Econometrica, 29 (1961), pp.~111--138.

\bibitem{hubsUS}
{\sc P.~B. Slater}, {\em Hubs and clusters in the evolving united states
  internal migration network}, ISBER, University of California, Santa Barbara,
  (2008).
\newblock arXiv:0809.2768v3.

\bibitem{luxburg}
{\sc U.~von Luxburg}, {\em A tutorial on spectral clustering}, Statistics and
  Computing, Springer Netherlands, 17 (2007), pp.~395--416.
\newblock ISSN 0960-3174.

\end{thebibliography}

\end{document}